\documentclass[fleqn,usenatbib]{mnras}

\usepackage{newtxtext,newtxmath}

\usepackage[T1]{fontenc}
\usepackage{orcidlink}
\usepackage{color,soul}
\DeclareRobustCommand{\VAN}[3]{#2}
\let\VANthebibliography\thebibliography

\def\thebibliography{\DeclareRobustCommand{\VAN}[3]{##3}\VANthebibliography}

\usepackage{graphicx}	
\usepackage{amsmath}	
\usepackage{mathtools}

\title[A Rubin eye on AT\,2025agpz]{AT\,2025agpz in Rubin commissioning data: distinguishing a luminous interacting supernova from nuclear transients in compact galaxies}

\author[C.~R.~Angus et al.]{C.~R.~Angus$^{1}$\orcidlink{0000-0002-4269-7999}\thanks{E-mail: c.angus@qub.ac.uk (CRA)},
M.~Quilt$^{2}$\orcidlink{0009-0001-7000-7406},
H.~F.~Stevance$^{3}$\orcidlink{0000-0002-0504-4323},
M.~Nicholl$^{1}$\orcidlink{0000-0002-2555-3192},
S.~J.~Smartt$^{3,1}$\orcidlink{0000-0002-8229-1731},
P.~Wiseman$^{2}$\orcidlink{0000-0002-3073-1512},
A.~M\"oller$^{4,5}$\orcidlink{0000-0001-8211-8608},
\newauthor
C.~T.~Murphey$^{6,7}$\orcidlink{0009-0006-5214-0736},
P.~J.~Pessi$^{8}$\orcidlink{0000-0002-8041-8559},
K.~Auchettl$^{9,10}$\orcidlink{0000-0002-4449-9152},
M.~Dennefeld$^{11}$\orcidlink{0000-0002-8197-5410},
M.~Dominik$^{12}$\orcidlink{0000-0002-3202-0343},
C.~Frohmaier$^{13}$\orcidlink{0000-0001-9553-4723},
\newauthor
P.~Francis$^{14}$\orcidlink{0000-0001-9835-1404},
M.~Gromadzki$^{15}$\orcidlink{0000-0002-1650-1518},
A.~Lawrence$^{16}$\orcidlink{0000-0002-3134-6093},
G.~Leloudas$^{17}$\orcidlink{0000-0002-8597-0756},
C.~Lidman$^{18,19}$\orcidlink{0000-0003-1731-0497},
D.~Magill$^{1}$\orcidlink{0009-0000-6521-8842},
\newauthor
I.~Mandel$^{20,21}$\orcidlink{0000-0002-6134-8946},
B.~Martin$^{18}$\orcidlink{0009-0006-4963-3206},
S.~Mattila$^{22,23}$\orcidlink{0000-0001-7497-2994},
G.~Narayan$^{6,7,24}$\orcidlink{0000-0001-6022-0484},
F.~Onori$^{25}$\orcidlink{0000-0001-6286-1744},
S. R.~Oates$^{26}$\orcidlink{0000-0001-9309-7873},
\newauthor
H.~M.~L.~Perkins$^{6,7,27}$\orcidlink{0009-0000-5561-9116},
L.~Rauf$^{18,19}$\orcidlink{0000-0003-0824-669X},
A.~Rest$^{28,29}$\orcidlink{0000-0002-4410-5387},
S.~Romagnoli$^{9}$\orcidlink{0009-0003-8153-9576},
R.~Roy$^{30}$\orcidlink{0000-0002-9711-6207},
S.~Schulze$^{31}$\orcidlink{0000-0001-6797-1889},
X.~Sheng$^{1}$\orcidlink{0000-0002-6527-1368},
\newauthor
K.~W.~Smith$^{3,1}$\orcidlink{0000-0001-9535-3199},
K.~de Soto$^{32}$\orcidlink{0000-0002-9886-2834},
P.~M.~Veres$^{33}$\orcidlink{0000-0002-9553-2987},
M.~E. Verrico$^{6,7}$\orcidlink{0000-0003-1535-4277},
A.~Wasserman$^{6,7,24}$\orcidlink{0000-0003-1535-4277},
R.~Williams$^{16}$\orcidlink{ORCID 0009-0006-9214-4520},
\newauthor
{\L}.~Wyrzykowski$^{34,8}$\orcidlink{0000-0002-9658-6151} and
D.~R.~Young$^{1}$\orcidlink{0000-0002-1229-2499}
\\
$^{1}$ Astrophysics Research Centre, School of Mathematics and Physics, Queen’s University Belfast, Belfast BT7 1NN, UK\\
$^{2}$ School of Physics and Astronomy, University of Southampton, Southampton, SO17 1BJ, UK\\
$^{3}$ Department of Physics, University of Oxford, Keble Road, Oxford, OX1 3RH, UK \\
$^{4}$ Centre for Astrophysics \& Supercomputing, Swinburne University of Technology, John Street, Melbourne, 3122, Victoria, Australia \\
$^{5}$ ARC Centre of Excellence for Gravitational Wave Discovery (OzGrav), John Street, Hawthorn, 3122, Victoria, Australia \\
$^{6}$ Department of Astronomy, University of Illinois, Urbana-Champaign, Urbana, IL 61820, USA \\
$^{7}$ Center for AstroPhysical Surveys, National Center for Supercomputing Applications, Urbana, IL 61820, USA \\
$^{8}$ Astrophysics Division, National Centre for Nuclear Research, Pasteura 7, 02-093 Warsaw, Poland \\
$^{9}$ OzGrav, School of Physics, The University of Melbourne, Parkville, VIC, Australia \\
$^{10}$ Department of Astronomy and Astrophysics, University of California, Santa Cruz, CA, USA \\
$^{11}$ Sorbonne University \& CNRS, Institut d’Astrophysique de Paris (IAP), F-75014 Paris, France \\
$^{12}$ Scottish Universities Physics Alliance, University of St Andrews, School of Physics and Astronomy, North Haugh, KY16 9SS, St Andrews, UK \\
$^{13}$ Institute of Cosmology and Gravitation, University of Portsmouth, Burnaby Rd, Portsmouth PO1 3FX, UK \\
$^{14}$ EPCC, University of Edinburgh, The Bayes Centre, 47 Potterrow, Edinburgh, EH8 9BT, UK \\
$^{15}$ Astronomical Observatory, University of Warsaw, Al. Ujazdowskie 4, 00-478 Warszawa, Poland   \\
$^{16}$ Institute for Astronomy, University of Edinburgh, Royal Observatory, Blackford Hill, Edinburgh EH9 3HJ, UK \\  
$^{17}$ DTU Space, Department of Space Research and Space Technology, Technical University of Denmark, Elektrovej 327, DK-2800 Kgs. Lyngby, Denmark \\   
$^{18}$ The Research School of Astronomy and Astrophysics, Mt Stromlo Observatory, The Australian National University, Canberra, ACT 2611, Australia \\
$^{19}$ Centre for Gravitational Astrophysics, College of Science, The Australian National University, ACT 2601, Australia \\   
$^{20}$ School of Physics and Astronomy, Monash University, Clayton VIC 3800, Australia \\
$^{21}$ ARC centre of Excellence for Gravitational Wave Discovery (OzGrav), Clayton VIC 3800, Australia \\
$^{22}$ Tuorla Observatory, Department of Physics and Astronomy, University of Turku, 20014 Turku, Finland \\
$^{23}$ School of Sciences, European University Cyprus, Diogenes Street, Engomi, 1516 Nicosia, Cyprus \\
$^{24}$ NSF-Simons SkAI Institute, 875 North Michigan Avenue, Chicago, IL 60611, USA \\
$^{25}$ INAF Osservatorio Astronomico di Roma, via Frascati 33. 00078, Monte Porzio Catone, Rome, Italy \\
$^{26}$ School of Physics and Astronomy, Lancaster University, Lancashire, LA1 4YB, UK \\
$^{27}$ Illinois Center for Advanced Studies of the Universe, Urbana, IL 61820, USA \\
$^{28}$ Space Telescope Science Institute, 3700 San Martin Drive, Baltimore, MD 21218, USA \\
$^{29}$ Department of Physics and Astronomy, The Johns Hopkins University, Baltimore, MD 21218, USA \\
$^{30}$ Institute of Astronomy Space and Earth Science (IASES), P 177, CIT Road, Scheme 7m, Kolkata-700054, West Bengal, India \\
$^{31}$ Center for Interdisciplinary Exploration and Research in Astrophysics (CIERA), Northwestern University, 1800 Sherman Avenue, Evanston, IL 60201, USA \\
$^{32}$ Harvard-Smithsonian Center for Astrophysics, 60 Garden Street, Cambridge, MA 02138, USA \\
$^{33}$ Ruhr University Bochum, Faculty of Physics and Astronomy, Astronomical Institute (AIRUB), Universit\"atsstra\ss e 150, 44801 Bochum, Germany\\
$^{34}$ Astronomical Observatory, University of Warsaw, Al. Ujazdowskie 4, 00-478 Warsaw, Poland \\
}
\date{Accepted XXX. Received YYY; in original form ZZZ}

\pubyear{\the\year{}}

\begin{document}
\label{firstpage}
\pagerange{\pageref{firstpage}--\pageref{lastpage}}
\maketitle
\clearpage

\begin{abstract}
The Vera C. Rubin Observatory's Legacy Survey of Space and Time (LSST) will discover unprecedented numbers of rare, long-lived optical transients, which will be too numerous and many too faint for comprehensive spectroscopic follow-up. As a result, we should make efforts to characterise objects whose observational properties span the boundaries between established transient classes to improve photometric classification. We present AT\,2025agpz, a luminous transient at a redshift of $z=0.147$, discovered around peak by ATLAS, but with extensive monitoring from Rubin data during commissioning observations of the Euclid Deep Field South, including deep detections within days of explosion, which exemplifies this challenge. AT\,2025apgz has a projected location coincident with the nucleus of a faint dwarf galaxy, exhibits a long rest-frame rise time of $77.9\pm1.3$\,d, and reaches a peak bolometric luminosity of $6.3\times10^{43}$\,erg\,s$^{-1}$. Follow-up low-resolution spectra revealed a remarkably slow spectroscopic evolution dominated by narrow Balmer emission. The depth and cadence of the Rubin commissioning observations in this Deep Drilling Field, combined with complementary imaging from DECam, tightly constrain the explosion epoch and yield an early-time power-law rise of $n=3.33^{+0.33}_{-0.28}$. Higher-resolution X-shooter spectroscopy reveals broad electron-scattering wings and multiple H$\alpha$ emission components characteristic of interaction-powered supernovae, while spectral energy distribution modelling indicates a low-mass, moderately star-forming host galaxy typical of superluminous supernovae. Although AT\,2025agpz occupies observational parameter space shared by the recently identified Ambiguous Nuclear Transient population, its colour evolution, host-galaxy environment, and line-profile morphology favour an interaction-powered luminous supernova interpretation. More broadly, this event demonstrates the growing observational overlap between luminous interacting supernovae and nuclear transients, illustrating the challenges and opportunities for transient classification in the Rubin era.
\end{abstract}




\section{Introduction}

The field of time-domain astrophysics has undergone a profound transformation over the past two decades with the advent of untargeted, wide-field imaging surveys. These facilities have dramatically expanded the accessible parameter space in luminosity, duration, and cadence, leading to the discovery of numerous classes of rare and extreme transients that challenge traditional models of stellar evolution and accretion-powered variability. Among these are superluminous supernovae (SLSNe), whose peak luminosities exceed those of canonical core-collapse supernovae by more than an order of magnitude \citep{gal-yamLuminousSupernovae2012,quimbyHydrogenpoorSuperluminousStellar2011,inserraSuperluminousTypeIc2013}; fast blue optical transients (FBOTs), characterised by their rapid evolution and high luminosities \citep{droutRapidlyEvolvingLuminous2014,pursiainenRapidlyEvolvingTransients2018,hoSearchExtragalacticFast2023}; and an increasingly diverse population of nuclear transients, including tidal disruption events \citep[TDEs;][]{gezariTidalDisruptionEvents2021a,hammersteinFinalSeasonReimagined2023a,charalampopoulosDetailedSpectroscopicStudy2022,somalwarVLASSTidalDisruption2025,grotovaPopulationTidalDisruption2025}, large-amplitude variability in active galactic nuclei \citep[AGN;][]{grahamUnderstandingExtremeQuasar2017,rumbaughExtremeVariabilityQuasars2018,trakhtenbrotNewClassFlares2019,trakhtenbrot1ES1927+654AGN2019,frederickFamilyTreeOptical2021,ricciChanginglookActiveGalactic2022,dennefeldAGNGaiaAlerts2026}, and the recently recognised class of Ambiguous Nuclear Transients \citep[ANTs;][]{kankarePopulationHighlyEnergetic2017,holoienInvestigatingNatureLuminous2022,petrushevskaRiseFallIronstrong2023,oatesSwiftUVOTDiscovery2024,wisemanSystematicallySelectedSample2025}. The most luminous members of this population have recently been termed Extreme Nuclear Transients \citep[ENTs;][]{grahamUnderstandingExtremeQuasar2017,hinkleMostEnergeticTransients2025}, exemplified by the prototype AT\,2021lwx \citep{wisemanMultiwavelengthObservationsExtraordinary2023,subrayanScaryBarbieExtremely2024}.

Many of the slower-evolving members of these populations are intrinsically rare \citetext{$\sim10^{-8}$--$10^{-7}$\,Mpc$^{-3}$\,yr$^{-1}$ for SLSNe: \citealp{quimbyRatesSuperluminousSupernovae2013,frohmaierCoreCollapseSuperluminous2021}}; $\sim3\times10^{-7}$\,Mpc$^{-3}$\,yr$^{-1}$ for TDEs \citep{yaoTidalDisruptionEvent2023}, and $\gtrsim3\times10^{-11}$\,Mpc$^{-3}$\,yr$^{-1}$ for ANTs  \citep{wisemanSystematicallySelectedSample2025}. Combined with the need to identify these objects over their characteristically long rise times, early discoveries were typically limited to isolated, often enigmatic events, making it difficult to determine whether they represented outlying extremes of known transient populations (e.g. the TDE interpretation of AT\,2021lwx; \citealt{subrayanScaryBarbieExtremely2024}) or members of genuinely new classes. However, deeper optical surveys such as the Zwicky Transient Facility \citep[ZTF;][]{bellmZwickyTransientFacility2019}, the Panoramic Survey Telescope and Rapid Response System \citep[Pan-STARRS;][]{huberPanSTARRSSurveyTransients2015}, and the Dark Energy Survey Supernova Programme \citep[DES;][]{bernsteinSupernovaSimulationsStrategies2012,smithFirstCosmologyResults2020} have substantially increased the discovery rate of these populations, gradually enabling statistical studies of their photometric, spectroscopic, and host-galaxy properties. As time-domain surveys continue to increase in depth, cadence, and sky coverage, they are revealing both the intrinsic diversity within established transient classes and a growing population of objects that blur the boundaries between traditional classifications.

This evolution is particularly well illustrated by SLSNe. Originally identified as a small population of exceptionally luminous explosions with $M_{V}<-21$\,mag \citep{gal-yamLuminousSupernovae2012}, the class has gradually evolved from a purely luminosity-based definition to one motivated primarily by spectroscopic properties. Broadly, SLSNe are divided into hydrogen-poor (SLSN-I) and hydrogen-rich (SLSN-II) subclasses. SLSN-I events exhibit hot, blue continua at early times that evolve to resemble stripped-envelope Type Ic supernovae several weeks after maximum light \citep{pastorelloUltrabrightOpticalTransients2010,quimbyHydrogenpoorSuperluminousStellar2011}, whereas SLSNe-II display prominent hydrogen features \citep{inserraNatureHydrogenrichSuperluminous2018,kangasZwickyTransientFacility2022}. Despite their spectroscopic diversity, both subclasses are widely thought to originate from the explosions of massive stars, owing to their strong preference for low-mass, star-forming host galaxies with sub-solar metallicities \citep{neillExtremeHostsExtreme2011,lunnanHydrogenpoorSuperluminousSupernovae2014,angusHubbleSpaceTelescope2016,schulzeCosmicEvolutionMetal2018,taggartCorecollapseSuperluminousGammaray2021,leloudasSpectroscopySuperluminousSupernova2015,clelandMetallicityBeatsSSFR2023}.

As uniformly selected samples have grown, however, the observed luminosity function \citep{deciaLightCurvesHydrogenpoor2018,lunnanHydrogenpoorSuperluminousSupernovae2018,angusSuperluminousSupernovaeDark2019,gomezLuminousSupernovaeUnveiling2022,pessiSampleHydrogenrichSuperluminous2025}, spectroscopic diversity \citep{inserraNatureHydrogenrichSuperluminous2018,kangasZwickyTransientFacility2022,aamerTypeSuperluminousSupernova2025,nichollSuperluminousSupernovaeDiverse2026}, and photometric evolution \citep{gomezTypeSuperluminousSupernova2024,pessiSampleHydrogenrichSuperluminous2025} have expanded considerably beyond those represented by the original discovery samples. Increasingly, both hydrogen-poor and hydrogen-rich SLSNe appear to form a continuum with more ordinary core-collapse supernovae rather than representing physically distinct populations. For SLSNe-I, this is reflected in their smooth spectroscopic evolution towards Type Ic supernovae \citep{pastorelloUltrabrightOpticalTransients2010,aamerTypeSuperluminousSupernova2025}. Recent work further suggests that much of the observed spectroscopic diversity within the SLSN-I population is driven by differences in photospheric temperature at maximum light, rather than by fundamentally distinct explosion channels \citep{nichollSuperluminousSupernovaeDiverse2026}. Likewise, hydrogen-rich events show no compelling luminosity gap with ordinary SNe IIn, suggesting that the term `superluminous’ is primarily an observational description rather than a physically distinct subclass \citep{hiramatsuTypeIInSupernovae2026a}.

This ambiguity is particularly important for narrow-line SLSNe-II (hereafter SLSNe-IIn). Their spectra are dominated by narrow, often Lorentzian Balmer emission that is largely indistinguishable from ordinary SNe IIn, differing primarily in their luminosities and light-curve timescales. In the simplest interpretation, this continuum reflects differences in the properties of the circumstellar material and the efficiency with which kinetic energy is converted into radiation \citep[e.g.][]{smithInteractingSupernovaeTypes2017,pessiSampleHydrogenrichSuperluminous2025}. However, some hydrogen-rich SLSNe instead develop broad Balmer features and have been suggested to share a central-engine power source with SLSNe-I \citep{gezariDiscoveryUltraBrightType2009,inserraNatureHydrogenrichSuperluminous2018,kangasZwickyTransientFacility2022,kangasSN2023gpwExploring2026}, illustrating the growing diversity within the hydrogen-rich population.

While circumstellar interaction accounts for both the spectra and luminosity evolution of SLSNe-IIn, it also obscures the nature of the underlying explosion, making it unclear whether SLSNe-IIn simply represent the luminous extension of the normal Type IIn population or if some have a different explosion mechanism, as has been proposed for events such as SN\,2006gy \citep{jerkstrandTypeIaSupernova2020}. In the former case, if purely interaction driven, the extreme luminosities imply mass-loss rates well beyond those typically inferred for ordinary interacting supernovae, potentially requiring unusually powerful stellar winds \citep[e.g.][]{lucyMassLossHot1970,smithMassLossIts2014}, eruptive mass loss associated with pulsational pair instability \citep[e.g.][]{woosleyPulsationalPairinstabilitySupernovae2017}, or extensive binary mass transfer prior to explosion \citep[e.g.][]{yoonTypeIbIIb2017,laplaceExpansionStrippedenvelopeStars2020}.

The observational properties of SLSNe-IIn increasingly overlap with those of several classes of nuclear transient. Long-duration flares from AGN \citep{trakhtenbrotNewClassFlares2019,ridleyTimevaryingDoublepeakedEmission2024,sanchez-saezSDSS1335+0728Awakening1062024}, some tidal disruption events \citep[e.g. AT\,2019dsg;][]{cannizzaroAccretionDiscCooling2021a}, and the recently identified populations of ANTs and ENTs can all exhibit blue continua together with strong, often Lorentzian Balmer emission \citep{wisemanSystematicallySelectedSample2025,hinkleMostEnergeticTransients2025}, producing remarkably similar optical spectra. This overlap is compounded by the observational circumstances under which many SLSNe-IIn are discovered. Their preference for faint dwarf host galaxies, combined with the modest angular resolution of wide-field transient surveys and the typically larger distances at which they are observed, means that genuinely offset explosions can appear consistent with the nuclei of their host galaxies. Indeed, of the 118 hydrogen-rich SLSNe presented by \citet{pessiSampleHydrogenrichSuperluminous2025}, eleven were classified as having ambiguous nuclear locations, precluding a secure distinction between an interacting supernova and a nuclear flare. More broadly, several transients initially classified as interacting supernovae have only been reinterpreted as accretion-powered nuclear transients following continued photometric and spectroscopic monitoring \citep[e.g.][]{yaoTidalDisruptionEvent2023,pessiAmbiguousAT2022rzeChanginglook2025}.

The emergence of the ANT population has brought this overlap into sharper focus. Initially identified as long-duration, smoothly evolving nuclear flares with exceptionally large radiated energies \citep{neustadtTDENotTDE2020,hinkleCuriousCaseASASSN20hx2022,holoienInvestigatingNatureLuminous2022,petrushevskaRiseFallIronstrong2023,oatesSwiftUVOTDiscovery2024,wisemanMultiwavelengthObservationsExtraordinary2023,subrayanScaryBarbieExtremely2024,hinkleMostEnergeticTransients2025}, subsequent studies have demonstrated that these events span a much broader range of luminosities ($-20<M_r<-25$), evolutionary timescales, and colour evolution than first appreciated \citep{wisemanSystematicallySelectedSample2025}. Their physical origin likewise remains uncertain, with proposed explanations including changes to underlying AGN accretion \citep{frederickNewClassChanginglook2019}, extreme tidal disruption events \citetext{\citealp{subrayanScaryBarbieExtremely2024}, sometimes occurring within pre-existing accretion discs \citealp[e.g.][]{frederickFamilyTreeOptical2021,petrushevskaRiseFallIronstrong2023}}, and episodes of super-Eddington accretion following the tidal disruption of dense gas clouds \citep{wisemanMultiwavelengthObservationsExtraordinary2023}.

Several recent events have highlighted the difficulty of assigning unique classifications. This issue is best illustrated in the transient ASASSN-15lh, whose extreme luminosity ($L\sim10^{45}$\,erg\,s$^{-1}$) has been interpreted as originating from both a hydrogen-poor SLSN \citep{dongASASSN15lhHighlySuperluminous2016}, and from the tidal disruption of a star by a Kerr black hole \citep{leloudasSuperluminousTransientASASSN15lh2016a,kruhlerSupermassiveBlackHole2018}. Similar examples exist for hydrogen rich events; AT\,2019fdr has been interpreted as a tidal disruption event, an interacting SLSN, and an AGN flare by different authors \citep{frederickFamilyTreeOptical2021,reuschCandidateTidalDisruption2022,pitikHighenergyNeutrinoEvent2022}, while luminous interacting supernovae such as SN\,2016aps \citep{nichollExtremelyEnergeticSupernova2020} and  SN\,2016gsd \citep{reynoldsSN2016gsdUnusually2020} have further expanded the observed parameter space occupied by hydrogen-rich stellar explosions. Collectively, these discoveries suggest that current classification schemes do not necessarily map cleanly onto distinct physical progenitor channels. Even locations of transients within their apparent host galaxies can no longer be used to confidently discriminate between transient classes, with increasing numbers of \lq off-nuclear\rq\, TDEs now being identified in optical surveys \citep{yaoMassiveBlackHole2025,steinTDE2025abcrTidal2026,patraJWSTKeckObservations2026}, thought to originate from wandering supermassive black holes. 

These challenges become particularly acute in the Rubin era. The Vera C. Rubin Observatory's ten-year Legacy Survey of Space and Time \citep[LSST;][]{ivezicLSSTScienceDrivers2019a} will discover many thousands of transients each night, increasing sample sizes by orders of magnitude while simultaneously outstripping the available spectroscopic follow-up resources. Consequently, photometric classification and prioritisation will become increasingly important for constructing statistically robust samples of rare transients. Further to this, the depth of its imaging makes it particularly sensitive to discovering more distant, high-redshift transients, whose angular offsets from their host nuclei will be small, increasing the number of apparently nuclear transients. Establishing reliable observational diagnostics capable of distinguishing interacting supernovae from nuclear transients is therefore essential for constraining the progenitor populations of both classes while minimising sample contamination. Borderline events provide a valuable opportunity to test these diagnostics, exposing the limitations of existing classification frameworks and informing the development of future machine-learning classifiers and follow-up strategies.

Here we present one such object: AT\,2025agpz, a long-rising luminous transient discovered during Rubin commissioning whose nuclear location, photometric evolution, and spectroscopic properties place it at the boundary between the SLSN-IIn and ANT populations. Using densely sampled Rubin commissioning observations together with complementary photometric and spectroscopic follow-up, we investigate its nature and assess the evidence for both interpretations, showing that an interacting SN is the more likely. In doing so, we explore the observational overlap between interacting supernovae and luminous nuclear transients, and consider the implications for transient classification in the Rubin era.

The paper is organized as follows. In Section~\ref{sec:discovery} we describe the discovery of AT\,2025agpz, while the photometric and spectroscopic observations are presented in Section~\ref{sec:data}. The transient and its host galaxy are characterised in Section~\ref{sec:properties}, before we discuss the implications of the event in the context of current SLSN-IIn and ANT populations in Section~\ref{sec:discussion}. Our conclusions are presented in Section~\ref{sec:conclusions}. Throughout this paper we assume a standard $\Lambda$CDM cosmology with $H_0=70$\,km\,s$^{-1}$\,Mpc$^{-1}$, $\Omega_{\rm M}=0.3$, and $\Omega_\Lambda=0.7$.

\section{Discovery and Rubin early detections}\label{sec:discovery}

AT\,2025agpz (ATLAS25pny) was discovered by the Asteroid Terrestrial-impact Last Alert System \citep[ATLAS;][]{tonryATLASHighcadenceAllsky2018} in the $c$ band on 2025 December 10.75 UT (MJD 61\,019.90) at coordinates $\alpha=04^{\rm h}\,07^{\rm m}\,51.557^{\rm s}$, $\delta=-48^{\circ}\,42'\,48.40''$ (J2000) 
and magnitude $c=19.74\pm0.19$
\citep{smithDesignOperationATLAS2020,at2025agpzTNSdiscreport}. 

The Rubin Observatory \citep{ivezicLSSTScienceDrivers2019a} was undergoing its commissioning period and observing a number of selected sky areas repeatedly in late 2025.  The transient is located within the Euclid Deep Field South (EDFS), one of the Rubin Observatory LSST Deep Drilling Fields \citep{biancoRubinObservingCadence2022,jones_2025_15128504}, which was observed extensively during Rubin commissioning between 2025 November 24 and 2026 March 08 (MJD 61\,003–61\,107). Observations of EDFS began again on 
2026 June 26 (MJD 61\,217), as the field reappeared from solar conjunction and the LSST officially began its 10\,yr survey. 

Transient alerts from the Rubin commissioning observations were ingested into the Lasair alert broker\footnote{https://lasair.lsst.ac.uk/} \citep{williamsEnablingScienceRubin2024} throughout commissioning and science verification. AT\,2025agpz was subsequently identified
(as
LSST-P-DO-313761043604045880\footnote{https://lasair.lsst.ac.uk/objects/313761043604045880/})
during testing of the Lasair Virtual Research Assistant \citep{lvra2026}, which builds upon the methodology presented by \citet{stevanceATLASVirtualResearch2025}.
The ranking algorithm was trained on 411 events in the commissioning data, chosen using Active Learning method and labelled by visual examination. 
Upon testing the top ranked new alerts on 2026-01-17 it was recognised as a long rising transient; inspection of the commissioning data revealed that the transient had been detected significantly earlier in the 
deep Rubin difference imaging photometry than its ATLAS discovery, with the first $r$-band detection occurring on 2025 November 24.34 (MJD 61\,003.34) at $r = 23.76 \pm 0.18$. This event was also independently identified by \cite{perez-fournon2025agpzATLAS25pnyLSSTPDO3137610436040458802026} and the Fink broker \citep{mollerFinkNewGeneration2021}, who reported it as part of the first public alerts to the Transient Name Server (TNS).

The EDFS was also monitored contemporaneously with the Dark Energy Camera (DECam) in the $g$, $r$, and $i$ bands as part of the DECam survey of the Young Supernova Experiment Collaboration (YSE DECam; \citealt{yse-decam}; PropID 2025A-388357), beginning on 2025 September 01 with an approximately three-night cadence in \textit{g} and six-night cadence in \textit{r} and \textit{i}. No source is detected at the transient position in the DECam images obtained between MJD 60\,919 and MJD 60\,997. The first DECam detection is recorded on MJD 61\,003.19. This happened to be the same Chilean night as the first LSST detection, again  preceding the public ATLAS discovery by almost three weeks, with a measured brightness of $g = 23.51 \pm 0.18$. We report all transient phases in the rest-frame of the event relative to this earliest discovery epoch. The near-simultaneous initial detections in both DECam and Rubin commissioning data coupled with its deep prior non-detections place strong constraints upon the rise time of AT\,2025agpz, which we explore further in Section \ref{sec:properties}.

A classification spectrum taken with the FOcal Reducer and low dispersion Spectrograph 2 \citep[FORS2;][]{appenzellerSuccessfulCommissioningFORS11998} mounted on Unit Telescope 1 (UT1) of the ESO Very Large Telescope (VLT) was obtained on 2026 January 20 \citep{angusSpectroscopicObservations2025agpz2026}, which revealed narrow Balmer emission lines at a redshift of $z=0.147$, corresponding to a luminosity distance of 697.6 Mpc under our assumed cosmology.

\subsection{Astrometric Location}\label{sec:astrometry}
From the Rubin alerts, the transient is spatially associated by the Lasair Sherlock contextual cross-matching algorithm \citep{youngSherlockContextualClassification2023} with a faint  ($m_{r}=23.75$) galaxy identified in the DESI Legacy Imaging Surveys catalogue, with an angular separation of only a few tenths of an arcsecond. This association motivated its initial classification as a candidate nuclear transient. 

\begin{figure}
    \centering
    \includegraphics[width=\linewidth]{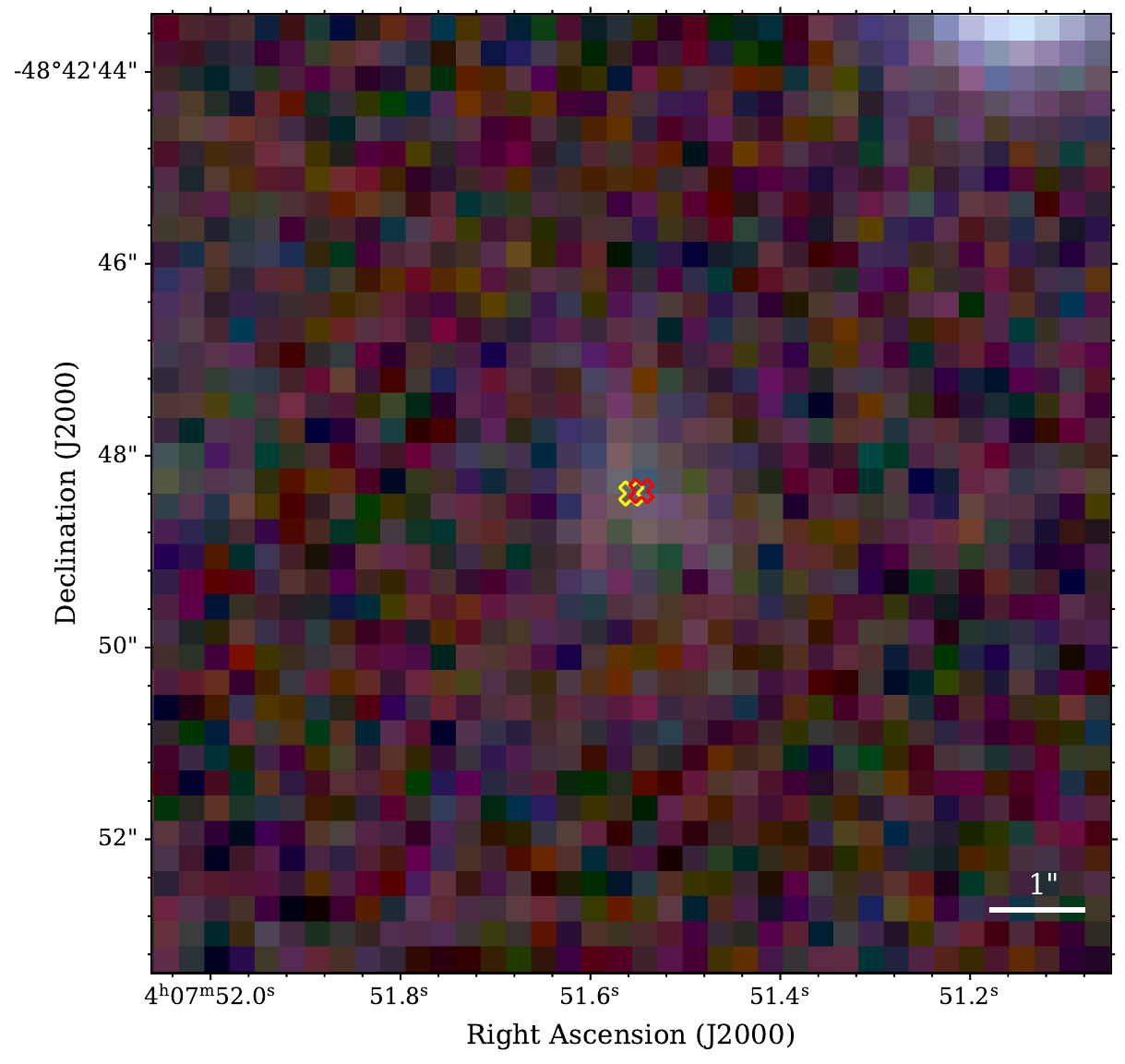}
    \caption{DESI Legacy Survey $grz$ colour-composite image of the host galaxy of AT\,2025agpz. The yellow cross marks the centroid of the host galaxy determined from the Legacy Survey imaging, while the red cross shows the astrometric position of the transient measured from LCO follow-up imaging and transformed onto the reference frame. The measured offset of $0.16\pm0.19$ arcsec ($0.4\pm0.5$\,kpc projected) is consistent with the transient originating from the nucleus of its host galaxy. }
    \label{fig:hostastrometry}
\end{figure}

To determine the location of AT 2025agpz within its host, we perform astrometry using pre-explosion imaging from the DESI Legacy Survey DR10 \citep{deyOverviewDESILegacy2019}. As we do not yet have access to the full LSST frame, we perform astrometry using $r$-band follow-up imaging obtained on MJD 61\,124 with the Las Cumbres Observatory (LCO; see Section~\ref{sec:data} for details of the observations and data reduction). We use routine {\tt{iraf}} tasks to determine the coordinates of cross-matched point sources in both transient and $r$-band Legacy survey reference images, then map and transform these coordinates to find the transient location in the reference image. We measure the location of AT\,2025agpz to be at $\alpha=04^{\rm h}\,07^{\rm m}\,51.53^{\rm s}$,
$\delta=-48^{\circ}\,42'\,48.3''$ ($\pm0.19''$). The location corresponds to a nuclear separation of 0.16$\pm$0.19'' from the host galaxy centre (or a projected separation of 0.4$\pm$0.5\,kpc), broadly consistent with the transient originating from the nucleus. This position is also coincident to 0.26'' of the quoted source position on Lasair, which averages the transient location across all detections. We indicate the location of AT\,2025agpz relative to the nucleus of the host galaxy in Figure \ref{fig:hostastrometry}.

\section{Data}\label{sec:data}
\subsection{Photometry}

We retrieve $c$ and $o-$band photometry from the ATLAS forced photometry server \citep{smithDesignOperationATLAS2020}, binning the photometry at nightly intervals and report the values in AB mags in the ATLAS system \citep{tonryATLASHighcadenceAllsky2018,TonryRefCat2018}. 
The difference photometry of SN\,2025agpz started in the Rubin alert data on MJD 61\,003.34 (2025-11-24), but no forced photometry is available before 61\,088.10 (2026-02-17) as forced photometry data from the Rubin commissioning data only became available in mid-February.  
All Rubin data were obtained directly from the alert stream produced by the LSST Science Pipelines \citep{10.71929/rubin/2570545} ingested into Lasair. The difference image analysis on LSSTCam data is calibrated with `The Monster' reference catalogue \citep{MonsterRefCatFerguson2025} which is based on Gaia DR3 for astrometry and a combination of DES Y6, PS1, SkyMapper and the VST ATLAS survey for photometric calibration. In this paper we use the LSST measured flux values of \texttt{psfFlux}$\pm$\texttt{psfFluxerr} for each \texttt{DIASource} detected on each night, averaged in nightly bins and converted to AB mags. No forced photometry is used in the analysis and  no additional photometric calibrations are applied by Lasair. The photometry we present is based on the difference images created by the Rubin Alert Production pipeline \citep{10.71929/rubin/2570545}. 

Comparison with contemporaneous DECam photometry reveals an approximately constant systematic offset between the commissioning Rubin and DECam photometry of $\Delta g = -0.20$ mag and $\Delta i = -0.25$ mag (in the sense that the Rubin photometry is fainter), while the $r$-band measurements remain consistent within the uncertainties. The origin of this offset is presently unclear, as the commissioning reference images used for the Rubin difference imaging are not yet publicly available. It may arise from residual transient flux in the reference images, a commissioning calibration offset, or a combination of these effects. We therefore apply constant corrections of +0.20 mag in $g$ and +0.25 mag in $i$ to align the Rubin commissioning photometry with DECam for the scientific analysis presented in this work. Future Rubin data releases, including the reference images used for the alert production, should enable the origin of this offset to be determined.

We perform reduction, image differencing, and forced photometry of the DECam images using {\tt{photpipe}} \citep{restTestingLMCMicrolensing2005,RestPS1cosmology}. We take our templates from single exposures taken on 2025 September 25 (\textit{g}), 2025 September 16 (\textit{r}), and 2025 September 12 (\textit{iz}). {\tt Photpipe} handles the data reduction, calibrates the EDFS fields against the SkyMapper stellar catalog \citep{skymapper}, performs A\&L difference imaging \citep{Alard1998} via HOTPANTS\footnote{\url{https://github.com/acbecker/hotpants}}, and lastly forced photometry via DOPHOT \citep{dophot}. In total, there were 139 exposures of the region around AT 2025agpz between September 2025 and April 2026, with 72 total detections starting November 24.

One epoch of LCO photometry in the $g,r,i$ bands was obtained as part of public photometric follow-up of time-domain phenomena by BHTOM \citep{wyrzykowskiPowerManyBHTOM2024,mikolajczykBlackHoleTOM2025} on MJD 61\,097 (+82\,d from first detection). These data were reduced with the BHTOM pipeline, calibrated using standard software packages included in CCD-Phot \citep{zielinskiAutomaticProcessingCCD2020}: {\tt{SExtractor}} \citep{bertinSExtractorSoftwareSource1996}, Software for Calibrating AstroMetry and Photometry \citep[{\tt{SCAMP}};][]{bertinAutomaticAstrometricPhotometric2006}, Dominion Astrophysical Observatory Photometry \citep[{\tt{DAOPHOT II}}][]{stetsonDAOPHOTComputerProgram1987}, and {\tt{WCSTools}}. PSF photometric calibration is performed using Gaia Synthetic Photometry \citep[GaiaSP;][]{gaiacollaborationGaiaDataRelease2023}. The uncertainty assigned to each measurement corresponds to the formal PSF fitting error returned by {\tt{DAOPHOT II}}.
 
We also obtained 5 additional epochs of $u,g,r,i$ LCO photometry via the Global Supernova Experiment, using the Sinistro 1\,m telescopes at the Las Cumbres Observatory. These data span 105 -- 126 days post explosion in the rest frame. We reduced the $g,r,i$ photometry using {\tt{psf}} routine \citep{nicholl2022aedmNewClass2023}, extracting using an optimized aperture and archival imaging from the DESI Legacy Survey DR10 \citep{deyOverviewDESILegacy2019} for template subtraction and calibration. As no suitable pre-explosion $u$-band templates are available for the field, we do not perform host subtraction to the $u$-band photometry.

\begin{figure*}
    \centering
    \includegraphics[width=\textwidth]{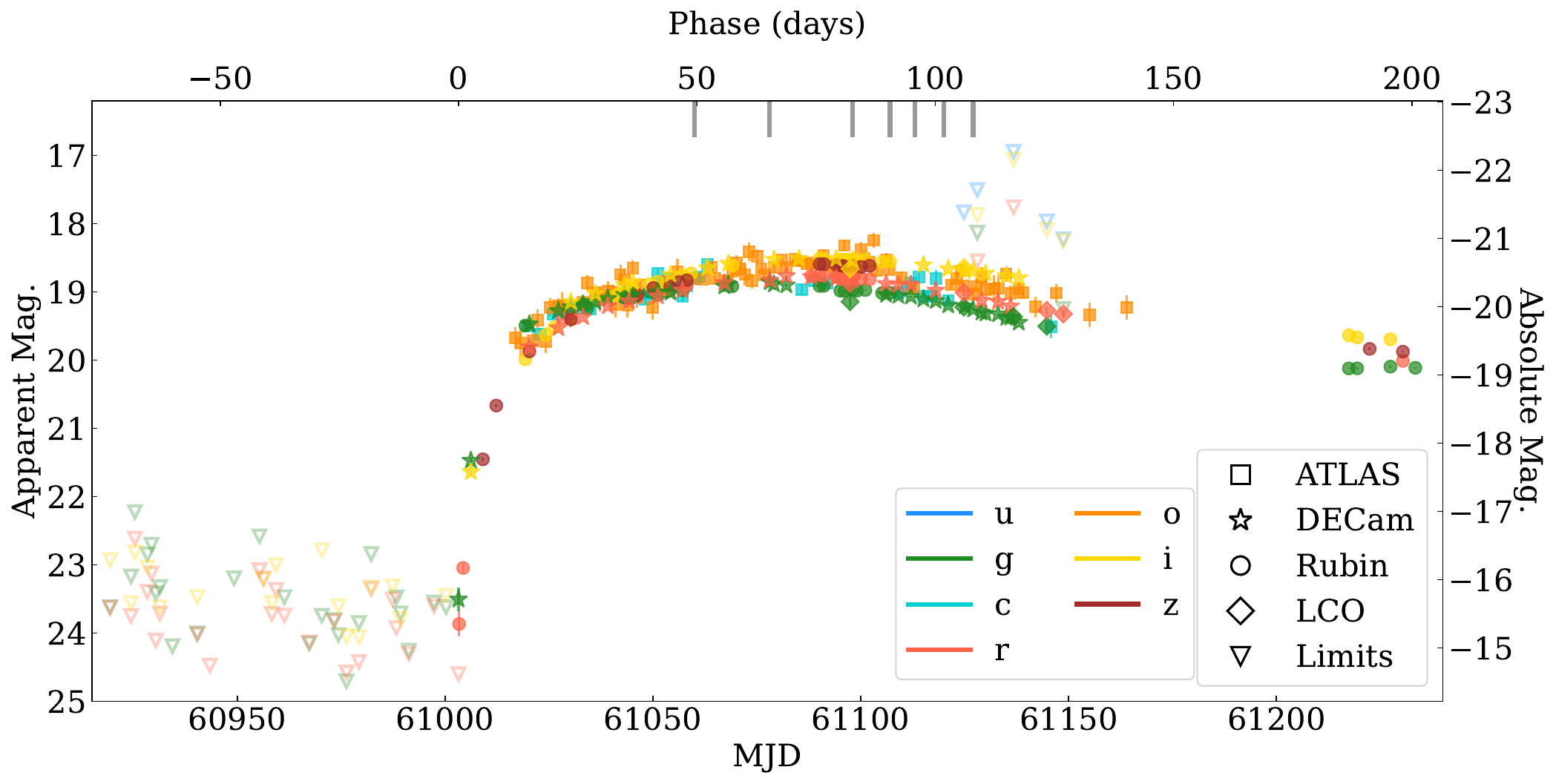}
    \caption{Multi-band light curve of AT\,2025agpz from DECam, Rubin commissioning observations, ATLAS, and LCO. Open downward triangles denote $3\sigma$ upper limits. The upper axis shows the rest-frame phase relative to the epoch of first detection with DECam (MJD 61\,003.19), while the right-hand axis gives the corresponding absolute magnitude assuming a redshift of $z=0.147$. Epochs of spectroscopic observations are marked with grey lines along the upper x-axis. The dense Rubin commissioning cadence, together with deep DECam pre-discovery imaging, provides unusually tight constraints on the rise time and colour evolution.}
    \label{fig:lc}
\end{figure*}

We correct all photometry for line of sight extinction using the dust maps of \cite{schlaflyMeasuringReddeningSloan2011}, assuming an $E\,(B\,-\,V)=0.011$ and $R_{V}=3.1$. We do not correct for extinction arising from the host galaxy. We present the light curve of AT\,2025agpz in Figure \ref{fig:lc}, spanning from $-77$\,d pre-detection to $+201$\,d. 

\subsection{Spectroscopy}

We obtained three epochs of optical spectroscopy of AT\,2025agpz with FORS2 on the VLT. Observations were acquired on 2026 January 20, February 07, and April 06 using the 300V+10 grism together with the GG435 order-sorting filter. All spectra were obtained with a 1.3 arcsec slit aligned at the parallactic angle to minimize differential atmospheric refraction, providing a spectral resolving power of $R \approx 440$ over the wavelength range $\sim4450$–8650\,\AA.

To investigate the kinematic structure of the narrow emission lines, we obtained a higher-resolution spectrum with the VLT X-shooter spectrograph \citep{vernetXshooterNewWide2011} on 2026 March 21. Observations were acquired in nodding mode using slit widths of 1.0, 0.9, and 0.9 arcsec in the UVB, VIS, and NIR arms, respectively, yielding spectral resolving powers of $R \approx 5400$, 8900, and 5600. The broad wavelength coverage of X-shooter ($\sim3000$–25,000\,\AA) additionally enables a search for diagnostic emission features beyond the optical spectral range.

Both the FORS2 and X-shooter data were reduced using the standard ESO Reflex workflows \citep{freudlingAutomatedDataReduction2013}, including bias subtraction, flat-fielding, wavelength calibration, sky subtraction, and one-dimensional optimal extraction. Flux calibration was performed using spectrophotometric standard stars observed on the same night as each science exposure. Telluric absorption was corrected for all FORS2 spectra and the X-shooter VIS arm using {\tt molecfit} \citep{smetteMolecfitGeneralTool2015,kauschMolecfitGeneralTool2015}.

An additional three epochs of optical spectroscopy of AT\,2025agpz were obtained with the Wide Field Spectrograph \citep[WiFeS][]{dopitaWideFieldSpectrograph2007,dopitaWideFieldSpectrograph2010} mounted on the Australian National University’s 2.3\,m telescope at Siding Spring Observatory \citep{priceConvertingANU232024}. Observations were acquired on 2026 February 28, March 14 and March 28 using the RT560 dichroic and the B3000/R3000 gratings, providing wavelength coverage over $3400 - 9500$\,\AA\, at a spectral resolving power of $R\sim$3000. Three Nod \& Shuffle (N \& S) exposures were obtained with 600s spent on the source and 600s spent on the sky in each exposure. A nearby region free of objects was used for the sky background.

The data were reduced using the updated {\tt{pyWiFeS}} reduction pipeline \citep{Price:2024pywifes,childressPyWiFeSRapidData2014}, which includes bias subtraction, flat-fielding, wavelength calibration, sky subtraction, and flux calibration against spectrophotometric standard stars, and correction for atmospheric telluric absorption using a telluric standard observed during the same night.

We present the spectroscopic evolution of AT\,2025agpz in Figure \ref{fig:specev}. A log of all spectroscopic observations is provided in the Appendix. 

\begin{figure*}
    \centering
    \includegraphics[width=\textwidth]{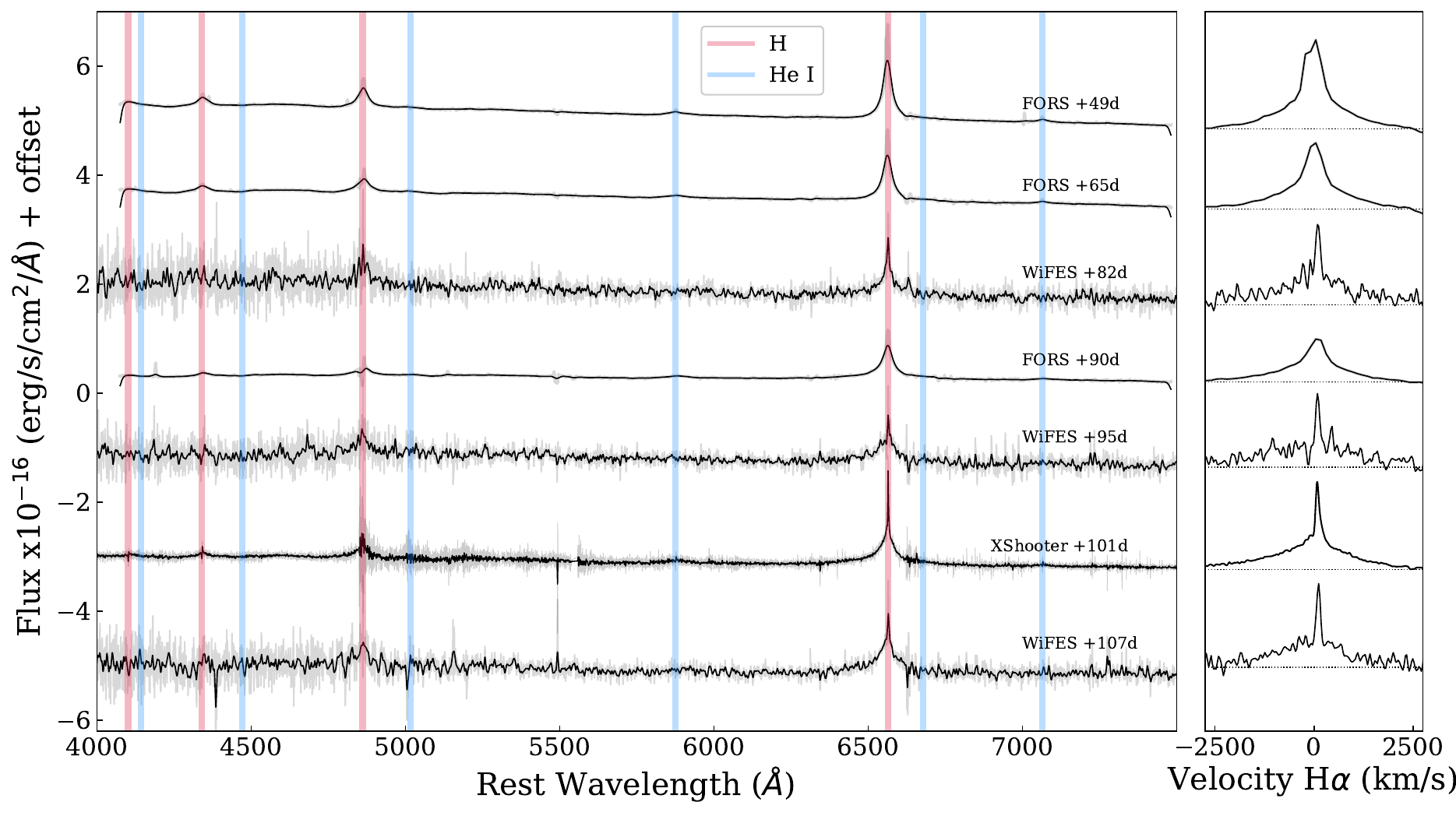}
    \caption{Spectroscopic evolution of AT\,2025agpz from +49 to +107 rest-frame days after discovery. Spectra have been corrected to the rest frame and offset in flux space for clarity. The positions of the principal Balmer and He\,\textsc{i} transitions are indicated by the shaded red and blue bands respectively. The right-hand panel shows the evolution of the continuum subtracted H$\alpha$ velocity profile (dotted lines mark the continuum level). The spectra exhibit slow evolution, dominated by narrow Balmer emission with broad electron-scattering wings becoming apparent in higher resolution spectra. From the right panel, we can see the blue wing of H$\alpha$ gradually becoming more extended with time.}
    \label{fig:specev}
\end{figure*}

\section{AT\,2025agpz Properties}\label{sec:properties}

Below we examine the spectroscopic, photometric and host-galaxy properties of AT\,2025agpz, and place its physical characteristics in the context of other classes of slow-evolving narrow-lined hydrogen rich transients; namely SLSNe-IIn and ANTs. 

\subsection{Photometric properties}\label{sec:photprop}

\begin{figure}
    \centering
    \includegraphics[width=\linewidth]{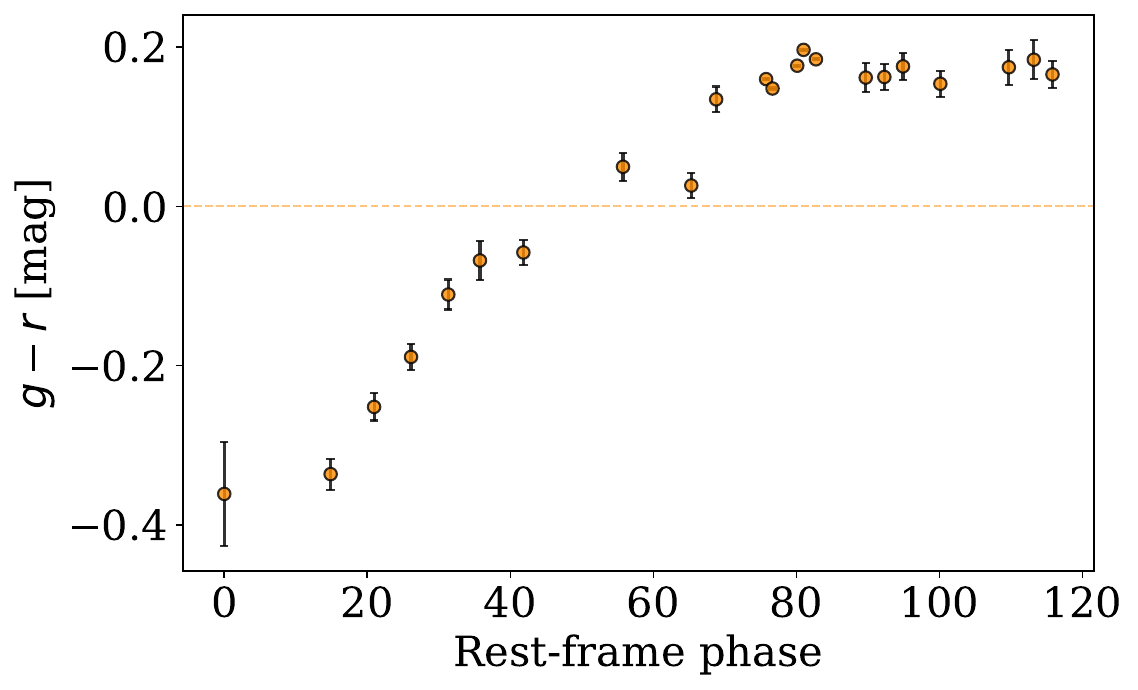}
    \caption{Observed $g-r$ colour evolution of AT\,2025agpz. The transient reddens steadily during the first $\sim90$ rest-frame days before reaching an approximately constant colour of $g-r\simeq0.16$, indicative of gradual cooling of the continuum. }
    \label{fig:color}
\end{figure}

The densely sampled Rubin commissioning observations, with an average cadence of 2.6\,days, together with complementary photometry from DECam, ATLAS, and LCO, provide an unusually complete view of the photometric evolution of AT\,2025agpz from shortly after first light through the decline from maximum. From Figure \ref{fig:lc} we see an apparent cooling of the light curve following its initial detection. In Figure~\ref{fig:color} we show the evolution of the observed $g-r$ colour from DECam photometry. At discovery, AT\,2025agpz is very blue, with $g-r=-0.36\pm0.06$, then undergoes a gradual monotonic reddening over the first $\sim90$\,days of its evolution before reaching an approximately constant colour of $g-r\simeq0.16$. Such behaviour closely resembles that observed in the volume-limited sample of SLSNe-IIn presented by \citet{pessiSampleHydrogenrichSuperluminous2025}, where most events evolve from blue colours at early times before settling onto a slowly varying colour plateau approximately 100 days after peak (albeit with a redder average post-peak colour of $g-r\sim0.5$; see their Figure 7). 

In contrast, accretion-powered nuclear transients are expected to exhibit relatively stable optical colours owing to the approximately constant temperature of their accretion discs/reprocessing spheres \citep{bonningAccretionDiskTemperatures2007,rothXRAYOPTICALFLUXES2016}, which is indeed observed within most TDEs \citep{hammersteinFinalSeasonReimagined2023a}{\footnote{Although we note that colour evolution is anticipated in the case for TDEs occurring with AGN environments, \citep[e.g.][]{jordana-mitjansOpticalPolarisationStellarfed2025}.}}. However, notable reddening has also been observed in several well-studied ANT and ENT candidates, including AT\,202lwx \citep[][]{hinkleMostEnergeticTransients2025}, AT\,2019fdr \citep[][]{pitikHighenergyNeutrinoEvent2022}, AT\,2019kn \citep[][]{wisemanSystematicallySelectedSample2025} and AT\,2021loi \citep{makrygianni2021loiBowenFluorescence2023}. From the ZTF selected sample of ANTs presented in \cite{wisemanSystematicallySelectedSample2025}, on average their light curves appear to redden gradually, with an average $\Delta\,g-r\sim0.2$ over the first 250 days of their evolution (the timescale of our AT\,2025agpz observations), although with significant scatter in their colours, spanning $-0.1<g-r<0.5$ (Quilt et al. {\textit{in prep.}}). This behaviour falls broadly within the general distribution of SLSNe-IIn colour evolutions measured by \cite{pessiSampleHydrogenrichSuperluminous2025}, indicating that this evolution alone cannot uniquely distinguish between these populations (see also their Figure 14).

\begin{figure}
    \centering
    \includegraphics[width=\linewidth]{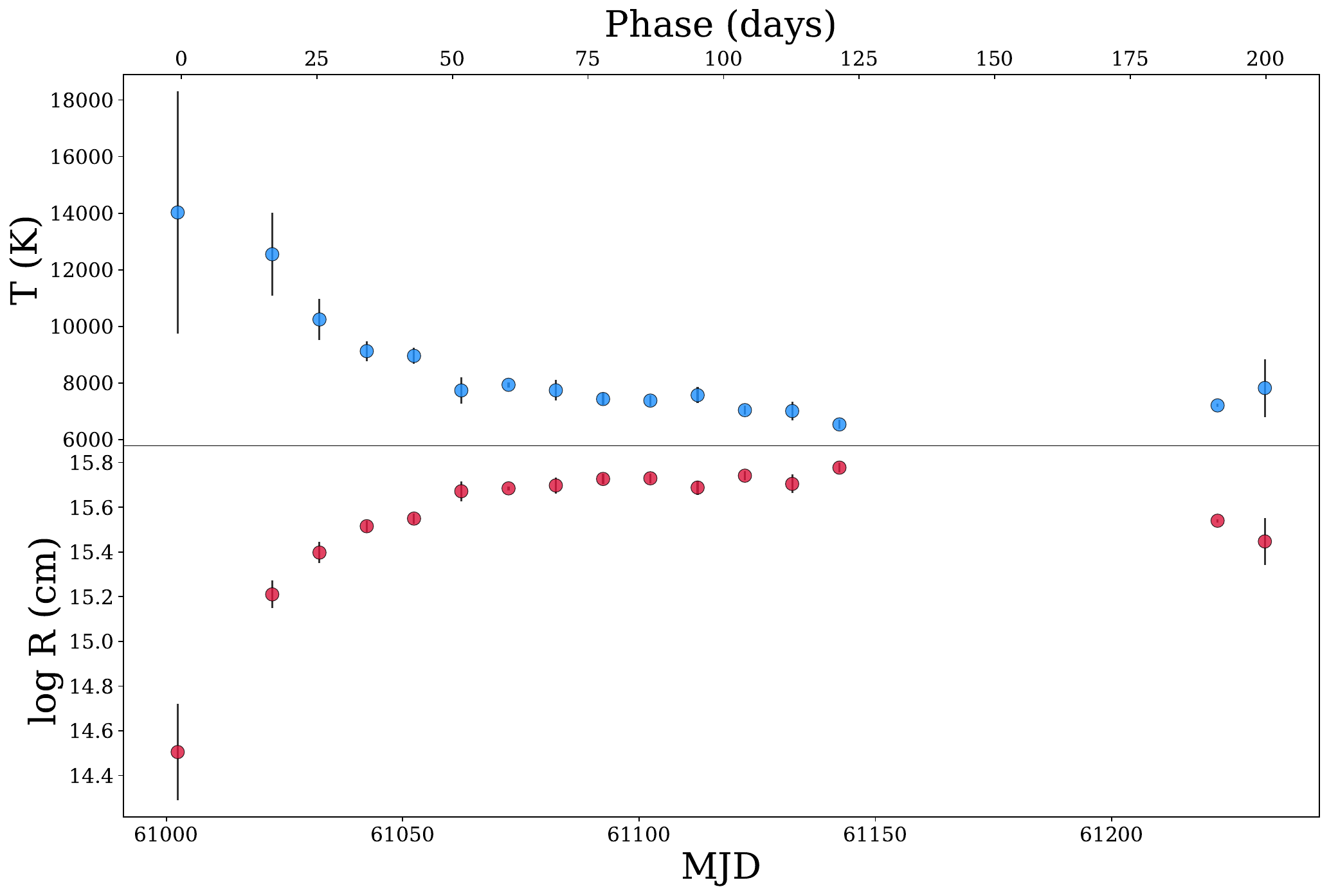}
    \caption{Blackbody evolution of AT\,2025agpz derived from multi-band photometry grouped into 10-day intervals. Phase on the upper axis is with respect to discoery epoch in the rest-frame of AT\,2025agpz. {\it{Top:}} Best-fitting blackbody temperature. {\it{Bottom:}} Inferred photospheric radius. The transient cools by approximately 5000\,K during the first $\sim100$ days while the inferred blackbody radius expands from $\sim10^{14.5}$ to $\sim10^{15.7}$\,cm before beginning to plateau.}
    \label{fig:bbfit}
\end{figure}

To gauge the level of cooling, we fit the light curve of AT\,2025agpz with a simple blackbody, from which we also construct the bolometric light curve. We group the foreground extinction corrected photometry at 10-day intervals, then fit a simple blackbody function to each epoch and integrate across the full blackbody curve to determine the bolometric luminosity. The resulting evolution of the black body temperature and radius are shown in Figure \ref{fig:bbfit}. We measure a peak bolometric luminosity of $6.3\pm1.4\times10^{43}$\,erg\,s$^{-1}$ and an integrated radiated energy of $7.2\pm0.1\times10^{50}$\,erg over the first 191 days following discovery. The inferred blackbody temperature declines smoothly by approximately $5000$\,K over the first $\sim100$ days, while the photospheric radius increases from $\sim10^{14.5}$ to $\sim10^{15.7}$\,cm before beginning to plateau (Figure~\ref{fig:bbfit}). Such behaviour is broadly consistent with expectations for interaction-powered supernovae, where the continuum-forming radius expands through optically thick circumstellar material \citep[e.g.][]{chevalierSHOCKBREAKOUTDENSE2011,chatzopoulosHYDROGENPOORCIRCUMStelLARSHELLS2012}. The inferred radii also overlap those measured for luminous ANTs, which exhibit characteristic blackbody radii of $10^{15}$–$10^{16}$\,cm throughout much of their evolution \citep{wisemanSystematicallySelectedSample2025,hinkleMostEnergeticTransients2025}, although we note that single blackbody fits have been shown to poorly describe other classes of accreting nuclear transient, such as TDEs \citep[e.g.][]{guoloSizeAccretionDisks2025}, so may not truly reflect the photospheric temperatures and radii of the ANT population. Similarly, the peak luminosity of AT\,2025agpz lies comfortably within the observed distributions of hydrogen-rich SLSNe \citep{pessiSampleHydrogenrichSuperluminous2025} and just within the fainter-tail of the distribution of ANT luminosities \citep{wisemanSystematicallySelectedSample2025}, illustrating the considerable overlap between these populations.

The deep pre-discovery DECam imaging, together with the near-simultaneous initial detections from DECam and Rubin, tightly constrains the explosion epoch, enabling a robust measurement of the rise time. As a model-independent estimate, we first use the discovery bracket defined by the final non-detection and the first significant DECam $g$-band detection, adopting the midpoint of this interval as the explosion epoch. Measuring from this epoch to maximum light yields a total $g$-band rise time of $89.4\pm1.5$\,d in the observer frame, corresponding to $77.9\pm1.3$\,d in the rest frame at $z=0.147$. We also characterise the early rise using an exponential model, obtaining an $e$-folding rise time of $3.09\pm0.16$\,d in the observer frame.

This places AT\,2025agpz amongst the slowest-rising hydrogen-rich interacting supernovae identified by ZTF \citep{pessiSampleHydrogenrichSuperluminous2025}. Similarly, the inferred rise time falls comfortably within the broad distribution of ANTs, many of which exhibit rest-frame rise times of $\sim50$–200\,d \citep{wisemanSystematicallySelectedSample2025}. The rise time alone therefore does not distinguish between the two populations, but instead reinforces the substantial overlap in their observed photometric properties.

In interacting supernovae, such long rise times are commonly interpreted as the photon diffusion time through an optically thick CSM, often implying very large ($\gtrsim100\,M_\odot$) CSM masses under simple diffusion models \citep{smithShellshockedDiffusionModel2007,chevalierSHOCKBREAKOUTDENSE2011}. However, more recent radiative-transfer calculations have shown that similarly long rise times can be reproduced with substantially smaller CSM masses ($\sim10\,M_\odot$) provided the circumstellar density profile is sufficiently shallow \citep{moriyaNatureSlowlyRising2023}. Such configurations may naturally arise from eruptive pre-supernova mass loss, for example through pulsational pair-instability episodes \citep{woosleyPulsationalPairinstabilitySupernovae2017}.

\begin{figure}
    \centering
    \includegraphics[width=\linewidth]{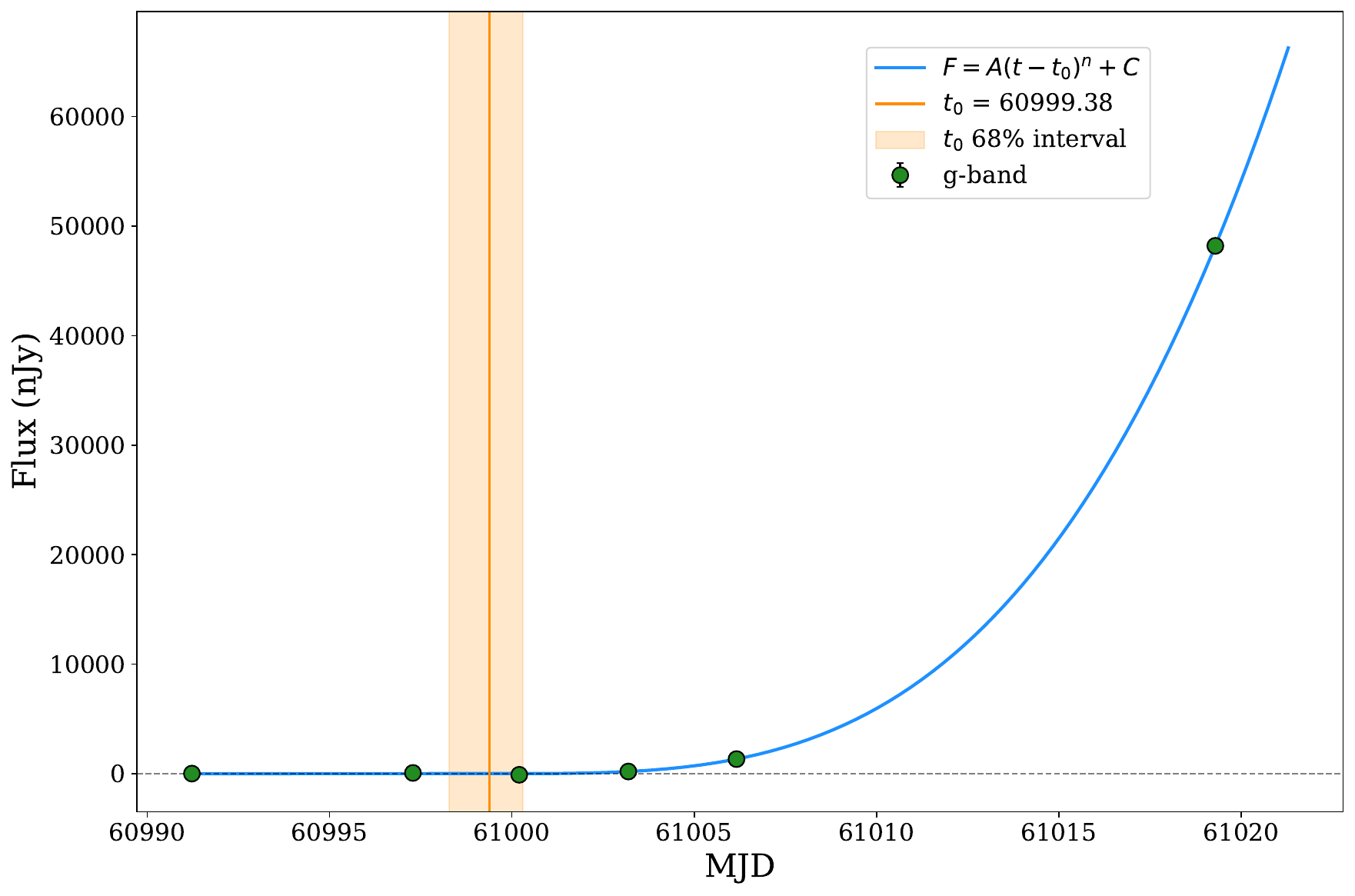}
    \caption{Power-law fit to the early DECam and Rubin photometry $g$-band light curve. 
    The shaded region indicates the 68\% confidence interval on the inferred explosion epoch, $t_0$, obtained from bootstrap resampling. The fit yields a rise index of $n=3.33^{+0.33}_{-0.28}$ and an explosion epoch (MJD) of $t_{0}=60\,999.19_{-1.14}^{+2.02}$.}
    \label{fig:powerlaw}
\end{figure}

Given the depth and cadence of the pre-explosion imaging from Rubin and DECam, we also fit the early $g$-band rise directly in flux space with a power law model, shown in Figure~\ref{fig:powerlaw}, parametrized as:
\begin{equation}
    F(t)= 
\begin{cases}
    A[(t-t_0)/(1+z)]^n+C,& \text{if $t>t_0$} \\
    C,              & \text{if $t<t_0$}
\end{cases}
\end{equation}

where $t_0$ is the explosion epoch, $A$ is a normalization, $n$ is the power-law index, and $C$ accounts for any residual baseline offset. All forced-flux measurements within the early-time fitting window were included, including non-detections and negative-flux measurements, with each point weighted by its flux uncertainty. The fit was restricted to the rising portion of the light curve, ending at approximately half of the peak $g$-band flux, to avoid bias from curvature near maximum light. We perform bootstrap resampling of the flux measurements to estimate uncertainties on $t_0$ and $n$. 

Fitting the early DECam and Rubin $g$-band forced photometry in flux space yields a power-law rise with an index of $n=3.33^{+0.33}_{-0.28}$. We note the inferred pseudo-$g$ measurement from the converted Rubin $r$-band photometry appears to be systematically brighter than the surrounding DECam photometry, which may be a result of residual transient flux in the $r$-band templates.

This is steeper than the canonical $n=2$ power law expected for homologous expansion with approximately constant temperature and photospheric velocity (i.e. the expanding `fireball’ model). While significant diversity about this simple model has long been recognised for $^{56}$Ni-powered supernovae \citep{firthRisingLightCurves2015,liuDecodingEarlyTimeLight2026}, values of $n\gtrsim3$ remain relatively uncommon. For comparison, \citet{gonzalez-gaitanRisetimeTypeII2015} measured a median rise index of $n\approx0.9$ for normal Type II supernovae, and using TESS data \citealt{vallelyHighcadenceEarlytimeObservations2021} measure median rise of $n=1.04 \pm 0.14$ when fitting a single power law, highlighting the unusually steep early rise of AT\,2025agpz. In interaction-powered transients, a steep rise may reflect a rapid increase in radiative efficiency as the expanding ejecta encounter dense circumstellar material \citep{suzukiSystematicStudyRise2020}. Although the power-law fit provides a robust estimate of the explosion epoch and rise time, at present the value of $n$ alone does not uniquely discriminate between progenitor scenarios. However we note that, whilst this has yet to be measured for the ANT population, there are cases for other classes of accreting transients, where a shallower rise of $n\sim2$ has been found \citep{holoienDiscoveryEarlyEvolution2019a,nichollOutflowPowersOptical2020a,hinkleDiscoveryFollowupASASSN19dj2021}. 

From the power-law model we infer an \lq explosion\rq\,\footnote{Here \lq explosion epoch\rq\, refers to the epoch of first optical brightening inferred from the power-law fit, rather than implying a particular physical origin for the transient.} epoch (MJD) of $t_0=60\,999.19^{+2.02}_{-1.14}$, corresponding to just $\sim3.3$ rest-frame days before the first detection. Previous studies of slowly evolving interacting transients have generally relied on shallower surveys, often leaving the explosion epoch poorly constrained and limiting measurements of the initial rise. The depth of the Rubin and DECam imaging enabled AT\,2025agpz to be detected $\gtrsim$4 magnitudes below its peak brightness, and over 2 weeks before its discovery with ATLAS, permitting strong constraints upon the explosion epoch. Rubin’s combination of depth and cadence therefore offers the opportunity to characterise the early rise of much larger samples of luminous transients in a homogeneous manner, allowing the intrinsic distribution of rise indices within populations such as SLSNe-IIn and ANTs to be established.

Overall, the general photometric properties of AT\,2025agpz occupy regions of parameter space common to both SLSN-II and ANT populations, making it difficult to uniquely identify the physical nature of the transient from photometry alone.

\subsection{Spectroscopic properties}\label{sec:specprop}

The spectroscopic evolution of AT\,2025agpz is shown in Figure~\ref{fig:specev}. Across the seven spectroscopic epochs, spanning +49 to +107 rest-frame days after discovery, the spectra are dominated by strong Balmer emission lines superposed on a blue continuum, exhibiting remarkably little evolution over this period. Weak He\,\textsc{i} emission is also detected, most prominently at $5876$ and $7065$\,\AA.

\begin{figure*}
    \centering
    \includegraphics[width=\textwidth]{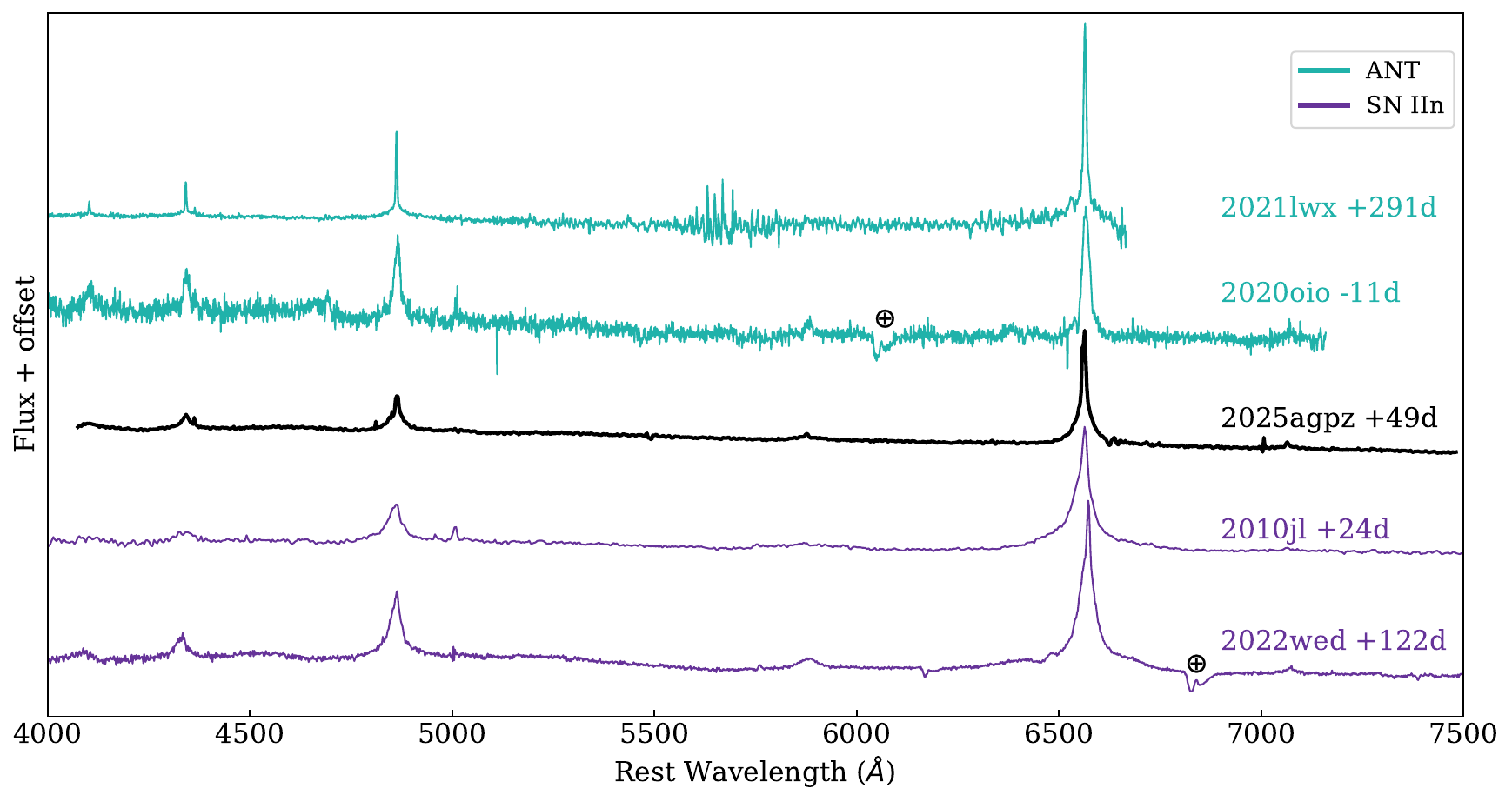}
    \caption{Comparison of the early (+49\,d) spectrum of AT\,2025agpz (black) with representative hydrogen-rich interacting supernovae (purple; SN\,2010jl and SN\,2022wed \citep{zhangTYPEIInSUPERNOVA2012,salmasoDiversityStronglyInteracting2025}) and Ambiguous Nuclear Transients (cyan; AT\,2020oio and AT\,2021lwx \citep{wisemanSystematicallySelectedSample2025}). Circle crosses mark uncorrected telluric absorption features. Spectra have been corrected to the rest frame, continuum normalised, and vertically offset for clarity. AT\,2025agpz exhibits the blue continuum, strong Balmer emission, and weak He\,\textsc{i} features common to both populations, illustrating the substantial spectroscopic overlap between interaction-powered supernovae and slowly evolving nuclear transients.}
    \label{fig:speccomp}
\end{figure*}

Throughout the sequence, the H$\alpha$ profile is characterised by a narrow emission core superposed on broad, symmetric wings. Such profiles are commonly observed in interaction-powered supernovae, where the narrow emission arises from slowly moving photoionised circumstellar material and the broad wings are produced through multiple electron scatterings in an optically thick CSM \citep{dessartNumericalSimulationsSuperluminous2015,smithInteractingSupernovaeTypes2017}. Similar broad winged Balmer emission is also observed in some AGN, caused by turbulent motions within the line-emitting gas in combination with rotational motion of the accretion disc \citep{kollatschnyShapeBroadlineProfiles2013}. Recent studies have shown that comparable line profiles are an emerging observational characteristic of ANTs and ENTs, which are likewise characterised by strong, relatively narrow Balmer emission superposed on blue continua \citep{wisemanSystematicallySelectedSample2025, hinkleMostEnergeticTransients2025}, and in some cases He \textsc{ii} and Mg\,\textsc{ii} emission. However, 
given the infancy of these spectroscopic classes and the late epochs at which they are often identified, cadenced spectroscopic observations of ANT/ENT evolution are lacking, so it is unclear how the features evolve throughout the transient lifetime, and whether these can be used to cleanly discriminate progenitor scenarios \citep[e.g.][]{pessiAmbiguousAT2022rzeChanginglook2025}. 

Figure~\ref{fig:speccomp} compares the early spectrum of AT\,2025agpz with representative SLSNe-IIn and ANTs. The overall spectral morphology—including the blue continuum, narrow Balmer emission, and weak He\,\textsc{i} features—closely resembles both classes, illustrating the observational overlap between luminous interacting supernovae and some slowly evolving nuclear transients. Although the Balmer emission in the comparison SLSNe-IIn appears broader, comparison of the FORS2, WiFeS, and X-shooter spectra in Figure~\ref{fig:specev} demonstrates that the apparent width of the narrow component is strongly influenced by instrumental resolution. A similar effect may complicate comparisons with nuclear transients, where higher-resolution spectroscopy has resolved more complex Balmer emission in individual objects such as PS16dtm \citep{blanchardPS16dtmTidalDisruption2017,petrushevskaRiseFallIronstrong2023}. In that case, the emission is interpreted as originating from pre-existing AGN line-emitting material illuminated during the transient, further illustrating the difficulty in differentiating between physical origins with low-resolution spectra. No single spectroscopic feature uniquely distinguishes the two classes at the resolution of typical classification spectra.

Despite the overall stability of the spectra, subtle evolution is evident in the H$\alpha$ profile (Figure~\ref{fig:specev}, right panel). The peak line flux gradually declines with time, while the blue wing becomes progressively more extended, reaching increasingly negative velocities in the final X-shooter and WiFeS spectra. In contrast, the red wing evolves comparatively little, resulting in a modest increase in the asymmetry of the profile while preserving the narrow central component. This behaviour suggests that the principal spectroscopic evolution is confined to the broad emission component.

\begin{figure}
    \centering
    \includegraphics[width=\linewidth]{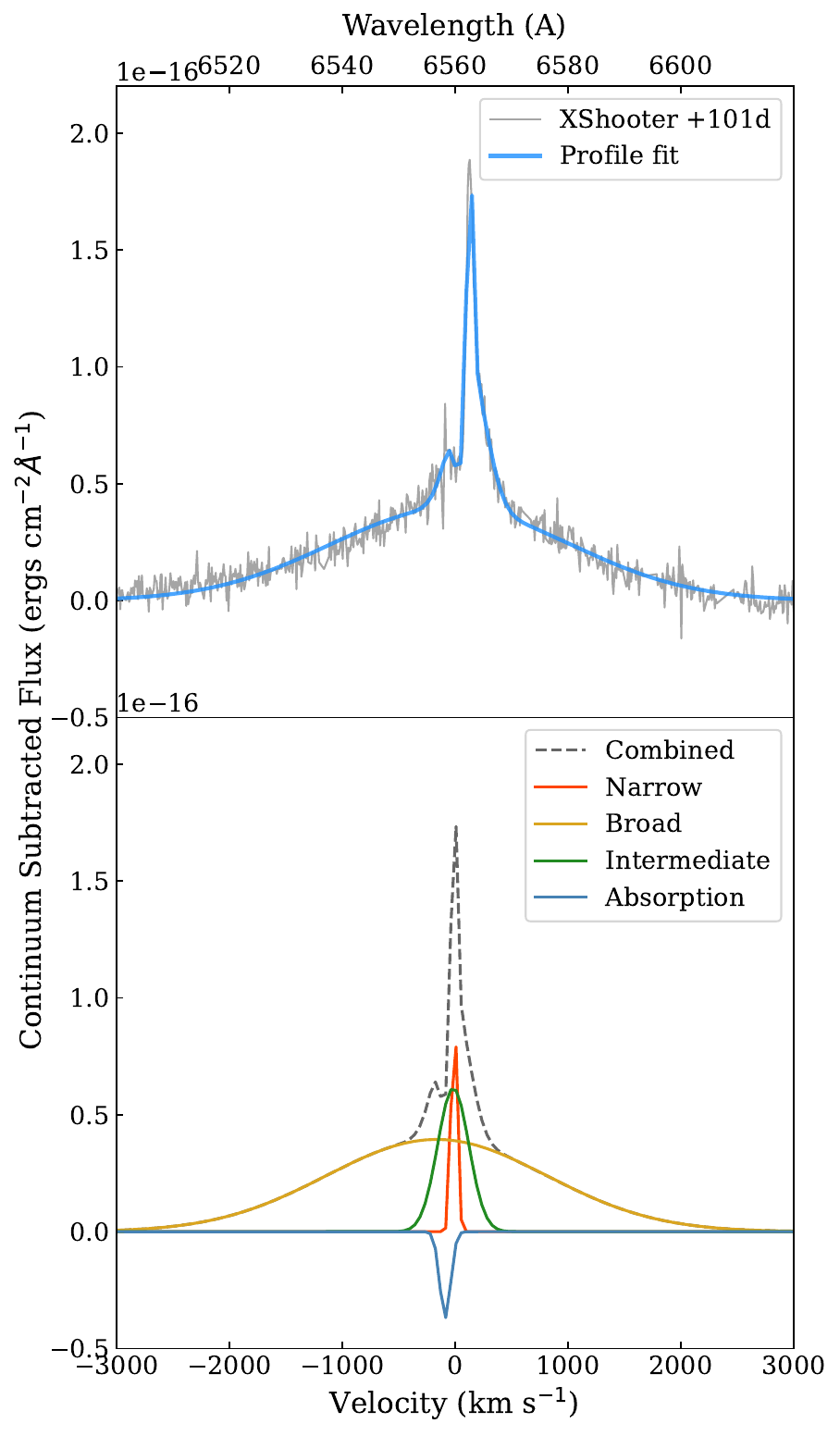}
    \caption{Multi-component decomposition of the H$\alpha$ emission profile in the +101\,d X-shooter spectrum. {\it{Top:}} continuum-subtracted spectrum (grey) together with the best-fitting combined model (blue). {\it{Bottom:}} individual Gaussian components comprising the fit, including narrow, intermediate-width, and broad emission components together with a weak blueshifted absorption feature. The X-shooter resolution resolves multiple kinematic components that are blended together in lower-resolution spectra.}
    \label{fig:halpha}
\end{figure}

To investigate the H$\alpha$ morphology in greater detail, we exploit the higher spectral resolution of the X-shooter spectrum (FWHM $\approx40$ km\,s$^{-1}$). Using the {\tt{specutils}} package, we fit the continuum-subtracted profile with a model comprising three Gaussian emission components (narrow, intermediate, and broad) together with a narrow blueshifted absorption component. For each component we fit the central wavelength ($\lambda_{\rm c}$), FWHM ($\sigma_{\lambda}$) and peak amplitude. We present the best fitting parameters for each of these profile in Table \ref{tab:halpha_fit} and show our combined fitted profile alongside the data in Figure \ref{fig:halpha}. 

\begin{table}
\centering
\caption{Best-fitting parameters for the four Gaussian components used to model the H$\alpha$ profile from the +101d XShooter spectrum. Velocity offsets are measured relative to the rest-frame H$\alpha$ wavelength.}
\label{tab:halpha_fit}
\begin{tabular}{lcccc}
\hline
Component & $\lambda_{\rm c}$& $\sigma_{\lambda}$& $v_{\rm{offset}}$& Amplitude \\
& (\AA) & (\AA) & (km\,s$^{-1}$) & $\times10^{-17}$\\
& & & & erg\,s$^{-1}$\,cm$^{-2}$\\
\hline
Narrow &
$6562.6 \pm 0.6$ & $0.6 \pm 0.5$ & $29 \pm 26$ & $9.8 \pm 6.1$\\
Broad&
$6559.2 \pm 2.5$ & $21.4 \pm 2.8$ & $-128 \pm 115$ & $3.95 \pm 0.58$ \\
Intermediate&
$6562.5 \pm 1.0$ & $3.0 \pm 1.0$ & $21 \pm 47$ & $6.1 \pm 3.9$ \\
Absorption &
$6560.9 \pm 3.1$ & $1.1 \pm 1.6$ & $-50 \pm 142$ & $-3.7 \pm 7.7$ \\
\hline
\end{tabular}
\end{table}

The X-shooter spectrum cleanly resolves the H$\alpha$ emission into a very narrow core (${\rm FWHM}\approx70$ km\,s$^{-1}$), an intermediate-width component (${\rm FWHM}\approx320$ km\,s$^{-1}$), and a broad component with ${\rm FWHM}\approx2300$ km\,s$^{-1}$, and potentially a weak blueshifted absorption feature. All components are centred close to the H$\alpha$ rest frame, with offsets below $\sim150$ km\,s$^{-1}$.

The morphology of the resolved H$\alpha$ profile is qualitatively similar to those observed in several well-studied interacting luminous IIn. Higher-resolution spectroscopy of events such as SN\,2010jl \citep{franssonHighdensityCircumstellarInteraction2014,gallRapidFormationLarge2014}, SN\,2021adxl \citep{brennanSN2021adxlLuminous2024} and ASASSN-15ua \citep{dickinsonSuperluminousTypeIIn2024} likewise reveal a narrow emission core superposed on broad, approximately Lorentzian wings interpreted as electron scattering within an optically thick circumstellar medium. The inferred narrow ($\sim70$ km\,s$^{-1}$) and broad ($\sim2300$ km\,s$^{-1}$) components fall comfortably within the range expected for interaction-powered supernovae. The presence of a narrow P-Cygni profile (represented by the narrow emission and blueshifted absorption components) further supports an interacting-supernova interpretation, as such features are generally attributed to slowly expanding, unshocked circumstellar material surrounding the progenitor star \citep{smithInteractingSupernovaeTypes2017}. However, the inferred width of this absorption is comparable to the instrumental resolution of the X-shooter spectrum (FWHM $\approx40$ km\,s$^{-1}$), and is only weakly constrained by the fit (${\rm FWHM}\approx120\pm170$ km\,s$^{-1}$). Consequently, while the absorption is consistent with an interacting-supernova interpretation, we cannot confidently infer a progenitor wind velocity from these observations.

Whilst the origin of the narrow Balmer emission cannot be established unambiguously from a single high-resolution spectrum, as the measured FWHM ($\sim70$ km\,s$^{-1}$) is only modestly broader than the X-shooter instrumental resolution ($\sim40$ km\,s$^{-1}$), the inferred narrow component lies towards the upper end of the line widths typically observed in the emission-line spectra of dwarf galaxies \citep[e.g.][]{ismVelMoiseev2021,ismDwarfIrrARAA} and individual star-forming regions, which are more commonly in the range $\sim20$–50 km\,s$^{-1}$ \citep{30DorTorresFlores}. Combined with the absence of narrow forbidden emission such as [O\,{\sc iii}] and [N\,{\sc ii}], together with the overall H$\alpha$ morphology and its evolution between epochs, this suggests that, despite the uncertainty in the intrinsic width of the narrow component, the narrow Balmer component is more likely associated with the transient than with the underlying host galaxy. While we cannot exclude some contribution from host-galaxy emission, the available evidence favours an origin in circumstellar material photoionised by the transient.

We also searched for comparable kinematic structure in the He\,\textsc{i} $\lambda5876$ emission. Although weak, broad ($\sigma_{\lambda}\sim30\AA$) He\,\textsc{i} emission is detected in the X-shooter spectrum, the signal-to-noise ratio of this line is insufficient to robustly decompose the profile into multiple velocity components analogous to those identified in H$\alpha$. In particular, we find no statistically significant evidence for narrow or intermediate-width He\,\textsc{i} emission, and therefore cannot determine whether the helium emission originates from the same kinematic regions as the Balmer lines. Likewise, while a comparison with the H$\beta$ profile and the corresponding Balmer decrement would provide a useful diagnostic of the excitation mechanism, H$\beta$ falls close to the join between the UVB and VIS arms of X-shooter where the spectrum is comparatively noisy and potentially affected by inter-arm flux-calibration uncertainties. We therefore do not attempt to decompose the H$\beta$ profile, leaving H$\alpha$ as the most reliable probe of the line-forming region.

Conversely, the resolved broad component we measure for H$\alpha$ is also somewhat narrower than the characteristic H$\beta$ widths reported for ANTs \citep[$\sim2900$ km\,s$^{-1}$;][]{wisemanSystematicallySelectedSample2025}, although differences in spectral resolution, line decomposition, and the use of H$\alpha$ versus H$\beta$ make a direct comparison difficult. Similar Balmer line morphologies have also been reported in the luminous nuclear transient PS1-10adi, whose early spectra exhibit a blue continuum and Balmer emission with broad electron-scattering wings \citep{kankarePopulationHighlyEnergetic2017}. However, the Balmer emission in PS1-10adi is substantially broader (FWHM $\sim900$ km\,s$^{-1}$) and develops a pronounced red shoulder at later epochs, neither of which are evident in AT\,2025agpz. 
At the resolution of the FORS2 spectra, the narrow and intermediate components blend into a single unresolved emission peak, producing an apparently broader profile. Similar effects are likely to influence comparisons with published ANT spectra, many of which were obtained at lower spectral resolution. High-resolution spectroscopy therefore provides an important means of separating genuinely distinct velocity components from instrumental broadening, giving crucial physical information that helps to distinguish SLSNe-IIn and nuclear transients. 

\subsection{Host galaxy properties}\label{sec:host}

The host galaxy of AT\,2025agpz provides an additional observational diagnostic of the transient origin. We first model the host spectral energy distribution (SED) using archival optical photometry to infer its stellar mass, star-formation rate, and stellar population properties.

\begin{figure}
    \centering
    \includegraphics[width=\linewidth]{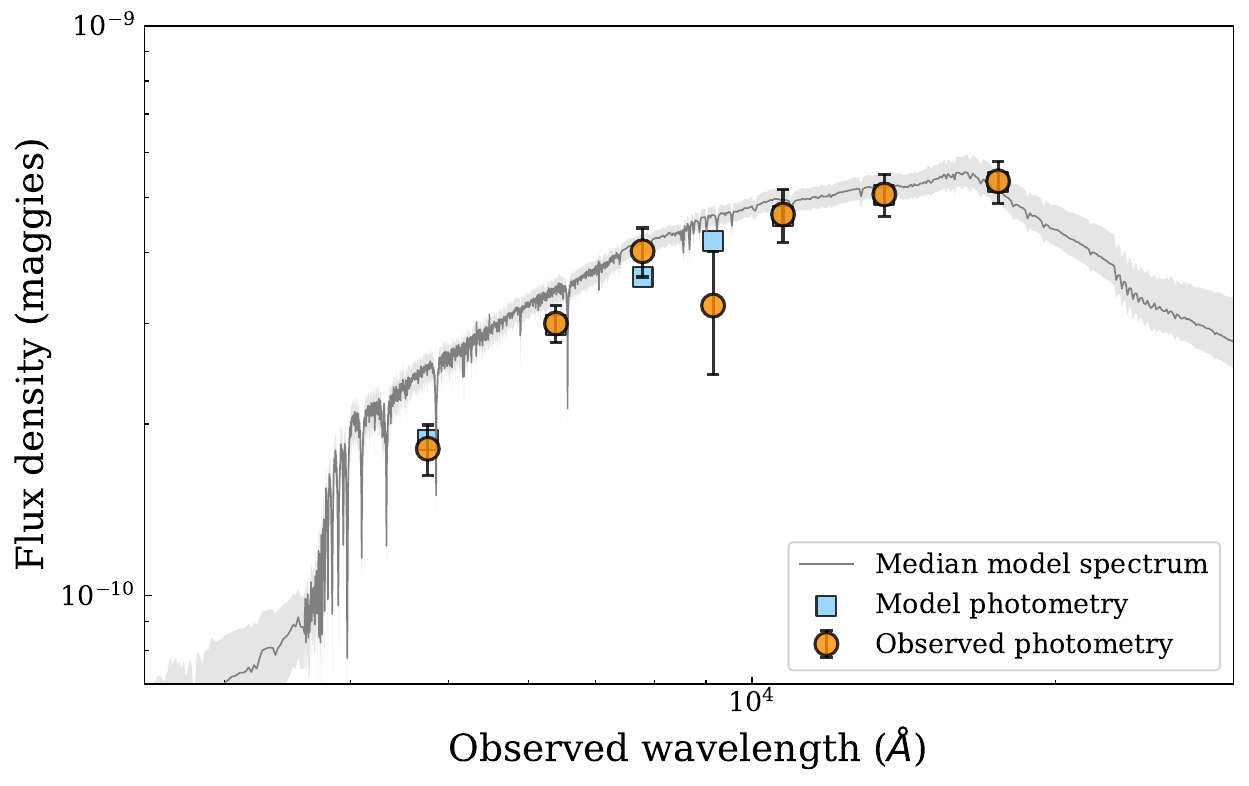}
    \caption{Prospector spectral energy distribution fit to the host galaxy of AT\,2025agpz using archival optical and NIR photometry. The grey curve shows the median posterior model spectrum, shading indicates the 68\% confidence interval, coloured squares denote the model photometry, and circles show the observed fluxes from survey data.}
    \label{fig:hostSED}
\end{figure}

We perform SED fitting using optical--near infrared (NIR) photometry from the Euclid Q1 MER catalogue \citep{euclidconsortiumEuclidQ1MER2025,euclidcollaborationEuclidQuickData2026}, which provides aperture-matched photometry of Euclid sources on PSF-matched images for external ground based surveys. We adopt the template-fitting ({\tt{TEMPLFIT}}) fluxes for both the Euclid $Y$, $H$, $J$, and external DECam bands $g$, $r$, $i$, and $z$ bands. These measurements use the higher resolution Euclid detection image to define a morphological template, which is convolved to the PSF of each band and fit to the corresponding image. This provides PSF-matched, deblended multi-band photometry on a consistent morphological model and avoids aperture-matching issues between the Euclid and ground-based imaging. To account for systematic uncertainties in both the photometry and the stellar population models, we adopt a conservative 10\% error floor in all bands. The photometry is modelled using the \textsc{Prospector} SED fitting framework \citep{lejaDerivingPhysicalProperties2017,johnsonStellarPopulationInference2021}. We adopt a non-parametric continuity star formation history with three age bins, allowing the stellar mass, stellar metallicity, dust attenuation, and relative star formation rates between adjacent bins to vary freely. Dust re-emission is included in the model. Although we do not have mid-infrared photometry to further constrain the dust emission, studies of nearby galaxies have shown that stellar masses for local galaxies ($z<0.25$) derived from optical SED fitting remain robust even with only optical-NIR wavelength coverage \citep{paulino-afonsoSystematicErrorsOpticalSED2022}. Whilst the lack of mid infrared data prevents us from cleanly fitting for the presence of an AGN, we also explore models including an AGN dust component to assess whether including it would significantly bias our derived parameters from fitting the optical and NIR data. Posterior probability distributions are sampled using \textsc{dynesty} \citep{speagleDYNESTYDynamicNested2020}. The best-fitting SED is shown in Figure~\ref{fig:hostSED}, while the inferred galaxy properties are listed in Table~\ref{tab:prospector}.

\begin{table}
\centering
\caption{Prospector-derived host galaxy properties for AT\,2025agpz from fits with and without an AGN dust component.}
\label{tab:prospector}
\begin{tabular}{lcc}
\hline
Parameter & No AGN & AGN Included \\
\hline
$\log(M_\star/M_\odot)$ & $8.33^{+0.12}_{-0.12}$ & $8.33^{+0.12}_{-0.12}$ \\
$\log(Z_\star/Z_\odot)$ & $-1.3^{+0.6}_{-0.5}$ & $-1.3^{+0.6}_{-0.5}$ \\
$\tau_V$ (dust2) & $0.6^{+0.4}_{-0.3}$ & $0.6^{+0.4}_{-0.3}$ \\
SFR$_{100\,\mathrm{Myr}}$ [$M_\odot\,\mathrm{yr}^{-1}$] & $0.014^{+0.02}_{-0.01}$ & $0.015^{+0.024}_{-0.010}$ \\
$\log(\mathrm{sSFR}/\mathrm{yr}^{-1})$ & $-10.18$ & $-10.15$ \\
$f_{\rm AGN}$ & --- & $10^{-4}$\\
$\tau_{\rm AGN}$ & --- & 5.0 \\
\hline
\end{tabular}
\end{table}

The inferred host properties are largely insensitive to the inclusion of an AGN component, indicating that the available optical$-$NIR broadband photometry is dominated by stellar emission. We infer a stellar mass of $\log(M_\star/M_\odot)=8.33^{+0.12}_{-0.12}$, a recent star-formation rate of $0.014^{+0.02}_{-0.01}\,M_\odot\,{\rm yr}^{-1}$, and a specific star-formation rate of $\log({\rm sSFR}/{\rm yr}^{-1})\approx-10.19$, consistent with a moderately star-forming dwarf galaxy. The inferred stellar metallicity is sub-solar ($\log(Z_\star/Z_\odot)\approx-1.3$), while the best-fitting dust attenuation is moderate ($A_V\lesssim0.6$). We search the X-shooter spectrum for the presence of narrow absorption lines from the host-galaxy interstellar medium, but find no convincing evidence of these. Combined with the apparent blue colour of AT\,2025agpz, this suggests there is no substantial internal host galaxy extinction.

These properties are broadly consistent with those of SLSN-II host galaxies. Hydrogen-rich SLSNe are preferentially found in low-mass, star-forming galaxies, with a relative scarcity of hosts above $\sim10^{10}\,M_\odot$, although they span a broader range of host properties than their hydrogen-poor counterparts \citep{angusHubbleSpaceTelescope2016,schulzeCosmicEvolutionMetal2018,taggartCorecollapseSuperluminousGammaray2021}. In contrast, the presently known ANT population appears to inhabit a more diverse range of host galaxies, with stellar masses spanning approximately $10^{9}$–$10^{11}\,M_\odot$. Several ANT hosts exhibit narrow emission lines indicative of ongoing star formation and/or low-level AGN activity, suggesting that the population may arise in a variety of galactic environments \citep{blanchardPS16dtmTidalDisruption2017,petrushevskaRiseFallIronstrong2023,wisemanSystematicallySelectedSample2025}. However, several ANTs lack an identified host galaxy in current ground-based imaging, leaving the low-mass end of the ANT host-galaxy population largely unconstrained. 

An accretion-powered interpretation for AT\,2025agpz would nevertheless imply an unusually low central black-hole mass. Applying the stellar-mass scaling relation of \citet{reinesRelationsCentralBlack2015} yields $\log(M_{\rm BH}/M_\odot)\sim4.6$, substantially below the black-hole-mass distributions inferred for current samples of ANTs \citep{hinkleMidinfraredEchoesAmbiguous2024} and TDEs \citep{yaoTidalDisruptionEvent2023}. This comparison should be treated cautiously, however, given the intrinsic scatter in low-mass black-hole scaling relations and the incompleteness of existing ANT host samples.

Taken together, the host properties mildly favour an interacting-supernova interpretation: its low stellar mass, ongoing star formation, and sub-solar metallicity are fully consistent with environments known to host SLSNe-II, while an accretion-powered origin would require a black hole at the extreme low-mass end of those presently inferred for ANTs. Nevertheless, given the diversity and incomplete characterisation of ANT environments, the host galaxy alone does not provide a definitive discriminator between the two scenarios.

\section{Discussion}\label{sec:discussion}

\subsection{AT2025agpz as a likely supernova}

AT\,2025agpz occupies an observational regime shared by hydrogen-rich SLSNe-IIn and the recently recognised population of ANTs. Its long rise time, slowly evolving light curve, blue continuum, narrow Balmer emission, and nuclear location all resemble the defining characteristics of ANTs, while its luminosity, gradual colour evolution, dwarf star-forming host galaxy, and evidence for multiple kinematic components within the H$\alpha$ profile are typical of strongly interacting supernovae. These observables, when considered individually, cannot uniquely distinguish between the two interpretations.

The photometric evolution most closely resembles that of hydrogen-rich SLSNe. The rest-frame rise time of $78$\,d is significantly longer than that of the majority of Type IIn supernovae, but is comparable to several well-studied SLSNe-IIn including SN\,2006gy \citep{smithSN2006gyDiscovery2007} and SN\,2016aps \citep{nichollExtremelyEnergeticSupernova2020}, as well as the population presented by \citet{pessiSampleHydrogenrichSuperluminous2025}. Likewise, the gradual reddening of the optical colours and the cooling blackbody temperature over the first $\sim100$ days are typical of interaction-powered supernovae, where the continuum photosphere recedes through an optically thick circumstellar medium. Although similar colour evolution has been observed in some ANTs \citep{wisemanSystematicallySelectedSample2025}, the colour evolution of these events seem to be much more diverse, with many events presenting more stable colours.

The spectroscopy likewise supports an interacting supernova origin. The X-shooter observations resolve the H$\alpha$ emission into multiple velocity components superposed upon broad wings, naturally interpreted as emission from distinct circumstellar regions together with electron scattering in an optically thick CSM. While Lorentzian Balmer profiles are also observed in narrow-line AGN and several ANTs, the resolved velocity structure and the gradual evolution of the blue wing are more readily explained within the framework of ongoing ejecta--CSM interaction. Though at the instrumental resolution limit, the apparent P Cygni profile also supports an interacting supernova interpretation, if originating from CSM associated with a massive progenitor star. The inferred host galaxy further supports this interpretation. With a stellar mass of $\log(M_\star/M_\odot)\approx8.3$, ongoing star formation, and sub-solar metallicity, the host closely resembles those typically associated with hydrogen-rich SLSNe, although these properties also fall comfortably within the range currently occupied by ANT hosts.

Taken together, we therefore favour an interpretation in which AT\,2025agpz is a hydrogen-rich interacting superluminous supernova. Nevertheless, the similarity of its observational properties to the currently defined ANT population illustrates the substantial overlap between these classes and highlights the difficulty of assigning unique classifications based solely on photometric or low-resolution spectroscopic observations.

\subsection{Implications for the ANT population}

The discovery of ANTs has highlighted the existence of a population of luminous, slowly evolving nuclear transients that do not fit comfortably within traditional classes of tidal disruption events or AGN variability \citep{kankarePopulationHighlyEnergetic2017,grahamUnderstandingExtremeQuasar2017,holoienInvestigatingNatureLuminous2022,oatesSwiftUVOTDiscovery2024,wisemanMultiwavelengthObservationsExtraordinary2023,subrayanScaryBarbieExtremely2024,hinkleMostEnergeticTransients2025}. However, the physical nature of these objects remains uncertain \citep[][]{petrushevskaRiseFallIronstrong2023,wisemanMultiwavelengthObservationsExtraordinary2023,subrayanScaryBarbieExtremely2024}, and current samples are necessarily defined using relatively broad observational criteria \citep{wisemanSystematicallySelectedSample2025,hinkleMostEnergeticTransients2025}. As a result, some contamination from neighbouring transient classes is likely, making it unclear whether ANTs represent a single astrophysical population or an observational grouping comprising multiple physical channels. It is plausible that this diversity reflects genuine differences in physical origin. While the extreme luminosities and radiated energies of the brightest ANTs are most naturally explained by accretion onto a massive black hole,   \citep{wisemanMultiwavelengthObservationsExtraordinary2023,subrayanScaryBarbieExtremely2024,hinkleMostEnergeticTransients2025}, the physical nature of lower-luminosity events remains less certain and may encompass multiple classes of transient.

AT\,2025agpz demonstrates that at least some interaction-powered supernovae may occupy this observational parameter space. The combination of a long-duration light curve, narrow Balmer emission, a faint dwarf host galaxy, and a projected nuclear location produces a transient that is difficult to distinguish from a luminous nuclear flare using photometry and low-resolution spectroscopy alone. We emphasize that our results do not imply that a significant fraction of ANTs are interacting supernovae. Rather, they demonstrate that the present observational definition of the ANT population encompasses a region of parameter space also occupied by extreme interacting supernovae. If AT\,2025agpz were instead powered by accretion onto a central black hole, the inferred host stellar mass would imply a black hole mass of only $\log(M_{\rm BH}/M_\odot)\sim4.6$, substantially below the masses inferred for the majority of currently known ANTs. While accretion-powered transients have been identified around comparably low-mass black holes, such as the TDE candidate AT\,2022zod (\citealt{dageAT2022zodUnusualTidal2025}) and the fast-rising TDE AT\,2020neh (\citealt{angusFastrisingTidalDisruption2022a}), these events evolve on much shorter timescales than AT\,2025agpz. Although an accretion-powered origin cannot be firmly excluded, the combination of its host properties and unusually slow evolution provides additional support for an interaction-powered supernova interpretation.

As ANT samples continue to grow, detailed studies of borderline events such as AT\,2025agpz will be essential for identifying which observed properties genuinely trace physical origin and which simply reflect shared characteristics. Ultimately, establishing robust observational discriminants will determine whether ANTs emerge as a physically distinct class or are found to comprise multiple channels of extreme accretion- and interaction-powered transients.

\subsection{Photometric classification in the Rubin era}

We have demonstrated that long-lived, strongly interacting supernovae like AT\,2025agpz occupy a similar region of parameter space to ANTs in many respects, and they may also appear in the centres of galaxies -- particularly at higher redshifts where projected angular offsets of all transients will be smaller. High resolution spectroscopy appears to be the most robust discriminant between these classes, but this will be available only for a minority of objects. To build a robust population of rare ANTs with Rubin, photometric classification will be essential.

To investigate how AT\,2025agpz would be classified within Rubin alert streams, we applied the {\tt{ANTEATER}} machine-learning classifier (Quilt et al., {\it{in prep.}}) to several subsets of the observed Rubin light curve. {\tt{ANTEATER}} (Ambiguous Nuclear Transient EArly Time ExtractoR) is a random forest classifier trained using a range of public and private simulations of ANT, AGN, TDE and SN IIn LSST light curves. These light curves are fit using Gaussian process regression and features such as current colour, band/colour change in the last 5, 10, 20, 50 days and the linearity of evolution in each band are extracted for training and testing. Each light curve is cut off at a random phase to train across ages, with a 90$\%$ weighting to ages $<$ 100 days to prioritise early classification ability. {\tt{ANTEATER}} can take an LSST unique object identifier, the diaObjectId, as input and can evaluate the transient at any chosen epoch using only the photometry available up to that time. By default, the classification is performed using the complete light curve available at the time of the query. {\tt{ANTEATER}} then returns predicted classification probabilities for each class. The planned primary use case is annotating promising ANT candidates in the Lasair broker and flagging these for spectroscopic follow-up. At the time of writing, {\tt{ANTEATER}} is still in active development, but full details will be given in Quilt et al., {\it{in prep}}.

\begin{figure}
    \centering
    \includegraphics[width=\linewidth]{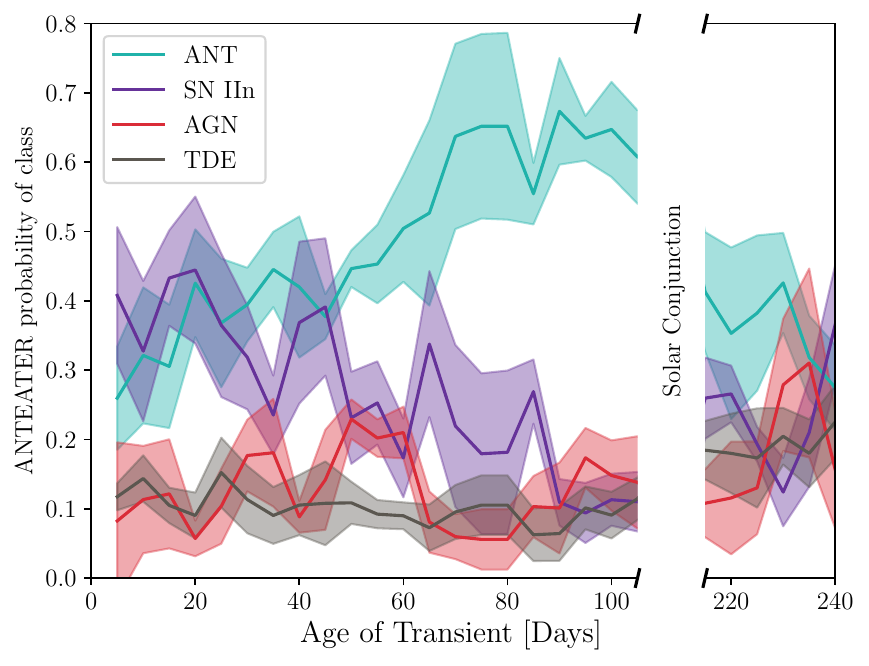}
    \caption{The {\tt{ANTEATER}} classification probabilities for AT\,2025agpz across its evolution, using the full LSST data from Lasair. The five best performing models are used to generate a median probability estimate with 1$\sigma$ error. While there is heavy confusion between an ANT and a SN IIn at early times, an ANT classification is favoured until late times.}
    \label{fig:ANTEATER}
\end{figure}

We evaluate {\tt{ANTEATER}} classification of AT\,2025agpz under four observing scenarios designed to mimic both Rubin and current-generation survey data. First, we use the complete Rubin $griz$ light curve, representative of the information expected for well-sampled LSST transients. Second, we impose a magnitude limit of $m\simeq20.5$, approximating the depth of present wide-field transient surveys while retaining the full wavelength coverage. Third, we restrict the data to the $g$ and $r$ bands, reproducing the limited colour information available from surveys such as ZTF and ATLAS. Finally, we combine both limitations to simulate a ZTF-like dataset (the survey in which the majority of ANTs have been identified). To test how the classification changes with the evolution of AT\,2025agpz, we evaluate in 5 day intervals. The resulting classification probabilities at 25, 50 and 100 days (where {\tt{ANTEATER}} has the most training) are presented for the LSST baseline and ZTF simulation in Table~\ref{tab:anteater_tests}, while the full time resolution is presented for the LSST baseline in Figure~\ref{fig:ANTEATER}.

\begin{table}
\centering
\caption{{\tt{ANTEATER}} classification probabilities for AT\,2025agpz under different photometric data configurations at different times. These probabilities are the median of the top 5 high-performing models. The LSST baseline shows an initial strong confusion between ANT and SN IIn while the ZTF simulation initially favours a SN IIn scenario. Both configurations converge to favouring an ANT classification at 100 days, but there is greater confidence with LSST.}
\begin{tabular}{lcccc}
\hline
Configuration & ANT & AGN & TDE & SN IIn \\
\hline
LSST $griz$ (baseline): 25 days& 0.368 & 0.104 & 0.152 &0.365\\
LSST $griz$ (baseline): 50 days& 0.447 & 0.229 & 0.109 &0.231\\
LSST $griz$ (baseline): 100 days& 0.647 & 0.148 & 0.091 &0.113\\
\hline
Full ZTF-like simulation: 25 days& 0.280 & 0.158 & 0.110 &0.467\\
Full ZTF-like simulation: 50 days& 0.377 & 0.210 & 0.116 &0.242 \\
Full ZTF-like simulation: 100 days& 0.464& 0.110& 0.198 &0.276\\
\end{tabular}
\label{tab:anteater_tests}
\end{table}

Using the full Rubin $griz$ light curve, {\tt{ANTEATER}} initially struggles to discern between an ANT and SN IIn scenario at early times, but past $\sim$ 50 days becomes confident with an ANT label; it would then be flagged for spectroscopic follow-up automatically. There is a gap in the cadence which includes the solar conjunction and once the extra data showing a decline is added, the confusion reappears. While this could be due to the decline timescale being quite short for an ANT, it is possibly also the case that this is an artefact of {\tt{ANTEATER}} having limited training at later epochs-- the tool is prioritised for early classification.

The ZTF-like data set set shows a similar classification trend, although it is initially more confident of a SN IIn classification and is overall more confused without the extra information available in LSST observations, with a smaller spread in the probabilities. The magnitude-only and band-only cut simulations reveal that the lack of extra bands is the primary driver of the confusion.

The behaviour of {\tt ANTEATER} for AT\,2025agpz highlights an important limitation of the current training sample for photometric classification. Although our spectroscopic analysis favours an interaction-powered SLSN-IIn interpretation, the classifier assigns the highest probability to the ANT class once the transient has been rising for $\gtrsim50$\,d, only returning to a more balanced ANT/SN\,IIn classification after $\sim200$\,d when both the rise and decline have been observed (Figure~\ref{fig:ANTEATER}). This behaviour suggests that the exceptionally long rise time is the principal feature driving the ANT classification. AT\,2025agpz lies towards the extreme slow end of the observed SLSN-IIn rise-time distribution \citep[e.g.][]{pessiSampleHydrogenrichSuperluminous2025,hiramatsuTypeIInSupernovae2026a}, indicating that such events are likely under-represented within the present training set. Expanding the sample of spectroscopically confirmed, long-rising SLSNe-IIn will therefore be important for mitigating this bias and improving the discrimination between slowly evolving interacting supernovae and luminous nuclear transients.

The training samples available to future classifiers will improve substantially during the Rubin era. Large spectroscopic programmes such as the Time Domain Extragalactic Survey (TiDES) on 4MOST \citep{frohmaierTiDES4MOSTTime2025}, together with rapid spectroscopic follow-up from higher-resolution facilities including WiFeS and SOXS \citep{schipaniSOXSWideBand2018,claudiSonXShooterMultiband2019}, will provide larger and more representative samples of spectroscopically classified ANTs, SLSNe-IIn, and related transients. Combined with Rubin's superior depth and multi-band cadence, these data should significantly improve the discriminatory power of future photometric classifiers.

\section{Conclusions}\label{sec:conclusions}

We have presented AT\,2025agpz, a luminous, long-rising transient discovered by ATLAS and identified by the Lasair Virtual Research Assistant in Rubin Observatory commissioning observations of the Euclid Deep Field South. Its projected nuclear location, slow photometric evolution, and interaction-dominated spectrum place it at the observational boundary between hydrogen-rich superluminous supernovae and the recently recognised class of Ambiguous Nuclear Transients. Our principal conclusions are:

\begin{itemize}

\item AT\,2025agpz exhibits a long rest-frame rise time of $\sim$78\,d, reaches a peak bolometric luminosity of $\sim6\times10^{43}$\,erg\,s$^{-1}$, and radiates $\sim7\times10^{50}$\,erg over the first $\sim$190 days. These photometric properties overlap those of both SLSNe-IIn and lower-luminosity ANTs.

\item The dense pre-discovery imaging from DECam and Rubin commissioning tightly constrains the explosion epoch to within only a few days, allowing a robust measurement of the early rise. A power-law fit yields a steep rise index of $n=3.33^{+0.33}_{-0.28}$, illustrating the potential of Rubin to characterise the earliest evolution of slowly rising transients.

\item Spectroscopically, AT\,2025agpz is dominated by narrow Balmer emission with broad electron-scattering wings that evolve only slowly over $\sim$100 rest-frame days. High-resolution X-shooter spectroscopy resolves narrow, intermediate, and broad H$\alpha$ components with widths consistent with interaction-powered supernovae, although comparisons with published ANT spectra remain limited by differences in spectral resolution.

\item The host galaxy is a low-mass ($\log M_\star/M_\odot\simeq8.3$), moderately star-forming dwarf galaxy with sub-solar metallicity, consistent with environments commonly hosting hydrogen-rich SLSNe, while remaining within the presently observed diversity of ANT host galaxies.

\item Taken together, the colour evolution, host-galaxy properties, and resolved H$\alpha$ profile favour an interaction-powered SLSN-IIn interpretation. Nevertheless, AT\,2025agpz demonstrates that luminous interacting supernovae can occupy the same observational parameter space currently used to define ANTs.

\end{itemize}

AT\,2025agpz represents precisely the type of object that will become increasingly important in the Rubin era. By occupying the observational boundary between interacting supernovae and nuclear transients, it provides a valuable test case for both phenomenological classification schemes and machine-learning approaches. Continued characterisation of such borderline events will refine the observational definitions of both ANTs and SLSNe-IIn, while ultimately determining whether these populations represent genuinely distinct physical channels or partially overlapping manifestations of extreme stellar explosions and accretion-powered variability. The exceptional cadence and depth of Rubin observations, when combined with complementary follow-up, provide exquisitely sampled light curves that can place stringent constraints on the underlying physical mechanisms. Robust classification of these rare events is therefore essential not only for constructing clean samples, but also for fully exploiting the diagnostic power of Rubin-quality datasets. 

\section*{Acknowledgements}

CRA and MN are supported by the European Research Council (ERC) under the European Union’s Horizon 2020 research and innovation programme (grant agreement No.~948381).
HFS thanks the Schmidt Sciences Foundation. 
SJS acknowledges funding from STFC Grants ST/Y001605/1, ST/X001253/1, a Royal Society Research Professorship and the Hintze Family Charitable Foundation. 
PW is grateful for the support from STFC grant ST/Z510269/1.
AM is supported by the ARC Discovery Early Research Award (DE230100055). 
DM acknowledges a studentship funded by the Leverhulme Interdisciplinary Network on Algorithmic Solutions.
SM acknowledges financial support from the Research Council of Finland project 350458.
IM acknowledges support from the Australian Research Council (ARC) center of Excellence for Gravitational Wave Discovery (OzGrav), through project number CE230100016.
KWS acknowledges funding from the Royal Society.
{\L}W acknowledges support from the Polish National Science Centre DAINA grant No 2024/52/L/ST9/00210 and funding from the European Union's Horizon Europe Research and Innovation programme ACME under grant agreement No 101131928.
PMV acknowledges the support from the DFG via the Collaborative Research Center SFB1491 \textit{Cosmic Interacting Matters - From Source to Signal}.
Parts of this research were supported by the Australian Research Council Centre of Excellence for Gravitational Wave Discovery (OzGrav), through project number CE230100016.

This material is based upon work supported in part by the National Science Foundation through Cooperative Agreements AST-1258333 and AST-2241526 and Cooperative Support Agreements AST-1202910 and 2211468 managed by the Association of Universities for Research in Astronomy (AURA), and the Department of Energy under Contract No. DE-AC02-76SF00515 with the SLAC National Accelerator Laboratory managed by Stanford University. Additional Rubin Observatory funding comes from private donations, grants to universities, and in-kind support from LSST-DA Institutional Members.
Lasair is supported by the UKRI Science and Technology Facilities Council as part of the LSST:UK Science Centre and is a collaboration between the University of Edinburgh, Queen’s University Belfast and the University of Oxford, funded by grants ST/X001334/1 and ST/X001253/1. 

The Asteroid Terrestrial-impact Last Alert System (ATLAS) project is primarily funded to search for near earth asteroids through NASA grants NN12AR55G, 80NSSC18K0284, and 80NSSC18K1575; byproducts of the NEO search include images and catalogs from the survey area. This work was partially funded by Kepler/K2 grant J1944/80NSSC19K0112 and HST GO-15889, and STFC grants ST/T000198/1 and ST/S006109/1. The ATLAS science products have been made possible through the contributions of the University of Hawaii Institute for Astronomy, the Queen’s University Belfast, the Space Telescope Science Institute, the South African Astronomical Observatory, and The Millennium Institute of Astrophysics (MAS), Chile and the University of Oxford. 

BHTOM is based on the open-source TOM Toolkit by LCO and has been developed with funding from the OPTICON-RadioNet Pilot (ORP) of the European Union's Horizon 2020 research and innovation programme under grant agreement No 101004719 (2021-2025). This project has received funding from the European Union's Horizon Europe Research and Innovation programme ACME under grant agreement No 101131928 (2024-2028).

This work is based in part on observations collected at the European Organisation for Astronomical Research in the Southern Hemisphere under ESO programmes 116.2959.001 and 116.296A.001 (PI: Angus).

This work is based in part on data acquired at the ANU 2.3-metre telescope. The automation of the telescope was made possible through an initial grant provided by the Centre of Gravitational Astrophysics and the Research School of Astronomy and Astrophysics at the Australian National University and through a grant provided by the Australian Research Council through LE230100063. The Lens proposal system is maintained by the AAO Research Data \& Software team as part of the Data Central Science Platform. We acknowledge the traditional custodians of the land on which the telescope stands, the Gamilaraay people, and pay our respects to elders past and present.

This work has made use of the Quick Release (Q1) data from the Euclid mission of the European Space Agency (ESA).

This work makes use of observations from the Las Cumbres Observatory network under programme ANU2026A-002 (PI: Auchettl).

The Legacy Surveys consist of three individual and complementary projects: the Dark Energy Camera Legacy Survey (DECaLS; Proposal ID \#2014B-0404; PIs: David Schlegel and Arjun Dey), the Beijing-Arizona Sky Survey (BASS; NOAO Prop. ID \#2015A-0801; PIs: Zhou Xu and Xiaohui Fan), and the Mayall z-band Legacy Survey (MzLS; Prop. ID \#2016A-0453; PI: Arjun Dey). DECaLS, BASS and MzLS together include data obtained, respectively, at the Blanco telescope, Cerro Tololo Inter-American Observatory, NSF's NOIRLab; the Bok telescope, Steward Observatory, University of Arizona; and the Mayall telescope, Kitt Peak National Observatory, NOIRLab. Pipeline processing and analyses of the data were supported by NOIRLab and the Lawrence Berkeley National Laboratory (LBNL). The Legacy Surveys project is honored to be permitted to conduct astronomical research on Iolkam Du'ag (Kitt Peak), a mountain with particular significance to the Tohono O'odham Nation.

NOIRLab is operated by the Association of Universities for Research in Astronomy (AURA) under a cooperative agreement with the National Science Foundation. LBNL is managed by the Regents of the University of California under contract to the U.S. Department of Energy.

This project used data obtained with the Dark Energy Camera (DECam), which was constructed by the Dark Energy Survey (DES) collaboration. Funding for the DES Projects has been provided by the U.S. Department of Energy, the U.S. National Science Foundation, the Ministry of Science and Education of Spain, the Science and Technology Facilities Council of the United Kingdom, the Higher Education Funding Council for England, the National Center for Supercomputing Applications at the University of Illinois at Urbana-Champaign, the Kavli Institute of Cosmological Physics at the University of Chicago, Center for Cosmology and Astro-Particle Physics at the Ohio State University, the Mitchell Institute for Fundamental Physics and Astronomy at Texas A\&M University, Financiadora de Estudos e Projetos, Fundacao Carlos Chagas Filho de Amparo, Financiadora de Estudos e Projetos, Fundacao Carlos Chagas Filho de Amparo a Pesquisa do Estado do Rio de Janeiro, Conselho Nacional de Desenvolvimento Cientifico e Tecnologico and the Ministerio da Ciencia, Tecnologia e Inovacao, the Deutsche Forschungsgemeinschaft and the Collaborating Institutions in the Dark Energy Survey. The Collaborating Institutions are Argonne National Laboratory, the University of California at Santa Cruz, the University of Cambridge, Centro de Investigaciones Energeticas, Medioambientales y Tecnologicas-Madrid, the University of Chicago, University College London, the DES-Brazil Consortium, the University of Edinburgh, the Eidgenossische Technische Hochschule (ETH) Zurich, Fermi National Accelerator Laboratory, the University of Illinois at Urbana-Champaign, the Institut de Ciencies de l'Espai (IEEC/CSIC), the Institut de Fisica d’Altes Energies, Lawrence Berkeley National Laboratory, the Ludwig Maximilians Universitat Munchen and the associated Excellence Cluster Universe, the University of Michigan, NSF’s NOIRLab, the University of Nottingham, the Ohio State University, the University of Pennsylvania, the University of Portsmouth, SLAC National Accelerator Laboratory, Stanford University, the University of Sussex, and Texas A\&M University.

BASS is a key project of the Telescope Access Program (TAP), which has been funded by the National Astronomical Observatories of China, the Chinese Academy of Sciences (the Strategic Priority Research Program``The Emergence of Cosmological Structures'' Grant \#XDB09000000), and the Special Fund for Astronomy from the Ministry of Finance. The BASS is also supported by the External Cooperation Program of Chinese Academy of Sciences (Grant \#114A11KYSB20160057), and Chinese National Natural Science Foundation (Grant \#12120101003, \#11433005).

The Legacy Survey team makes use of data products from the Near-Earth Object Wide-field Infrared Survey Explorer (NEOWISE), which is a project of the Jet Propulsion Laboratory/California Institute of Technology. NEOWISE is funded by the National Aeronautics and Space Administration.

The Legacy Surveys imaging of the DESI footprint is supported by the Director, Office of Science, Office of High Energy Physics of the U.S. Department of Energy under Contract No. DE-AC02-05CH1123, by the National Energy Research Scientific Computing Center, a DOE Office of Science User Facility under the same contract; and by the U.S. National Science Foundation, Division of Astronomical Sciences under Contract No. AST-0950945 to NOAO.


\section*{Data Availability}



\bibliographystyle{mnras}
\bibliography{refs} 

@ARTICLE{ismVelMoiseev2021,
       author = {{Moiseev}, Alexei V. and {Lozinskaya}, Tatiana A.},
        title = "{Ionized gas velocity dispersion in nearby dwarf galaxies: looking at supersonic turbulent motions}",
      journal = {\mnras},
         year = 2012,
        month = jun,
       volume = {423},
       number = {2},
        pages = {1831-1844},
          doi = {10.1111/j.1365-2966.2012.21005.x},
archivePrefix = {arXiv},
       eprint = {1203.6213},
 primaryClass = {astro-ph.CO},
       adsurl = {https://ui.adsabs.harvard.edu/abs/2012MNRAS.423.1831M}
}

@ARTICLE{30DorTorresFlores,
       author = {{Torres-Flores}, S. and {Barb{\'a}}, R. and {Ma{\'\i}z Apell{\'a}niz}, J. and {Rubio}, M. and {Bosch}, G. and {H{\'e}nault-Brunet}, V. and {Evans}, C.~J.},
        title = "{Studying the kinematics of the giant star-forming region 30 Doradus . I. The data}",
      journal = {\aap},
         year = 2013,
        month = jul,
       volume = {555},
          eid = {A60},
        pages = {A60},
          doi = {10.1051/0004-6361/201220474},
archivePrefix = {arXiv},
       eprint = {1305.0042},
 primaryClass = {astro-ph.CO},
       adsurl = {https://ui.adsabs.harvard.edu/abs/2013A&A...555A..60T}
}

@ARTICLE{ismDwarfIrrARAA,
       author = {{Hunter}, Deidre A. and {Elmegreen}, Bruce G. and {Madden}, Suzanne C.},
        title = "{The Interstellar Medium in Dwarf Irregular Galaxies}",
      journal = {\araa},
         year = 2024,
        month = sep,
       volume = {62},
       number = {1},
        pages = {113-155},
          doi = {10.1146/annurev-astro-052722-104109},
archivePrefix = {arXiv},
       eprint = {2402.17004},
 primaryClass = {astro-ph.GA},
       adsurl = {https://ui.adsabs.harvard.edu/abs/2024ARA&A..62..113H}
}

@ARTICLE{TonryRefCat2018,
       author = {{Tonry}, J.~L. and {Denneau}, L. and {Flewelling}, H. and {Heinze}, A.~N. and {Onken}, C.~A. and {Smartt}, S.~J. and {Stalder}, B. and {Weiland}, H.~J. and {Wolf}, C.},
        title = "{The ATLAS All-Sky Stellar Reference Catalog}",
      journal = {\apj},
         year = 2018,
        month = nov,
       volume = {867},
       number = {2},
          eid = {105},
        pages = {105},
          doi = {10.3847/1538-4357/aae386},
archivePrefix = {arXiv},
       eprint = {1809.09157},
 primaryClass = {astro-ph.IM},
       adsurl = {https://ui.adsabs.harvard.edu/abs/2018ApJ...867..105T}
}

@TechReport{MonsterRefCatFerguson2025,
   author = {{Ferguson}, P. S. and {Rykoff}, E. S. and {Carlin}, J. L. and {Saunders}, C. and {Parejko}, J. K},
   title = "{The Monster: A reference catalog with synthetic ugrizy-band fluxes for the Vera C. Rubin observatory}",
   institution = "{Vera C. Rubin Observatory}",
   year = "2025",
   month = "August",
   handle = "DMTN-277",
   type = "{Data Management Technical Note}",
   number = "{DMTN-277}",
   doi = "10.71929/rubin/2583688",
   url = "https://dmtn-277.lsst.io"
}

@TechReport{10.71929/rubin/2570545,
   author = "{Vera C. Rubin Observatory Science Pipelines Developers}",
   title = "{The LSST Science Pipelines Software: Optical Survey Pipeline Reduction and Analysis Environment}",
   institution = "{Vera C. Rubin Observatory}",
   year = "2025",
   month = "June",
   handle = "PSTN-019",
   type = "{Project Science Technical Note}",
   number = "PSTN-019",
   doi = "10.71929/rubin/2570545",
   url = "https://pstn-019.lsst.io/"
}

@misc{jones_2025_15128504,
  author       = {Jones, Lynne and
                  Bianco, Federica Bettina and
                  Yoachim, Peter and
                  Neilsen, Eric},
  title        = {https://survey-strategy.lsst.io},
  month        = apr,
  year         = 2025,
  publisher    = {Zenodo},
  version      = {v0.1.0},
  doi          = {10.5281/zenodo.15128504},
  url          = {https://doi.org/10.5281/zenodo.15128504},
}

@ARTICLE{biancoRubinObservingCadence2022,
       author = {{Bianco}, Federica B. and {Ivezi{\'c}}, {\v{Z}}eljko and {Jones}, R. Lynne and {Graham}, Melissa L. and {Marshall}, Phil and {Saha}, Abhijit and {Strauss}, Michael A. and {Yoachim}, Peter and {Ribeiro}, Tiago and {Anguita}, Timo and {Bauer}, A.~E. and {Bauer}, Franz E. and {Bellm}, Eric C. and {Blum}, Robert D. and {Brandt}, William N. and {Brough}, Sarah and {Catelan}, M{\'a}rcio and {Clarkson}, William I. and {Connolly}, Andrew J. and {Gawiser}, Eric and {Gizis}, John E. and {Hlo{\v{z}}ek}, Ren{\'e}e and {Kaviraj}, Sugata and {Liu}, Charles T. and {Lochner}, Michelle and {Mahabal}, Ashish A. and {Mandelbaum}, Rachel and {McGehee}, Peregrine and {Neilsen}, Jr., Eric H. and {Olsen}, Knut A.~G. and {Peiris}, Hiranya V. and {Rhodes}, Jason and {Richards}, Gordon T. and {Ridgway}, Stephen and {Schwamb}, Megan E. and {Scolnic}, Dan and {Shemmer}, Ohad and {Slater}, Colin T. and {Slosar}, An{\v{z}}e and {Smartt}, Stephen J. and {Strader}, Jay and {Street}, Rachel and {Trilling}, David E. and {Verma}, Aprajita and {Vivas}, A.~K. and {Wechsler}, Risa H. and {Willman}, Beth},
        title = "{Optimization of the Observing Cadence for the Rubin Observatory Legacy Survey of Space and Time: A Pioneering Process of Community-focused Experimental Design}",
      journal = {\apjs},
         year = 2022,
        month = jan,
       volume = {258},
       number = {1},
          eid = {1},
        pages = {1},
          doi = {10.3847/1538-4365/ac3e72},
archivePrefix = {arXiv},
       eprint = {2108.01683},
 primaryClass = {astro-ph.IM},
       adsurl = {https://ui.adsabs.harvard.edu/abs/2022ApJS..258....1B}
}

@ARTICLE{at2025agpzTNSdiscreport,
       author = {{Tonry}, J. and {Denneau}, L. and {Weiland}, H. and {Erasmus}, N. and {Koorts}, W. and {Jordan}, A. and {Suc}, V. and {Alarc{\'o}n}, M.~R. and {Licandro}, J. and {Nichita}, P. and {Smartt}, S.~J. and {Smith}, K.~W. and {Young}, D.~R. and {Nicholl}, M. and {Fulton}, M. and {McCollum}, M. and {Moore}, T. and {Weston}, J. and {Sheng}, X. and {Angus}, C.~R. and {Wilson}, A. and {Aamer}, A. and {Magill}, D. and {Broda}, P.~J. and {Smith}, A.~J. and {Ramsden}, P. and {Shingles}, L. and {Srivastav}, S. and {Gillanders}, J.~H. and {Stevance}, H. and {Cooper}, A.~J. and {Stoppa}, F. and {Tweddle}, J. and {Eastman}, L. and {Rhodes}, L. and {Rest}, A. and {Chen}, T.~W. and {Stubbs}, C. and {Sommer}, J.~S. and {Schmidt}, B.~P.},
        title = "{ATLAS Transient Discovery Report for 2025-12-13}",
      journal = {Transient Name Server Discovery Report},
         year = 2025,
        month = dec,
       volume = {2025-4949},
        pages = {1},
       adsurl = {https://ui.adsabs.harvard.edu/abs/2025TNSTR4949....1T}
}

@ARTICLE{lvra2026,
       author = {{Stevance}, H.~F. and {Smartt}, S.~J.},
        title = "{The Lasair Virtual Research Assistant (LVRA) annotator for Rubin alerts - r0b.a1}",
      journal = {Transient Name Server AstroNote},
         year = 2026,
        month = apr,
       volume = {107},
        pages = {1},
       adsurl = {https://ui.adsabs.harvard.edu/abs/2026TNSAN.107....1S}
}

@article{deyOverviewDESILegacy2019,
    title = {Overview of the {DESI} {Legacy} {Imaging} {Surveys}},
    volume = {157},
    issn = {0004-6256},
    url = {https://ui.adsabs.harvard.edu/abs/2019AJ....157..168D},
    doi = {10.3847/1538-3881/ab089d},
    urldate = {2026-06-04},
    journal = {The Astronomical Journal},
    publisher = {IOP},
    author = {Dey, Arjun and Schlegel, David J. and Lang, Dustin and Blum, Robert and Burleigh, Kaylan and Fan, Xiaohui and Findlay, Joseph R. and Finkbeiner, Doug and Herrera, David and Juneau, Stéphanie and Landriau, Martin and Levi, Michael and McGreer, Ian and Meisner, Aaron and Myers, Adam D. and Moustakas, John and Nugent, Peter and Patej, Anna and Schlafly, Edward F. and Walker, Alistair R. and Valdes, Francisco and Weaver, Benjamin A. and Yèche, Christophe and Zou, Hu and Zhou, Xu and Abareshi, Behzad and Abbott, T. M. C. and Abolfathi, Bela and Aguilera, C. and Alam, Shadab and Allen, Lori and Alvarez, A. and Annis, James and Ansarinejad, Behzad and Aubert, Marie and Beechert, Jacqueline and Bell, Eric F. and BenZvi, Segev Y. and Beutler, Florian and Bielby, Richard M. and Bolton, Adam S. and Briceño, César and Buckley-Geer, Elizabeth J. and Butler, Karen and Calamida, Annalisa and Carlberg, Raymond G. and Carter, Paul and Casas, Ricard and Castander, Francisco J. and Choi, Yumi and Comparat, Johan and Cukanovaite, Elena and Delubac, Timothée and DeVries, Kaitlin and Dey, Sharmila and Dhungana, Govinda and Dickinson, Mark and Ding, Zhejie and Donaldson, John B. and Duan, Yutong and Duckworth, Christopher J. and Eftekharzadeh, Sarah and Eisenstein, Daniel J. and Etourneau, Thomas and Fagrelius, Parker A. and Farihi, Jay and Fitzpatrick, Mike and Font-Ribera, Andreu and Fulmer, Leah and Gänsicke, Boris T. and Gaztanaga, Enrique and George, Koshy and Gerdes, David W. and Gontcho, Satya Gontcho A. and Gorgoni, Claudio and Green, Gregory and Guy, Julien and Harmer, Diane and Hernandez, M. and Honscheid, Klaus and Huang, Lijuan Wendy and James, David J. and Jannuzi, Buell T. and Jiang, Linhua and Joyce, Richard and Karcher, Armin and Karkar, Sonia and Kehoe, Robert and Kneib, Jean-Paul and Kueter-Young, Andrea and Lan, Ting-Wen and Lauer, Tod R. and Le Guillou, Laurent and Le Van Suu, Auguste and Lee, Jae Hyeon and Lesser, Michael and Perreault Levasseur, Laurence and Li, Ting S. and Mann, Justin L. and Marshall, Robert and Martínez-Vázquez, C. E. and Martini, Paul and du Mas des Bourboux, Hélion and McManus, Sean and Meier, Tobias Gabriel and Ménard, Brice and Metcalfe, Nigel and Muñoz-Gutiérrez, Andrea and Najita, Joan and Napier, Kevin and Narayan, Gautham and Newman, Jeffrey A. and Nie, Jundan and Nord, Brian and Norman, Dara J. and Olsen, Knut A. G. and Paat, Anthony and Palanque-Delabrouille, Nathalie and Peng, Xiyan and Poppett, Claire L. and Poremba, Megan R. and Prakash, Abhishek and Rabinowitz, David and Raichoor, Anand and Rezaie, Mehdi and Robertson, A. N. and Roe, Natalie A. and Ross, Ashley J. and Ross, Nicholas P. and Rudnick, Gregory and Safonova, Sasha and Saha, Abhijit and Sánchez, F. Javier and Savary, Elodie and Schweiker, Heidi and Scott, Adam and Seo, Hee-Jong and Shan, Huanyuan and Silva, David R. and Slepian, Zachary and Soto, Christian and Sprayberry, David and Staten, Ryan and Stillman, Coley M. and Stupak, Robert J. and Summers, David L. and Sien Tie, Suk and Tirado, H. and Vargas-Magaña, Mariana and Vivas, A. Katherina and Wechsler, Risa H. and Williams, Doug and Yang, Jinyi and Yang, Qian and Yapici, Tolga and Zaritsky, Dennis and Zenteno, A. and Zhang, Kai and Zhang, Tianmeng and Zhou, Rongpu and Zhou, Zhimin},
    month = may,
    year = {2019},
    note = {ADS Bibcode: 2019AJ....157..168D},
    pages = {168},
}

@article{lejaDerivingPhysicalProperties2017,
    title = {Deriving {Physical} {Properties} from {Broadband} {Photometry} with {Prospector}: {Description} of the {Model} and a {Demonstration} of its {Accuracy} {Using} 129 {Galaxies} in the {Local} {Universe}},
    volume = {837},
    issn = {0004-637X},
    shorttitle = {Deriving {Physical} {Properties} from {Broadband} {Photometry} with {Prospector}},
    url = {https://ui.adsabs.harvard.edu/abs/2017ApJ...837..170L},
    doi = {10.3847/1538-4357/aa5ffe},
    urldate = {2025-11-20},
    journal = {The Astrophysical Journal},
    publisher = {IOP},
    author = {Leja, Joel and Johnson, Benjamin D. and Conroy, Charlie and van Dokkum, Pieter G. and Byler, Nell},
    month = mar,
    year = {2017},
    note = {ADS Bibcode: 2017ApJ...837..170L},
    pages = {170},
}

@article{johnsonStellarPopulationInference2021,
    title = {Stellar {Population} {Inference} with {Prospector}},
    volume = {254},
    issn = {0067-0049},
    url = {https://ui.adsabs.harvard.edu/abs/2021ApJS..254...22J},
    doi = {10.3847/1538-4365/abef67},
    urldate = {2025-11-20},
    journal = {The Astrophysical Journal Supplement Series},
    publisher = {IOP},
    author = {Johnson, Benjamin D. and Leja, Joel and Conroy, Charlie and Speagle, Joshua S.},
    month = jun,
    year = {2021},
    note = {ADS Bibcode: 2021ApJS..254...22J},
    pages = {22},
}

@article{speagleDYNESTYDynamicNested2020,
    title = {{DYNESTY}: a dynamic nested sampling package for estimating {Bayesian} posteriors and evidences},
    volume = {493},
    issn = {0035-8711},
    shorttitle = {{DYNESTY}},
    url = {https://ui.adsabs.harvard.edu/abs/2020MNRAS.493.3132S},
    doi = {10.1093/mnras/staa278},
    urldate = {2025-04-11},
    journal = {Monthly Notices of the Royal Astronomical Society},
    publisher = {OUP},
    author = {Speagle, Joshua S.},
    month = apr,
    year = {2020},
    note = {ADS Bibcode: 2020MNRAS.493.3132S},
    pages = {3132--3158},
}

@article{tonryATLASHighcadenceAllsky2018,
    title = {{ATLAS}: {A} {High}-cadence {All}-sky {Survey} {System}},
    volume = {130},
    issn = {0004-6280},
    shorttitle = {{ATLAS}},
    url = {https://ui.adsabs.harvard.edu/abs/2018PASP..130f4505T},
    doi = {10.1088/1538-3873/aabadf},
    urldate = {2025-04-13},
    journal = {Publications of the Astronomical Society of the Pacific},
    publisher = {IOP},
    author = {Tonry, J. L. and Denneau, L. and Heinze, A. N. and Stalder, B. and Smith, K. W. and Smartt, S. J. and Stubbs, C. W. and Weiland, H. J. and Rest, A.},
    month = jun,
    year = {2018},
    note = {ADS Bibcode: 2018PASP..130f4505T},
    pages = {064505},
}

@article{restTestingLMCMicrolensing2005,
    title = {Testing {LMC} {Microlensing} {Scenarios}: {The} {Discrimination} {Power} of the {SuperMACHO} {Microlensing} {Survey}},
    volume = {634},
    issn = {0004-637X},
    shorttitle = {Testing {LMC} {Microlensing} {Scenarios}},
    url = {https://ui.adsabs.harvard.edu/abs/2005ApJ...634.1103R},
    doi = {10.1086/497060},
    urldate = {2026-06-05},
    journal = {The Astrophysical Journal},
    publisher = {IOP},
    author = {Rest, A. and Stubbs, C. and Becker, A. C. and Miknaitis, G. A. and Miceli, A. and Covarrubias, R. and Hawley, S. L. and Smith, R. C. and Suntzeff, N. B. and Olsen, K. and Prieto, J. L. and Hiriart, R. and Welch, D. L. and Cook, K. H. and Nikolaev, S. and Huber, M. and Prochtor, G. and Clocchiatti, A. and Minniti, D. and Garg, A. and Challis, P. and Keller, S. C. and Schmidt, B. P.},
    month = dec,
    year = {2005},
    note = {ADS Bibcode: 2005ApJ...634.1103R},
    pages = {1103--1115},
}

@article{smithDesignOperationATLAS2020,
    title = {Design and {Operation} of the {ATLAS} {Transient} {Science} {Server}},
    volume = {132},
    issn = {0004-6280},
    url = {https://ui.adsabs.harvard.edu/abs/2020PASP..132h5002S},
    doi = {10.1088/1538-3873/ab936e},
    urldate = {2025-04-13},
    journal = {Publications of the Astronomical Society of the Pacific},
    publisher = {IOP},
    author = {Smith, K. W. and Smartt, S. J. and Young, D. R. and Tonry, J. L. and Denneau, L. and Flewelling, H. and Heinze, A. N. and Weiland, H. J. and Stalder, B. and Rest, A. and Stubbs, C. W. and Anderson, J. P. and Chen, T. -W. and Clark, P. and Do, A. and Förster, F. and Fulton, M. and Gillanders, J. and McBrien, O. R. and O'Neill, D. and Srivastav, S. and Wright, D. E.},
    month = aug,
    year = {2020},
    note = {ADS Bibcode: 2020PASP..132h5002S},
    pages = {085002},
}

@article{stevanceATLASVirtualResearch2025,
    title = {The {ATLAS} {Virtual} {Research} {Assistant}},
    volume = {990},
    issn = {0004-637X},
    url = {https://ui.adsabs.harvard.edu/abs/2025ApJ...990..201S},
    doi = {10.3847/1538-4357/adf2a1},
    urldate = {2026-06-05},
    journal = {The Astrophysical Journal},
    publisher = {IOP},
    author = {Stevance, H. F. and Smith, K. W. and Smartt, S. J. and Roberts, S. J. and Erasmus, N. and Young, D. R. and Clocchiatti, A.},
    month = sep,
    year = {2025},
    note = {ADS Bibcode: 2025ApJ...990..201S},
    pages = {201},
}

@article{williamsEnablingScienceRubin2024,
    title = {Enabling science from the {Rubin} alert stream with {Lasair}},
    volume = {3},
    url = {https://ui.adsabs.harvard.edu/abs/2024RASTI...3..362W},
    doi = {10.1093/rasti/rzae024},
    urldate = {2026-06-05},
    journal = {RAS Techniques and Instruments},
    author = {Williams, Roy D. and Francis, Gareth P. and Lawrence, Andy and Sloan, Terence M. and Smartt, Stephen J. and Smith, Ken W. and Young, David R.},
    month = jan,
    year = {2024},
    note = {ADS Bibcode: 2024RASTI...3..362W},
    pages = {362--371},
}

@article{freudlingAutomatedDataReduction2013,
    title = {Automated data reduction workflows for astronomy. {The} {ESO} {Reflex} environment},
    volume = {559},
    issn = {0004-6361},
    url = {https://ui.adsabs.harvard.edu/abs/2013A&A...559A..96F},
    doi = {10.1051/0004-6361/201322494},
    urldate = {2026-06-06},
    journal = {Astronomy and Astrophysics},
    publisher = {EDP},
    author = {Freudling, W. and Romaniello, M. and Bramich, D. M. and Ballester, P. and Forchi, V. and García-Dabló, C. E. and Moehler, S. and Neeser, M. J.},
    month = nov,
    year = {2013},
    note = {ADS Bibcode: 2013A\&A...559A..96F},
    pages = {A96},
}

@article{smetteMolecfitGeneralTool2015,
    title = {Molecfit: {A} general tool for telluric absorption correction. {I}. {Method} and application to {ESO} instruments},
    volume = {576},
    issn = {0004-6361},
    shorttitle = {Molecfit},
    url = {https://ui.adsabs.harvard.edu/abs/2015A&A...576A..77S},
    doi = {10.1051/0004-6361/201423932},
    urldate = {2026-06-06},
    journal = {Astronomy and Astrophysics},
    publisher = {EDP},
    author = {Smette, A. and Sana, H. and Noll, S. and Horst, H. and Kausch, W. and Kimeswenger, S. and Barden, M. and Szyszka, C. and Jones, A. M. and Gallenne, A. and Vinther, J. and Ballester, P. and Taylor, J.},
    month = apr,
    year = {2015},
    note = {ADS Bibcode: 2015A\&A...576A..77S},
    pages = {A77},
}

@article{kauschMolecfitGeneralTool2015,
    title = {Molecfit: {A} general tool for telluric absorption correction. {II}. {Quantitative} evaluation on {ESO}-{VLT}/{X}-{Shooterspectra}},
    volume = {576},
    issn = {0004-6361},
    shorttitle = {Molecfit},
    url = {https://ui.adsabs.harvard.edu/abs/2015A&A...576A..78K},
    doi = {10.1051/0004-6361/201423909},
    urldate = {2026-06-06},
    journal = {Astronomy and Astrophysics},
    publisher = {EDP},
    author = {Kausch, W. and Noll, S. and Smette, A. and Kimeswenger, S. and Barden, M. and Szyszka, C. and Jones, A. M. and Sana, H. and Horst, H. and Kerber, F.},
    month = apr,
    year = {2015},
    note = {ADS Bibcode: 2015A\&A...576A..78K},
    pages = {A78},
}

@article{nicholl2022aedmNewClass2023,
    title = {{AT} 2022aedm and a {New} {Class} of {Luminous}, {Fast}-cooling {Transients} in {Elliptical} {Galaxies}},
    volume = {954},
    issn = {0004-637X},
    url = {https://ui.adsabs.harvard.edu/abs/2023ApJ...954L..28N},
    doi = {10.3847/2041-8213/acf0ba},
    urldate = {2025-04-13},
    journal = {The Astrophysical Journal},
    publisher = {IOP},
    author = {Nicholl, M. and Srivastav, S. and Fulton, M. D. and Gomez, S. and Huber, M. E. and Oates, S. R. and Ramsden, P. and Rhodes, L. and Smartt, S. J. and Smith, K. W. and Aamer, A. and Anderson, J. P. and Bauer, F. E. and Berger, E. and de Boer, T. and Chambers, K. C. and Charalampopoulos, P. and Chen, T. -W. and Fender, R. P. and Fraser, M. and Gao, H. and Green, D. A. and Galbany, L. and Gompertz, B. P. and Gromadzki, M. and Gutiérrez, C. P. and Howell, D. A. and Inserra, C. and Jonker, P. G. and Kopsacheili, M. and Lowe, T. B. and Magnier, E. A. and McCully, C. and McGee, S. L. and Moore, T. and Müller-Bravo, T. E. and Newsome, M. and Gonzalez, E. Padilla and Pellegrino, C. and Pessi, T. and Pursiainen, M. and Rest, A. and Ridley, E. J. and Shappee, B. J. and Sheng, X. and Smith, G. P. and Terreran, G. and Tucker, M. A. and Vinkó, J. and Wainscoat, R. J. and Wiseman, P. and Young, D. R.},
    month = sep,
    year = {2023},
    note = {ADS Bibcode: 2023ApJ...954L..28N},
    pages = {L28},
}

@article{angusSpectroscopicObservations2025agpz2026,
    title = {Spectroscopic {Observations} of {AT} 2025agpz, a {Long}-{Rising} {Hydrogen}-{Rich} {Transient} in {Rubin}-{LSST} {Commissioning} {Data}},
    volume = {45},
    url = {https://ui.adsabs.harvard.edu/abs/2026TNSAN..45....1A},
    urldate = {2026-06-26},
    journal = {Transient Name Server AstroNote},
    author = {Angus, C. R. and Stevance, H. and Nicholl, M. and Smith, K. W. and Smartt, S. and Young, D. R. and Fulton, M. and McCollum, M. and Moore, T. and Weston, J. and Sheng, X. and Aamer, A. and Magill, D. and Smith, A. J. and Ramsden, P. and Shingles, L. and Srivastav, S. and Gillanders, J. and Cooper, A. J. and Stoppa, F. and Tweddle, J. and Eastman, L. and Rhodes, L. and Denneau, L. and Tonry, J. and Weiland, H. and Siverd, R. and Erasmus, N. and Koorts, W. and Jordan, A. and Suc, V. and Alarcón, M. R. and Licandro, J. and Nichita, P. and Rest, A. and Chen, T. W. and Stubbs, C. and Sommer, J. and Schmidt, B. P.},
    month = feb,
    year = {2026},
    note = {ADS Bibcode: 2026TNSAN..45....1A},
    pages = {1},
}

@article{schlaflyMeasuringReddeningSloan2011,
    title = {Measuring {Reddening} with {Sloan} {Digital} {Sky} {Survey} {Stellar} {Spectra} and {Recalibrating} {SFD}},
    volume = {737},
    issn = {0004-637X},
    url = {https://ui.adsabs.harvard.edu/abs/2011ApJ...737..103S},
    doi = {10.1088/0004-637X/737/2/103},
    urldate = {2025-12-01},
    journal = {The Astrophysical Journal},
    publisher = {IOP},
    author = {Schlafly, Edward F. and Finkbeiner, Douglas P.},
    month = aug,
    year = {2011},
    note = {ADS Bibcode: 2011ApJ...737..103S},
    pages = {103},
}

@article{pessiSampleHydrogenrichSuperluminous2025,
    title = {Sample of hydrogen-rich superluminous supernovae from the {Zwicky} {Transient} {Facility}},
    volume = {695},
    issn = {0004-6361},
    url = {https://ui.adsabs.harvard.edu/abs/2025A&A...695A.142P},
    doi = {10.1051/0004-6361/202452014},
    urldate = {2026-07-09},
    journal = {Astronomy and Astrophysics},
    publisher = {EDP},
    author = {Pessi, P. J. and Lunnan, R. and Sollerman, J. and Schulze, S. and Gkini, A. and Gangopadhyay, A. and Yan, L. and Gal-Yam, A. and Perley, D. A. and Chen, T.-W. and Hinds, K. R. and Brennan, S. J. and Hu, Y. and Singh, A. and Andreoni, I. and Cook, D. O. and Fremling, C. and Ho, A. Y. Q. and Sharma, Y. and van Velzen, S. and Kangas, T. and Wold, A. and Bellm, E. C. and Bloom, J. S. and Graham, M. J. and Kasliwal, M. M. and Kulkarni, S. R. and Riddle, R. and Rusholme, B.},
    month = mar,
    year = {2025},
    note = {ADS Bibcode: 2025A\&A...695A.142P},
    pages = {A142},
}

@article{wisemanSystematicallySelectedSample2025,
    title = {A systematically selected sample of luminous, long-duration, ambiguous nuclear transients},
    volume = {537},
    issn = {0035-8711},
    url = {https://ui.adsabs.harvard.edu/abs/2025MNRAS.537.2024W},
    doi = {10.1093/mnras/staf116},
    urldate = {2026-07-09},
    journal = {Monthly Notices of the Royal Astronomical Society},
    publisher = {OUP},
    author = {Wiseman, P. and Williams, R. D. and Arcavi, I. and Galbany, L. and Graham, M. J. and Hönig, S. and Newsome, M. and Subrayan, B. and Sullivan, M. and Wang, Y. and Ilić, D. and Nicholl, M. and Oates, S. and Petrushevska, T. and Smith, K. W.},
    month = feb,
    year = {2025},
    note = {ADS Bibcode: 2025MNRAS.537.2024W},
    pages = {2024--2045},
}

@article{frederickFamilyTreeOptical2021,
    title = {A {Family} {Tree} of {Optical} {Transients} from {Narrow}-line {Seyfert} 1 {Galaxies}},
    volume = {920},
    issn = {0004-637X},
    url = {https://doi.org/10.3847/1538-4357/ac110f},
    doi = {10.3847/1538-4357/ac110f},
    language = {en},
    number = {1},
    urldate = {2026-07-10},
    journal = {The Astrophysical Journal},
    publisher = {The American Astronomical Society},
    author = {Frederick, Sara and Gezari, Suvi and Graham, Matthew J. and Sollerman, Jesper and van Velzen, Sjoert and Perley, Daniel A. and Stern, Daniel and Ward, Charlotte and Hammerstein, Erica and Hung, Tiara and Yan, Lin and Andreoni, Igor and Bellm, Eric C. and Duev, Dmitry A. and Kowalski, Marek and Mahabal, Ashish A. and Masci, Frank J. and Medford, Michael and Rusholme, Ben and Smith, Roger and Walters, Richard},
    month = oct,
    year = {2021},
    pages = {56},
}

@article{hinkleMostEnergeticTransients2025,
    title = {The most energetic transients: {Tidal} disruptions of high-mass stars},
    volume = {11},
    shorttitle = {The most energetic transients},
    url = {https://ui.adsabs.harvard.edu/abs/2025SciA...11...74H},
    doi = {10.1126/sciadv.adt0074},
    urldate = {2026-07-10},
    journal = {Science Advances},
    author = {Hinkle, Jason T. and Shappee, Benjamin J. and Auchettl, Katie and Kochanek, Christopher S. and Neustadt, Jack M. M. and Polin, Abigail and Strader, Jay and Holoien, Thomas W.-S. and Huber, Mark E. and Tucker, Michael A. and Ashall, Christopher and de Jaeger, Thomas and Desai, Dhvanil D. and Do, Aaron and Hoogendam, Willem B. and Payne, Anna V.},
    month = jun,
    year = {2025},
    note = {ADS Bibcode: 2025SciA...11...74H},
    pages = {eadt0074},
}

@article{gal-yamLuminousSupernovae2012,
    title = {Luminous {Supernovae}},
    volume = {337},
    issn = {0036-8075},
    url = {https://ui.adsabs.harvard.edu/abs/2012Sci...337..927G},
    doi = {10.1126/science.1203601},
    urldate = {2026-06-26},
    journal = {Science},
    author = {Gal-Yam, Avishay},
    month = aug,
    year = {2012},
    note = {ADS Bibcode: 2012Sci...337..927G},
    pages = {927},
}

@article{quimbyHydrogenpoorSuperluminousStellar2011,
    title = {Hydrogen-poor superluminous stellar explosions},
    volume = {474},
    issn = {0028-0836},
    url = {https://ui.adsabs.harvard.edu/abs/2011Natur.474..487Q},
    doi = {10.1038/nature10095},
    urldate = {2026-07-13},
    journal = {Nature},
    author = {Quimby, R. M. and Kulkarni, S. R. and Kasliwal, M. M. and Gal-Yam, A. and Arcavi, I. and Sullivan, M. and Nugent, P. and Thomas, R. and Howell, D. A. and Nakar, E. and Bildsten, L. and Theissen, C. and Law, N. M. and Dekany, R. and Rahmer, G. and Hale, D. and Smith, R. and Ofek, E. O. and Zolkower, J. and Velur, V. and Walters, R. and Henning, J. and Bui, K. and McKenna, D. and Poznanski, D. and Cenko, S. B. and Levitan, D.},
    month = jun,
    year = {2011},
    note = {ADS Bibcode: 2011Natur.474..487Q},
    pages = {487--489},
}

@article{inserraSuperluminousTypeIc2013,
    title = {Super-luminous {Type} {Ic} {Supernovae}: {Catching} a {Magnetar} by the {Tail}},
    volume = {770},
    issn = {0004-637X},
    shorttitle = {Super-luminous {Type} {Ic} {Supernovae}},
    url = {https://ui.adsabs.harvard.edu/abs/2013ApJ...770..128I},
    doi = {10.1088/0004-637X/770/2/128},
    urldate = {2026-07-13},
    journal = {The Astrophysical Journal},
    publisher = {IOP},
    author = {Inserra, C. and Smartt, S. J. and Jerkstrand, A. and Valenti, S. and Fraser, M. and Wright, D. and Smith, K. and Chen, T.-W. and Kotak, R. and Pastorello, A. and Nicholl, M. and Bresolin, F. and Kudritzki, R. P. and Benetti, S. and Botticella, M. T. and Burgett, W. S. and Chambers, K. C. and Ergon, M. and Flewelling, H. and Fynbo, J. P. U. and Geier, S. and Hodapp, K. W. and Howell, D. A. and Huber, M. and Kaiser, N. and Leloudas, G. and Magill, L. and Magnier, E. A. and McCrum, M. G. and Metcalfe, N. and Price, P. A. and Rest, A. and Sollerman, J. and Sweeney, W. and Taddia, F. and Taubenberger, S. and Tonry, J. L. and Wainscoat, R. J. and Waters, C. and Young, D.},
    month = jun,
    year = {2013},
    note = {ADS Bibcode: 2013ApJ...770..128I},
    pages = {128},
}

@article{droutRapidlyEvolvingLuminous2014,
    title = {Rapidly {Evolving} and {Luminous} {Transients} from {Pan}-{STARRS1}},
    volume = {794},
    issn = {0004-637X},
    url = {https://ui.adsabs.harvard.edu/abs/2014ApJ...794...23D},
    doi = {10.1088/0004-637X/794/1/23},
    urldate = {2026-07-13},
    journal = {The Astrophysical Journal},
    publisher = {IOP},
    author = {Drout, M. R. and Chornock, R. and Soderberg, A. M. and Sanders, N. E. and McKinnon, R. and Rest, A. and Foley, R. J. and Milisavljevic, D. and Margutti, R. and Berger, E. and Calkins, M. and Fong, W. and Gezari, S. and Huber, M. E. and Kankare, E. and Kirshner, R. P. and Leibler, C. and Lunnan, R. and Mattila, S. and Marion, G. H. and Narayan, G. and Riess, A. G. and Roth, K. C. and Scolnic, D. and Smartt, S. J. and Tonry, J. L. and Burgett, W. S. and Chambers, K. C. and Hodapp, K. W. and Jedicke, R. and Kaiser, N. and Magnier, E. A. and Metcalfe, N. and Morgan, J. S. and Price, P. A. and Waters, C.},
    month = oct,
    year = {2014},
    note = {ADS Bibcode: 2014ApJ...794...23D},
    pages = {23},
}

@article{pursiainenRapidlyEvolvingTransients2018,
    title = {Rapidly evolving transients in the {Dark} {Energy} {Survey}},
    volume = {481},
    issn = {0035-8711},
    url = {https://ui.adsabs.harvard.edu/abs/2018MNRAS.481..894P},
    doi = {10.1093/mnras/sty2309},
    urldate = {2026-07-13},
    journal = {Monthly Notices of the Royal Astronomical Society},
    publisher = {OUP},
    author = {Pursiainen, M. and Childress, M. and Smith, M. and Prajs, S. and Sullivan, M. and Davis, T. M. and Foley, R. J. and Asorey, J. and Calcino, J. and Carollo, D. and Curtin, C. and D'Andrea, C. B. and Glazebrook, K. and Gutierrez, C. and Hinton, S. R. and Hoormann, J. K. and Inserra, C. and Kessler, R. and King, A. and Kuehn, K. and Lewis, G. F. and Lidman, C. and Macaulay, E. and Möller, A. and Nichol, R. C. and Sako, M. and Sommer, N. E. and Swann, E. and Tucker, B. E. and Uddin, S. A. and Wiseman, P. and Zhang, B. and Abbott, T. M. C. and Abdalla, F. B. and Allam, S. and Annis, J. and Avila, S. and Brooks, D. and Buckley-Geer, E. and Burke, D. L. and Carnero Rosell, A. and Carrasco Kind, M. and Carretero, J. and Castander, F. J. and Cunha, C. E. and Davis, C. and De Vicente, J. and Diehl, H. T. and Doel, P. and Eifler, T. F. and Flaugher, B. and Fosalba, P. and Frieman, J. and García-Bellido, J. and Gruen, D. and Gruendl, R. A. and Gutierrez, G. and Hartley, W. G. and Hollowood, D. L. and Honscheid, K. and James, D. J. and Jeltema, T. and Kuropatkin, N. and Li, T. S. and Lima, M. and Maia, M. A. G. and Martini, P. and Menanteau, F. and Ogando, R. L. C. and Plazas, A. A. and Roodman, A. and Sanchez, E. and Scarpine, V. and Schindler, R. and Smith, R. C. and Soares-Santos, M. and Sobreira, F. and Suchyta, E. and Swanson, M. E. C. and Tarle, G. and Tucker, D. L. and Walker, A. R. and {DES Collaboration}},
    month = nov,
    year = {2018},
    note = {ADS Bibcode: 2018MNRAS.481..894P},
    pages = {894--917},
}

@article{hoSearchExtragalacticFast2023,
    title = {A {Search} for {Extragalactic} {Fast} {Blue} {Optical} {Transients} in {ZTF} and the {Rate} of {AT2018cow}-like {Transients}},
    volume = {949},
    issn = {0004-637X},
    url = {https://ui.adsabs.harvard.edu/abs/2023ApJ...949..120H},
    doi = {10.3847/1538-4357/acc533},
    urldate = {2026-07-13},
    journal = {The Astrophysical Journal},
    publisher = {IOP},
    author = {Ho, Anna Y. Q. and Perley, Daniel A. and Gal-Yam, Avishay and Lunnan, Ragnhild and Sollerman, Jesper and Schulze, Steve and Das, Kaustav K. and Dobie, Dougal and Yao, Yuhan and Fremling, Christoffer and Adams, Scott and Anand, Shreya and Andreoni, Igor and Bellm, Eric C. and Bruch, Rachel J. and Burdge, Kevin B. and Castro-Tirado, Alberto J. and Dahiwale, Aishwarya and De, Kishalay and Dekany, Richard and Drake, Andrew J. and Duev, Dmitry A. and Graham, Matthew J. and Helou, George and Kaplan, David L. and Karambelkar, Viraj and Kasliwal, Mansi M. and Kool, Erik C. and Kulkarni, S. R. and Mahabal, Ashish A. and Medford, Michael S. and Miller, A. A. and Nordin, Jakob and Ofek, Eran and Petitpas, Glen and Riddle, Reed and Sharma, Yashvi and Smith, Roger and Stewart, Adam J. and Taggart, Kirsty and Tartaglia, Leonardo and Tzanidakis, Anastasios and Winters, Jan Martin},
    month = jun,
    year = {2023},
    note = {ADS Bibcode: 2023ApJ...949..120H},
    pages = {120},
}

@article{hammersteinFinalSeasonReimagined2023a,
    title = {The {Final} {Season} {Reimagined}: 30 {Tidal} {Disruption} {Events} from the {ZTF}-{I} {Survey}},
    volume = {942},
    issn = {0004-637X},
    shorttitle = {The {Final} {Season} {Reimagined}},
    url = {https://ui.adsabs.harvard.edu/abs/2023ApJ...942....9H},
    doi = {10.3847/1538-4357/aca283},
    urldate = {2026-07-13},
    journal = {The Astrophysical Journal},
    publisher = {IOP},
    author = {Hammerstein, Erica and van Velzen, Sjoert and Gezari, Suvi and Cenko, S. Bradley and Yao, Yuhan and Ward, Charlotte and Frederick, Sara and Villanueva, Natalia and Somalwar, Jean J. and Graham, Matthew J. and Kulkarni, Shrinivas R. and Stern, Daniel and Andreoni, Igor and Bellm, Eric C. and Dekany, Richard and Dhawan, Suhail and Drake, Andrew J. and Fremling, Christoffer and Gatkine, Pradip and Groom, Steven L. and Ho, Anna Y. Q. and Kasliwal, Mansi M. and Karambelkar, Viraj and Kool, Erik C. and Masci, Frank J. and Medford, Michael S. and Perley, Daniel A. and Purdum, Josiah and van Roestel, Jan and Sharma, Yashvi and Sollerman, Jesper and Taggart, Kirsty and Yan, Lin},
    month = jan,
    year = {2023},
    note = {ADS Bibcode: 2023ApJ...942....9H},
    pages = {9},
}

@article{somalwarVLASSTidalDisruption2025,
    title = {{VLASS} {Tidal} {Disruption} {Events} with {Optical} {Flares}. {I}. {The} {Sample} and a {Comparison} to {Optically} {Selected} {TDEs}},
    volume = {982},
    issn = {0004-637X},
    url = {https://doi.org/10.3847/1538-4357/adba4f},
    doi = {10.3847/1538-4357/adba4f},
    language = {en},
    number = {2},
    urldate = {2026-07-13},
    journal = {The Astrophysical Journal},
    publisher = {The American Astronomical Society},
    author = {Somalwar, Jean J. and Ravi, Vikram and Dong, Dillon Z. and Hammerstein, Erica and Hallinan, Gregg and Law, Casey and Miller, Jessie and Myers, Steven T. and Yao, Yuhan and Dekany, Richard and Graham, Matthew and Groom, Steven L. and Purdum, Josiah and Wold, Avery},
    month = mar,
    year = {2025},
    pages = {163},
}

@article{gezariTidalDisruptionEvents2021a,
    title = {Tidal {Disruption} {Events}},
    volume = {59},
    issn = {0066-4146},
    url = {https://ui.adsabs.harvard.edu/abs/2021ARA&A..59...21G},
    doi = {10.1146/annurev-astro-111720-030029},
    urldate = {2026-07-13},
    journal = {Annual Review of Astronomy and Astrophysics},
    author = {Gezari, Suvi},
    month = sep,
    year = {2021},
    note = {ADS Bibcode: 2021ARA\&A..59...21G},
    pages = {21--58},
}

@article{charalampopoulosDetailedSpectroscopicStudy2022,
    title = {A detailed spectroscopic study of tidal disruption events},
    volume = {659},
    issn = {0004-6361},
    url = {https://ui.adsabs.harvard.edu/abs/2022A&A...659A..34C},
    doi = {10.1051/0004-6361/202142122},
    urldate = {2025-04-11},
    journal = {Astronomy and Astrophysics},
    author = {Charalampopoulos, P. and Leloudas, G. and Malesani, D. B. and Wevers, T. and Arcavi, I. and Nicholl, M. and Pursiainen, M. and Lawrence, A. and Anderson, J. P. and Benetti, S. and Cannizzaro, G. and Chen, T. -W. and Galbany, L. and Gromadzki, M. and Gutiérrez, C. P. and Inserra, C. and Jonker, P. G. and Müller-Bravo, T. E. and Onori, F. and Short, P. and Sollerman, J. and Young, D. R.},
    month = mar,
    year = {2022},
    note = {ADS Bibcode: 2022A\&A...659A..34C},
    pages = {A34},
}

@article{grotovaPopulationTidalDisruption2025,
    title = {The population of tidal disruption events discovered with {eROSITA}},
    volume = {697},
    copyright = {https://creativecommons.org/licenses/by/4.0},
    issn = {0004-6361, 1432-0746},
    url = {https://www.aanda.org/10.1051/0004-6361/202553669},
    doi = {10.1051/0004-6361/202553669},
    urldate = {2026-07-13},
    journal = {Astronomy \& Astrophysics},
    author = {Grotova, I. and Rau, A. and Baldini, P. and Goodwin, A. J. and Liu, Z. and Merloni, A. and Salvato, M. and Anderson, G. E. and Arcodia, R. and Buchner, J. and Krumpe, M. and Malyali, A. and Masterson, M. and Miller-Jones, J. C. A. and Nandra, K. and Shirley, R.},
    month = may,
    year = {2025},
    pages = {A159},
}

@article{rumbaughExtremeVariabilityQuasars2018,
    title = {Extreme {Variability} {Quasars} from the {Sloan} {Digital} {Sky} {Survey} and the {Dark} {Energy} {Survey}},
    volume = {854},
    issn = {0004-637X},
    url = {https://ui.adsabs.harvard.edu/abs/2018ApJ...854..160R},
    doi = {10.3847/1538-4357/aaa9b6},
    urldate = {2026-07-13},
    journal = {The Astrophysical Journal},
    publisher = {IOP},
    author = {Rumbaugh, N. and Shen, Yue and Morganson, Eric and Liu, Xin and Banerji, M. and McMahon, R. G. and Abdalla, F. B. and Benoit-Lévy, A. and Bertin, E. and Brooks, D. and Buckley-Geer, E. and Capozzi, D. and Carnero Rosell, A. and Carrasco Kind, M. and Carretero, J. and Cunha, C. E. and D'Andrea, C. B. and da Costa, L. N. and DePoy, D. L. and Desai, S. and Doel, P. and Frieman, J. and García-Bellido, J. and Gruen, D. and Gruendl, R. A. and Gschwend, J. and Gutierrez, G. and Honscheid, K. and James, D. J. and Kuehn, K. and Kuhlmann, S. and Kuropatkin, N. and Lima, M. and Maia, M. A. G. and Marshall, J. L. and Martini, P. and Menanteau, F. and Plazas, A. A. and Reil, K. and Roodman, A. and Sanchez, E. and Scarpine, V. and Schindler, R. and Schubnell, M. and Sheldon, E. and Smith, M. and Soares-Santos, M. and Sobreira, F. and Suchyta, E. and Swanson, M. E. C. and Walker, A. R. and Wester, W. and {DES Collaboration}},
    month = feb,
    year = {2018},
    note = {ADS Bibcode: 2018ApJ...854..160R},
    pages = {160},
}

@misc{ricciChanginglookActiveGalactic2022,
    title = {Changing-look {Active} {Galactic} {Nuclei}},
    copyright = {arXiv.org perpetual, non-exclusive license},
    url = {https://arxiv.org/abs/2211.05132},
    doi = {10.48550/ARXIV.2211.05132},
    urldate = {2026-07-13},
    publisher = {arXiv},
    author = {Ricci, Claudio and Trakhtenbrot, Benny},
    year = {2022},
    note = {Version Number: 1},
}

@article{wisemanMultiwavelengthObservationsExtraordinary2023,
    title = {Multiwavelength observations of the extraordinary accretion event {AT2021lwx}},
    volume = {522},
    issn = {0035-8711},
    url = {https://ui.adsabs.harvard.edu/abs/2023MNRAS.522.3992W},
    doi = {10.1093/mnras/stad1000},
    urldate = {2026-07-13},
    journal = {Monthly Notices of the Royal Astronomical Society},
    publisher = {OUP},
    author = {Wiseman, P. and Wang, Y. and Hönig, S. and Castro-Segura, N. and Clark, P. and Frohmaier, C. and Fulton, M. D. and Leloudas, G. and Middleton, M. and Müller-Bravo, T. E. and Mummery, A. and Pursiainen, M. and Smartt, S. J. and Smith, K. and Sullivan, M. and Anderson, J. P. and Acosta Pulido, J. A. and Charalampopoulos, P. and Banerji, M. and Dennefeld, M. and Galbany, L. and Gromadzki, M. and Gutiérrez, C. P. and Ihanec, N. and Kankare, E. and Lawrence, A. and Mockler, B. and Moore, T. and Nicholl, M. and Onori, F. and Petrushevska, T. and Ragosta, F. and Rest, S. and Smith, M. and Wevers, T. and Carini, R. and Chen, T.-W. and Chambers, K. and Gao, H. and Huber, M. and Inserra, C. and Magnier, E. and Makrygianni, L. and Toy, M. and Vincentelli, F. and Young, D. R.},
    month = jul,
    year = {2023},
    note = {ADS Bibcode: 2023MNRAS.522.3992W},
    pages = {3992--4002},
}

@article{petrushevskaRiseFallIronstrong2023,
    title = {The rise and fall of the iron-strong nuclear transient {PS16dtm}},
    volume = {669},
    copyright = {https://creativecommons.org/licenses/by/4.0},
    issn = {0004-6361, 1432-0746},
    url = {https://www.aanda.org/10.1051/0004-6361/202244623},
    doi = {10.1051/0004-6361/202244623},
    urldate = {2026-07-13},
    journal = {Astronomy \& Astrophysics},
    author = {Petrushevska, T. and Leloudas, G. and Ilić, D. and Bronikowski, M. and Charalampopoulos, P. and Jaisawal, G. K. and Paraskeva, E. and Pursiainen, M. and Rakić, N. and Schulze, S. and Taggart, K. and Wedderkopp, C. K. and Anderson, J. P. and De Boer, T. and Chambers, K. and Chen, T. W. and Damljanović, G. and Fraser, M. and Gao, H. and Gomboc, A. and Gromadzki, M. and Ihanec, N. and Maguire, K. and Marčun, B. and Müller-Bravo, T. E. and Nicholl, M. and Onori, F. and Reynolds, T. M. and Smartt, S. J. and Sollerman, J. and Smith, K. W. and Wevers, T. and Wyrzykowski, Ł.},
    month = jan,
    year = {2023},
    pages = {A140},
}

@article{oatesSwiftUVOTDiscovery2024,
    title = {Swift/{UVOT} discovery of {Swift} {J221951}-484240: a {UV} luminous ambiguous nuclear transient},
    volume = {530},
    issn = {0035-8711},
    shorttitle = {Swift/{UVOT} discovery of {Swift} {J221951}-484240},
    url = {https://ui.adsabs.harvard.edu/abs/2024MNRAS.530.1688O},
    doi = {10.1093/mnras/stae795},
    urldate = {2026-07-13},
    journal = {Monthly Notices of the Royal Astronomical Society},
    publisher = {OUP},
    author = {Oates, S. R. and Kuin, N. P. M. and Nicholl, M. and Marshall, F. and Ridley, E. and Boutsia, K. and Breeveld, A. A. and Buckley, D. A. H. and Cenko, S. B. and De Pasquale, M. and Edwards, P. G. and Gromadzki, M. and Gupta, R. and Laha, S. and Morrell, N. and Orio, M. and Pandey, S. B. and Page, M. J. and Page, K. L. and Parsotan, T. and Rau, A. and Schady, P. and Stevens, J. and Brown, P. J. and Evans, P. A. and Gronwall, C. and Kennea, J. A. and Klingler, N. J. and Siegel, M. H. and Tohuvavohu, A. and Ambrosi, E. and Barthelmy, S. D. and Beardmore, A. P. and Bernardini, M. G. and Bonnerot, C. and Campana, S. and Caputo, R. and Ciroi, S. and Cusumano, G. and D'Aì, A. and D'Avanzo, P. and D'Elia, V. and Giommi, P. and Hartmann, D. H. and Krimm, H. A. and Malesani, D. B. and Melandri, A. and Nousek, J. A. and O'Brien, P. T. and Osborne, J. P. and Pagani, C. and Palmer, D. M. and Perri, M. and Racusin, J. L. and Sakamoto, T. and Sbarufatti, B. and Schlieder, J. E. and Tagliaferri, G. and Troja, E. and Xu, D.},
    month = may,
    year = {2024},
    note = {ADS Bibcode: 2024MNRAS.530.1688O},
    pages = {1688--1710},
}

@article{holoienInvestigatingNatureLuminous2022,
    title = {Investigating the {Nature} of the {Luminous} {Ambiguous} {Nuclear} {Transient} {ASASSN}-17jz},
    volume = {933},
    issn = {0004-637X},
    url = {https://ui.adsabs.harvard.edu/abs/2022ApJ...933..196H},
    doi = {10.3847/1538-4357/ac74b9},
    urldate = {2026-07-13},
    journal = {The Astrophysical Journal},
    publisher = {IOP},
    author = {Holoien, Thomas W.-S. and Neustadt, Jack M. M. and Vallely, Patrick J. and Auchettl, Katie and Hinkle, Jason T. and Romero-Cañizales, Cristina and Shappee, Benjamin. J. and Kochanek, Christopher S. and Stanek, K. Z. and Chen, Ping and Dong, Subo and Prieto, Jose L. and Thompson, Todd A. and Brink, Thomas G. and Filippenko, Alexei V. and Zheng, WeiKang and Bersier, David and Bose, Subhash and Burgasser, Adam J. and Channa, Sanyum and de Jaeger, Thomas and Hestenes, Julia and Im, Myungshin and Jeffers, Benjamin and Jun, Hyunsung D. and Lansbury, George and Post, Richard S. and Ross, Timothy W. and Stern, Daniel and Tang, Kevin and Tucker, Michael A. and Valenti, Stefano and Yunus, Sameen and Zhang, Keto D.},
    month = jul,
    year = {2022},
    note = {ADS Bibcode: 2022ApJ...933..196H},
    pages = {196},
}

@article{quimbyRatesSuperluminousSupernovae2013,
    title = {Rates of superluminous supernovae at z ∼ 0.2},
    volume = {431},
    issn = {0035-8711},
    url = {https://ui.adsabs.harvard.edu/abs/2013MNRAS.431..912Q},
    doi = {10.1093/mnras/stt213},
    urldate = {2026-07-13},
    journal = {Monthly Notices of the Royal Astronomical Society},
    publisher = {OUP},
    author = {Quimby, Robert M. and Yuan, Fang and Akerlof, Carl and Wheeler, J. Craig},
    month = may,
    year = {2013},
    note = {ADS Bibcode: 2013MNRAS.431..912Q},
    pages = {912--922},
}

@article{frohmaierCoreCollapseSuperluminous2021,
    title = {From core collapse to superluminous: the rates of massive stellar explosions from the {Palomar} {Transient} {Factory}},
    volume = {500},
    issn = {0035-8711},
    shorttitle = {From core collapse to superluminous},
    url = {https://ui.adsabs.harvard.edu/abs/2021MNRAS.500.5142F},
    doi = {10.1093/mnras/staa3607},
    urldate = {2026-07-13},
    journal = {Monthly Notices of the Royal Astronomical Society},
    publisher = {OUP},
    author = {Frohmaier, C. and Angus, C. R. and Vincenzi, M. and Sullivan, M. and Smith, M. and Nugent, P. E. and Cenko, S. B. and Gal-Yam, A. and Kulkarni, S. R. and Law, N. M. and Quimby, R. M.},
    month = jan,
    year = {2021},
    note = {ADS Bibcode: 2021MNRAS.500.5142F},
    pages = {5142--5158},
}

@article{yaoTidalDisruptionEvent2023,
    title = {Tidal {Disruption} {Event} {Demographics} with the {Zwicky} {Transient} {Facility}: {Volumetric} {Rates}, {Luminosity} {Function}, and {Implications} for the {Local} {Black} {Hole} {Mass} {Function}},
    volume = {955},
    issn = {0004-637X},
    shorttitle = {Tidal {Disruption} {Event} {Demographics} with the {Zwicky} {Transient} {Facility}},
    url = {https://ui.adsabs.harvard.edu/abs/2023ApJ...955L...6Y},
    doi = {10.3847/2041-8213/acf216},
    urldate = {2025-04-11},
    journal = {The Astrophysical Journal},
    publisher = {IOP},
    author = {Yao, Yuhan and Ravi, Vikram and Gezari, Suvi and van Velzen, Sjoert and Lu, Wenbin and Schulze, Steve and Somalwar, Jean J. and Kulkarni, S. R. and Hammerstein, Erica and Nicholl, Matt and Graham, Matthew J. and Perley, Daniel A. and Cenko, S. Bradley and Stein, Robert and Ricarte, Angelo and Chadayammuri, Urmila and Quataert, Eliot and Bellm, Eric C. and Bloom, Joshua S. and Dekany, Richard and Drake, Andrew J. and Groom, Steven L. and Mahabal, Ashish A. and Prince, Thomas A. and Riddle, Reed and Rusholme, Ben and Sharma, Yashvi and Sollerman, Jesper and Yan, Lin},
    month = sep,
    year = {2023},
    note = {ADS Bibcode: 2023ApJ...955L...6Y},
    pages = {L6},
}

@article{grahamUnderstandingExtremeQuasar2017,
    title = {Understanding extreme quasar optical variability with {CRTS} – {I}. {Major} {AGN} flares},
    volume = {470},
    issn = {0035-8711, 1365-2966},
    url = {https://academic.oup.com/mnras/article-lookup/doi/10.1093/mnras/stx1456},
    doi = {10.1093/mnras/stx1456},
    language = {en},
    number = {4},
    urldate = {2026-07-13},
    journal = {Monthly Notices of the Royal Astronomical Society},
    author = {Graham, Matthew J. and Djorgovski, S. G. and Drake, Andrew J. and Stern, Daniel and Mahabal, Ashish A. and Glikman, Eilat and Larson, Steve and Christensen, Eric},
    month = oct,
    year = {2017},
    pages = {4112--4132},
}

@inproceedings{subrayanScaryBarbieExtremely2024,
    title = {Scary {Barbie}: {An} {Extremely} {Energetic}, {Long}-{Duration} {Tidal} {Disruption} {Event} {Candidate} {Without} a {Detected} {Host} {Galaxy} at z = 0.995},
    volume = {243},
    shorttitle = {Scary {Barbie}},
    url = {https://ui.adsabs.harvard.edu/abs/2024AAS...24334906S},
    urldate = {2026-07-14},
    author = {Subrayan, Bhagya and Milisavljevic, Dan and Chornock, Ryan and Margutti, Raffaella and Alexander, Kate and Ramakrishnan, Vandana and Duffell, Paul C. and Dickinson, Danielle A. and Lee, Kyoung-Soo and Giannios, Dimitrios and Lentner, Geoffrey and Linvill, Mark and Garretson, Braden and Graham, Matthew J. and Stern, Daniel and Brethauer, Daniel and Duong, Tien and Jacobson-Galan, Wynn and LeBaron, Natalie and Matthews, David and Sears, Huei and Venkatraman, Padma},
    month = feb,
    year = {2024},
    note = {ADS Bibcode: 2024AAS...24334906S},
    pages = {349.06},
}

@article{bellmZwickyTransientFacility2019,
    title = {The {Zwicky} {Transient} {Facility}: {System} {Overview}, {Performance}, and {First} {Results}},
    volume = {131},
    issn = {0004-6280},
    shorttitle = {The {Zwicky} {Transient} {Facility}},
    url = {https://ui.adsabs.harvard.edu/abs/2019PASP..131a8002B},
    doi = {10.1088/1538-3873/aaecbe},
    urldate = {2025-04-11},
    journal = {Publications of the Astronomical Society of the Pacific},
    publisher = {IOP},
    author = {Bellm, Eric C. and Kulkarni, Shrinivas R. and Graham, Matthew J. and Dekany, Richard and Smith, Roger M. and Riddle, Reed and Masci, Frank J. and Helou, George and Prince, Thomas A. and Adams, Scott M. and Barbarino, C. and Barlow, Tom and Bauer, James and Beck, Ron and Belicki, Justin and Biswas, Rahul and Blagorodnova, Nadejda and Bodewits, Dennis and Bolin, Bryce and Brinnel, Valery and Brooke, Tim and Bue, Brian and Bulla, Mattia and Burruss, Rick and Cenko, S. Bradley and Chang, Chan-Kao and Connolly, Andrew and Coughlin, Michael and Cromer, John and Cunningham, Virginia and De, Kishalay and Delacroix, Alex and Desai, Vandana and Duev, Dmitry A. and Eadie, Gwendolyn and Farnham, Tony L. and Feeney, Michael and Feindt, Ulrich and Flynn, David and Franckowiak, Anna and Frederick, S. and Fremling, C. and Gal-Yam, Avishay and Gezari, Suvi and Giomi, Matteo and Goldstein, Daniel A. and Golkhou, V. Zach and Goobar, Ariel and Groom, Steven and Hacopians, Eugean and Hale, David and Henning, John and Ho, Anna Y. Q. and Hover, David and Howell, Justin and Hung, Tiara and Huppenkothen, Daniela and Imel, David and Ip, Wing-Huen and Ivezić, Željko and Jackson, Edward and Jones, Lynne and Juric, Mario and Kasliwal, Mansi M. and Kaspi, S. and Kaye, Stephen and Kelley, Michael S. P. and Kowalski, Marek and Kramer, Emily and Kupfer, Thomas and Landry, Walter and Laher, Russ R. and Lee, Chien-De and Lin, Hsing Wen and Lin, Zhong-Yi and Lunnan, Ragnhild and Giomi, Matteo and Mahabal, Ashish and Mao, Peter and Miller, Adam A. and Monkewitz, Serge and Murphy, Patrick and Ngeow, Chow-Choong and Nordin, Jakob and Nugent, Peter and Ofek, Eran and Patterson, Maria T. and Penprase, Bryan and Porter, Michael and Rauch, Ludwig and Rebbapragada, Umaa and Reiley, Dan and Rigault, Mickael and Rodriguez, Hector and van Roestel, Jan and Rusholme, Ben and van Santen, Jakob and Schulze, S. and Shupe, David L. and Singer, Leo P. and Soumagnac, Maayane T. and Stein, Robert and Surace, Jason and Sollerman, Jesper and Szkody, Paula and Taddia, F. and Terek, Scott and Van Sistine, Angela and van Velzen, Sjoert and Vestrand, W. Thomas and Walters, Richard and Ward, Charlotte and Ye, Quan-Zhi and Yu, Po-Chieh and Yan, Lin and Zolkower, Jeffry},
    month = jan,
    year = {2019},
    note = {ADS Bibcode: 2019PASP..131a8002B},
    pages = {018002},
}

@article{huberPanSTARRSSurveyTransients2015,
    title = {The {Pan}-{STARRS} {Survey} for {Transients} ({PSST}) - first announcement and public release},
    volume = {7153},
    url = {https://ui.adsabs.harvard.edu/abs/2015ATel.7153....1H},
    urldate = {2025-07-16},
    journal = {The Astronomer's Telegram},
    author = {Huber, M. and Chambers, K. C. and Flewelling, H. and Willman, M. and Primak, N. and Schultz, A. and Gibson, B. and Magnier, E. and Waters, C. and Tonry, J. and Wainscoat, R. J. and Smith, K. W. and Wright, D. and Smartt, S. J. and Foley, R. J. and Jha, S. W. and Rest, A. and Scolnic, D.},
    month = feb,
    year = {2015},
    note = {ADS Bibcode: 2015ATel.7153....1H},
    pages = {1},
}

@incollection{smithInteractingSupernovaeTypes2017,
    address = {Cham},
    title = {Interacting {Supernovae}: {Types} {IIn} and {Ibn}},
    isbn = {978-3-319-21845-8 978-3-319-21846-5},
    shorttitle = {Interacting {Supernovae}},
    url = {http://link.springer.com/10.1007/978-3-319-21846-5_38},
    doi = {10.1007/978-3-319-21846-5_38},
    language = {en},
    urldate = {2026-07-14},
    booktitle = {Handbook of {Supernovae}},
    publisher = {Springer International Publishing},
    author = {Smith, Nathan},
    editor = {Alsabti, Athem W. and Murdin, Paul},
    year = {2017},
    doi = {10.1007/978-3-319-21846-5_38},
    pages = {403--429},
}

@article{kangasZwickyTransientFacility2022,
    title = {The {Zwicky} {Transient} {Facility} phase {I} sample of hydrogen-rich superluminous supernovae without strong narrow emission lines},
    volume = {516},
    issn = {0035-8711},
    url = {https://ui.adsabs.harvard.edu/abs/2022MNRAS.516.1193K},
    doi = {10.1093/mnras/stac2218},
    urldate = {2026-07-14},
    journal = {Monthly Notices of the Royal Astronomical Society},
    publisher = {OUP},
    author = {Kangas, T. and Yan, Lin and Schulze, S. and Fransson, C. and Sollerman, J. and Lunnan, R. and Omand, C. M. B. and Andreoni, I. and Burruss, R. and Chen, T.-W. and Drake, A. J. and Fremling, C. and Gal-Yam, A. and Graham, M. J. and Groom, S. L. and Lezmy, J. and Mahabal, A. A. and Masci, F. J. and Perley, D. and Riddle, R. and Tartaglia, L. and Yao, Y.},
    month = oct,
    year = {2022},
    note = {ADS Bibcode: 2022MNRAS.516.1193K},
    pages = {1193--1218},
}

@article{inserraNatureHydrogenrichSuperluminous2018,
    title = {On the nature of hydrogen-rich superluminous supernovae},
    volume = {475},
    issn = {0035-8711},
    url = {https://ui.adsabs.harvard.edu/abs/2018MNRAS.475.1046I},
    doi = {10.1093/mnras/stx3179},
    urldate = {2026-07-14},
    journal = {Monthly Notices of the Royal Astronomical Society},
    publisher = {OUP},
    author = {Inserra, C. and Smartt, S. J. and Gall, E. E. E. and Leloudas, G. and Chen, T.-W. and Schulze, S. and Jerkstrand, A. and Nicholl, M. and Anderson, J. P. and Arcavi, I. and Benetti, S. and Cartier, R. A. and Childress, M. and Della Valle, M. and Flewelling, H. and Fraser, M. and Gal-Yam, A. and Gutiérrez, C. P. and Hosseinzadeh, G. and Howell, D. A. and Huber, M. and Kankare, E. and Krühler, T. and Magnier, E. A. and Maguire, K. and McCully, C. and Prajs, S. and Primak, N. and Scalzo, R. and Schmidt, B. P. and Smith, M. and Smith, K. W. and Tucker, B. E. and Valenti, S. and Wilman, M. and Young, D. R. and Yuan, F.},
    month = mar,
    year = {2018},
    note = {ADS Bibcode: 2018MNRAS.475.1046I},
    pages = {1046--1072},
}

@article{gezariDiscoveryUltraBrightType2009,
    title = {Discovery of the {Ultra}-{Bright} {Type} {II}-{L} {Supernova} 2008es},
    volume = {690},
    issn = {0004-637X},
    url = {https://ui.adsabs.harvard.edu/abs/2009ApJ...690.1313G},
    doi = {10.1088/0004-637X/690/2/1313},
    urldate = {2026-07-14},
    journal = {The Astrophysical Journal},
    publisher = {IOP},
    author = {Gezari, S. and Halpern, J. P. and Grupe, D. and Yuan, F. and Quimby, R. and McKay, T. and Chamarro, D. and Sisson, M. D. and Akerlof, C. and Wheeler, J. C. and Brown, P. J. and Cenko, S. B. and Rau, A. and Djordjevic, J. O. and Terndrup, D. M.},
    month = jan,
    year = {2009},
    note = {ADS Bibcode: 2009ApJ...690.1313G},
    pages = {1313--1321},
}

@article{kangasSN2023gpwExploring2026,
    title = {{SN} 2023gpw: {Exploring} the diversity and power sources of hydrogen-rich superluminous supernovae},
    volume = {705},
    issn = {0004-6361},
    shorttitle = {{SN} 2023gpw},
    url = {https://ui.adsabs.harvard.edu/abs/2026A&A...705A..52K},
    doi = {10.1051/0004-6361/202555588},
    urldate = {2026-07-14},
    journal = {Astronomy and Astrophysics},
    publisher = {EDP},
    author = {Kangas, T. and Charalampopoulos, P. and Nagao, T. and Yan, L. and Stritzinger, M. and Schulze, S. and Das, K. and Elias-Rosa, N. and Fremling, C. and Perley, D. and Sollerman, J. and Müller-Bravo, T. E. and Galbany, L. and Groom, S. L. and Gutiérrez, C. P. and Kasliwal, M. M. and Kotak, R. and Laher, R. R. and Lundqvist, P. and Mattila, S. and Smith, R.},
    month = jan,
    year = {2026},
    note = {ADS Bibcode: 2026A\&A...705A..52K},
    pages = {A52},
}

@article{angusHubbleSpaceTelescope2016,
    title = {A {Hubble} {Space} {Telescope} survey of the host galaxies of {Superluminous} {Supernovae}},
    volume = {458},
    issn = {0035-8711},
    url = {https://ui.adsabs.harvard.edu/abs/2016MNRAS.458...84A},
    doi = {10.1093/mnras/stw063},
    urldate = {2026-07-14},
    journal = {Monthly Notices of the Royal Astronomical Society},
    publisher = {OUP},
    author = {Angus, C. R. and Levan, A. J. and Perley, D. A. and Tanvir, N. R. and Lyman, J. D. and Stanway, E. R. and Fruchter, A. S.},
    month = may,
    year = {2016},
    note = {ADS Bibcode: 2016MNRAS.458...84A},
    pages = {84--104},
}

@article{leloudasSpectroscopySuperluminousSupernova2015,
    title = {Spectroscopy of superluminous supernova host galaxies. {A} preference of hydrogen-poor events for extreme emission line galaxies},
    volume = {449},
    issn = {0035-8711},
    url = {https://ui.adsabs.harvard.edu/abs/2015MNRAS.449..917L},
    doi = {10.1093/mnras/stv320},
    urldate = {2026-07-14},
    journal = {Monthly Notices of the Royal Astronomical Society},
    publisher = {OUP},
    author = {Leloudas, G. and Schulze, S. and Krühler, T. and Gorosabel, J. and Christensen, L. and Mehner, A. and de Ugarte Postigo, A. and Amorín, R. and Thöne, C. C. and Anderson, J. P. and Bauer, F. E. and Gallazzi, A. and Hełminiak, K. G. and Hjorth, J. and Ibar, E. and Malesani, D. and Morell, N. and Vinko, J. and Wheeler, J. C.},
    month = may,
    year = {2015},
    note = {ADS Bibcode: 2015MNRAS.449..917L},
    pages = {917--932},
}

@article{clelandMetallicityBeatsSSFR2023,
    title = {Metallicity beats {sSFR}: the connection between superluminous supernova host galaxy environments and the importance of metallicity for their production},
    volume = {524},
    issn = {0035-8711},
    shorttitle = {Metallicity beats {sSFR}},
    url = {https://ui.adsabs.harvard.edu/abs/2023MNRAS.524.3559C},
    doi = {10.1093/mnras/stad2118},
    urldate = {2026-07-14},
    journal = {Monthly Notices of the Royal Astronomical Society},
    publisher = {OUP},
    author = {Cleland, Cressida and McGee, Sean L. and Nicholl, Matt},
    month = sep,
    year = {2023},
    note = {ADS Bibcode: 2023MNRAS.524.3559C},
    pages = {3559--3567},
}

@article{schulzeCosmicEvolutionMetal2018,
    title = {Cosmic evolution and metal aversion in superluminous supernova host galaxies},
    volume = {473},
    issn = {0035-8711},
    url = {https://ui.adsabs.harvard.edu/abs/2018MNRAS.473.1258S},
    doi = {10.1093/mnras/stx2352},
    urldate = {2026-07-14},
    journal = {Monthly Notices of the Royal Astronomical Society},
    publisher = {OUP},
    author = {Schulze, S. and Krühler, T. and Leloudas, G. and Gorosabel, J. and Mehner, A. and Buchner, J. and Kim, S. and Ibar, E. and Amorín, R. and Herrero-Illana, R. and Anderson, J. P. and Bauer, F. E. and Christensen, L. and de Pasquale, M. and de Ugarte Postigo, A. and Gallazzi, A. and Hjorth, J. and Morrell, N. and Malesani, D. and Sparre, M. and Stalder, B. and Stark, A. A. and Thöne, C. C. and Wheeler, J. C.},
    month = jan,
    year = {2018},
    note = {ADS Bibcode: 2018MNRAS.473.1258S},
    pages = {1258--1285},
}

@article{lunnanHydrogenpoorSuperluminousSupernovae2014,
    title = {Hydrogen-poor {Superluminous} {Supernovae} and {Long}-duration {Gamma}-{Ray} {Bursts} {Have} {Similar} {Host} {Galaxies}},
    volume = {787},
    issn = {0004-637X},
    url = {https://ui.adsabs.harvard.edu/abs/2014ApJ...787..138L},
    doi = {10.1088/0004-637X/787/2/138},
    urldate = {2026-07-14},
    journal = {The Astrophysical Journal},
    publisher = {IOP},
    author = {Lunnan, R. and Chornock, R. and Berger, E. and Laskar, T. and Fong, W. and Rest, A. and Sanders, N. E. and Challis, P. M. and Drout, M. R. and Foley, R. J. and Huber, M. E. and Kirshner, R. P. and Leibler, C. and Marion, G. H. and McCrum, M. and Milisavljevic, D. and Narayan, G. and Scolnic, D. and Smartt, S. J. and Smith, K. W. and Soderberg, A. M. and Tonry, J. L. and Burgett, W. S. and Chambers, K. C. and Flewelling, H. and Hodapp, K. W. and Kaiser, N. and Magnier, E. A. and Price, P. A. and Wainscoat, R. J.},
    month = jun,
    year = {2014},
    note = {ADS Bibcode: 2014ApJ...787..138L},
    pages = {138},
}

@article{neillExtremeHostsExtreme2011,
    title = {The {Extreme} {Hosts} of {Extreme} {Supernovae}},
    volume = {727},
    issn = {0004-637X},
    url = {https://ui.adsabs.harvard.edu/abs/2011ApJ...727...15N},
    doi = {10.1088/0004-637X/727/1/15},
    urldate = {2026-07-14},
    journal = {The Astrophysical Journal},
    publisher = {IOP},
    author = {Neill, James D. and Sullivan, Mark and Gal-Yam, Avishay and Quimby, Robert and Ofek, Eran and Wyder, Ted K. and Howell, D. Andrew and Nugent, Peter and Seibert, Mark and Martin, D. Christopher and Overzier, Roderik and Barlow, Tom A. and Foster, Karl and Friedman, Peter G. and Morrissey, Patrick and Neff, Susan G. and Schiminovich, David and Bianchi, Luciana and Donas, José and Heckman, Timothy M. and Lee, Young-Wook and Madore, Barry F. and Milliard, Bruno and Rich, R. Michael and Szalay, Alex S.},
    month = jan,
    year = {2011},
    note = {ADS Bibcode: 2011ApJ...727...15N},
    pages = {15},
}

@article{taggartCorecollapseSuperluminousGammaray2021,
    title = {Core-collapse, superluminous, and gamma-ray burst supernova host galaxy populations at low redshift: the importance of dwarf and starbursting galaxies},
    volume = {503},
    issn = {0035-8711},
    shorttitle = {Core-collapse, superluminous, and gamma-ray burst supernova host galaxy populations at low redshift},
    url = {https://ui.adsabs.harvard.edu/abs/2021MNRAS.503.3931T},
    doi = {10.1093/mnras/stab174},
    urldate = {2026-07-14},
    journal = {Monthly Notices of the Royal Astronomical Society},
    publisher = {OUP},
    author = {Taggart, K. and Perley, D. A.},
    month = may,
    year = {2021},
    note = {ADS Bibcode: 2021MNRAS.503.3931T},
    pages = {3931--3952},
}

@article{chatzopoulosHYDROGENPOORCIRCUMStelLARSHELLS2012,
    title = {{HYDROGEN}-{POOR} {CIRCUMS}℡{LAR} {SHELLS} {FROM} {PULSATIONAL} {PAIR}-{INSTABILITY} {SUPERNOVAE} {WITH} {RAPIDLY} {ROTATING} {PROGENITORS}},
    volume = {760},
    issn = {0004-637X},
    url = {https://doi.org/10.1088/0004-637X/760/2/154},
    doi = {10.1088/0004-637X/760/2/154},
    language = {en},
    number = {2},
    urldate = {2026-07-14},
    journal = {The Astrophysical Journal},
    publisher = {The American Astronomical Society},
    author = {Chatzopoulos, E. and Craig Wheeler, J.},
    month = nov,
    year = {2012},
    pages = {154},
}

@article{chevalierSHOCKBREAKOUTDENSE2011,
    title = {{SHOCK} {BREAKOUT} {IN} {DENSE} {MASS} {LOSS}: {LUMINOUS} {SUPERNOVAE}},
    volume = {729},
    issn = {2041-8205},
    shorttitle = {{SHOCK} {BREAKOUT} {IN} {DENSE} {MASS} {LOSS}},
    url = {https://doi.org/10.1088/2041-8205/729/1/L6},
    doi = {10.1088/2041-8205/729/1/L6},
    language = {en},
    number = {1},
    urldate = {2026-07-14},
    journal = {The Astrophysical Journal Letters},
    publisher = {The American Astronomical Society},
    author = {Chevalier, Roger A. and Irwin, Christopher M.},
    month = feb,
    year = {2011},
    pages = {L6},
}

@article{angusSuperluminousSupernovaeDark2019,
    title = {Superluminous supernovae from the {Dark} {Energy} {Survey}},
    volume = {487},
    issn = {0035-8711},
    url = {https://ui.adsabs.harvard.edu/abs/2019MNRAS.487.2215A},
    doi = {10.1093/mnras/stz1321},
    urldate = {2026-07-14},
    journal = {Monthly Notices of the Royal Astronomical Society},
    publisher = {OUP},
    author = {Angus, C. R. and Smith, M. and Sullivan, M. and Inserra, C. and Wiseman, P. and D'Andrea, C. B. and Thomas, B. P. and Nichol, R. C. and Galbany, L. and Childress, M. and Asorey, J. and Brown, P. J. and Casas, R. and Castander, F. J. and Curtin, C. and Frohmaier, C. and Glazebrook, K. and Gruen, D. and Gutierrez, C. and Kessler, R. and Kim, A. G. and Lidman, C. and Macaulay, E. and Nugent, P. and Pursiainen, M. and Sako, M. and Soares-Santos, M. and Thomas, R. C. and Abbott, T. M. C. and Avila, S. and Bertin, E. and Brooks, D. and Buckley-Geer, E. and Burke, D. L. and Carnero Rosell, A. and Carretero, J. and da Costa, L. N. and De Vicente, J. and Desai, S. and Diehl, H. T. and Doel, P. and Eifler, T. F. and Flaugher, B. and Fosalba, P. and Frieman, J. and García-Bellido, J. and Gruendl, R. A. and Gschwend, J. and Hartley, W. G. and Hollowood, D. L. and Honscheid, K. and Hoyle, B. and James, D. J. and Kuehn, K. and Kuropatkin, N. and Lahav, O. and Lima, M. and Maia, M. A. G. and March, M. and Marshall, J. L. and Menanteau, F. and Miller, C. J. and Miquel, R. and Ogando, R. L. C. and Plazas, A. A. and Romer, A. K. and Sanchez, E. and Schindler, R. and Schubnell, M. and Sobreira, F. and Suchyta, E. and Swanson, M. E. C. and Tarle, G. and Thomas, D. and Tucker, D. L. and {DES Collaboration}},
    month = aug,
    year = {2019},
    note = {ADS Bibcode: 2019MNRAS.487.2215A},
    pages = {2215--2241},
}

@article{gomezLuminousSupernovaeUnveiling2022,
    title = {Luminous {Supernovae}: {Unveiling} a {Population} between {Superluminous} and {Normal} {Core}-collapse {Supernovae}},
    volume = {941},
    issn = {0004-637X},
    shorttitle = {Luminous {Supernovae}},
    url = {https://ui.adsabs.harvard.edu/abs/2022ApJ...941..107G},
    doi = {10.3847/1538-4357/ac9842},
    urldate = {2026-07-14},
    journal = {The Astrophysical Journal},
    publisher = {IOP},
    author = {Gomez, Sebastian and Berger, Edo and Nicholl, Matt and Blanchard, Peter K. and Hosseinzadeh, Griffin},
    month = dec,
    year = {2022},
    note = {ADS Bibcode: 2022ApJ...941..107G},
    pages = {107},
}

@article{lunnanHydrogenpoorSuperluminousSupernovae2018,
    title = {Hydrogen-poor {Superluminous} {Supernovae} from the {Pan}-{STARRS1} {Medium} {Deep} {Survey}},
    volume = {852},
    issn = {0004-637X},
    url = {https://doi.org/10.3847/1538-4357/aa9f1a},
    doi = {10.3847/1538-4357/aa9f1a},
    language = {en},
    number = {2},
    urldate = {2026-07-14},
    journal = {The Astrophysical Journal},
    publisher = {The American Astronomical Society},
    author = {Lunnan, R. and Chornock, R. and Berger, E. and Jones, D. O. and Rest, A. and Czekala, I. and Dittmann, J. and Drout, M. R. and Foley, R. J. and Fong, W. and Kirshner, R. P. and Laskar, T. and Leibler, C. N. and Margutti, R. and Milisavljevic, D. and Narayan, G. and Pan, Y.-C. and Riess, A. G. and Roth, K. C. and Sanders, N. E. and Scolnic, D. and Smartt, S. J. and Smith, K. W. and Chambers, K. C. and Draper, P. W. and Flewelling, H. and Huber, M. E. and Kaiser, N. and Kudritzki, R. P. and Magnier, E. A. and Metcalfe, N. and Wainscoat, R. J. and Waters, C. and Willman, M.},
    month = jan,
    year = {2018},
    pages = {81},
}

@article{deciaLightCurvesHydrogenpoor2018,
    title = {Light {Curves} of {Hydrogen}-poor {Superluminous} {Supernovae} from the {Palomar} {Transient} {Factory}},
    volume = {860},
    issn = {0004-637X},
    url = {https://ui.adsabs.harvard.edu/abs/2018ApJ...860..100D},
    doi = {10.3847/1538-4357/aab9b6},
    urldate = {2026-07-14},
    journal = {The Astrophysical Journal},
    publisher = {IOP},
    author = {De Cia, Annalisa and Gal-Yam, A. and Rubin, A. and Leloudas, G. and Vreeswijk, P. and Perley, D. A. and Quimby, R. and Yan, Lin and Sullivan, M. and Flörs, A. and Sollerman, J. and Bersier, D. and Cenko, S. B. and Gal-Yam, M. and Maguire, K. and Ofek, E. O. and Prentice, S. and Schulze, S. and Spyromilio, J. and Valenti, S. and Arcavi, I. and Corsi, A. and Howell, D. A. and Mazzali, P. and Kasliwal, M. M. and Taddia, F. and Yaron, O.},
    month = jun,
    year = {2018},
    note = {ADS Bibcode: 2018ApJ...860..100D},
    pages = {100},
}

@article{gomezTypeSuperluminousSupernova2024,
    title = {The {Type} {I} superluminous supernova catalogue {I}: light-curve properties, models, and catalogue description},
    volume = {535},
    issn = {0035-8711},
    shorttitle = {The {Type} {I} superluminous supernova catalogue {I}},
    url = {https://ui.adsabs.harvard.edu/abs/2024MNRAS.535..471G},
    doi = {10.1093/mnras/stae2270},
    urldate = {2026-07-14},
    journal = {Monthly Notices of the Royal Astronomical Society},
    publisher = {OUP},
    author = {Gomez, Sebastian and Nicholl, Matt and Berger, Edo and Blanchard, Peter K. and Villar, V. Ashley and Rest, Sofia and Hosseinzadeh, Griffin and Aamer, Aysha and Ajay, Yukta and Athukoralalage, Wasundara and Coulter, David C. and Eftekhari, Tarraneh and Fiore, Achille and Franz, Noah and Fox, Ori and Gagliano, Alexander and Hiramatsu, Daichi and Howell, D. Andrew and Hsu, Brian and Karmen, Mitchell and Siebert, Matthew R. and Könyves-Tóth, Réka and Kumar, Harsh and McCully, Curtis and Pellegrino, Craig and Pierel, Justin and Rest, Armin and Wang, Qinan},
    month = nov,
    year = {2024},
    note = {ADS Bibcode: 2024MNRAS.535..471G},
    pages = {471--515},
}

@article{aamerTypeSuperluminousSupernova2025,
    title = {The {Type} {I} superluminous supernova catalogue ─ {II}. {Spectroscopic} evolution in the photospheric phase, velocity measurements, and constraints on diversity},
    volume = {541},
    issn = {0035-8711},
    url = {https://ui.adsabs.harvard.edu/abs/2025MNRAS.541.2674A},
    doi = {10.1093/mnras/staf1113},
    urldate = {2026-07-14},
    journal = {Monthly Notices of the Royal Astronomical Society},
    publisher = {OUP},
    author = {Aamer, Aysha and Nicholl, Matt and Gomez, Sebastian and Berger, Edo and Blanchard, Peter and Anderson, Joseph P. and Angus, Charlotte and Aryan, Amar and Ashall, Chris and Chen, Ting-Wan and Dimitriadis, Georgios and Galbany, Lluís and Gkini, Anamaria and Gromadzki, Mariusz and Gutiérrez, Claudia P. and Hiramatsu, Daichi and Hosseinzadeh, Griffin and Inserra, Cosimo and Kumar, Amit and Kumar, Harsh and Kuncarayakti, Hanindyo and Leloudas, Giorgos and Mazzali, Paolo and Medler, Kyle and Müller-Bravo, Tomás E. and Ramirez, Mauricio and Sankar K, Aiswarya and Schulze, Steve and Singh, Avinash and Sollerman, Jesper and Srivastav, Shubham and Terwel, Jacco H. and Young, David R.},
    month = aug,
    year = {2025},
    note = {ADS Bibcode: 2025MNRAS.541.2674A},
    pages = {2674--2706},
}

@article{hiramatsuTypeIInSupernovae2026a,
    title = {Type {IIn} {Supernovae}. {I}. {Uniform} {Light}-curve {Characterization} and a {Bimodality} in the {Radiated} {Energy} {Distribution}},
    volume = {1005},
    issn = {0004-637X},
    url = {https://ui.adsabs.harvard.edu/abs/2026ApJ..1005...82H},
    doi = {10.3847/1538-4357/ae743c},
    urldate = {2026-07-14},
    journal = {The Astrophysical Journal},
    publisher = {IOP},
    author = {Hiramatsu, Daichi and Berger, Edo and Gomez, Sebastian and Blanchard, Peter K. and Kumar, Harsh and Athukoralalage, Wasundara},
    month = jul,
    year = {2026},
    note = {ADS Bibcode: 2026ApJ..1005...82H},
    pages = {82},
}

@article{jerkstrandTypeIaSupernova2020,
    title = {A type {Ia} supernova at the heart of superluminous transient {SN} 2006gy},
    volume = {367},
    issn = {0036-8075},
    url = {https://ui.adsabs.harvard.edu/abs/2020Sci...367..415J},
    doi = {10.1126/science.aaw1469},
    urldate = {2026-07-14},
    journal = {Science},
    author = {Jerkstrand, Anders and Maeda, Keiichi and Kawabata, Koji S.},
    month = jan,
    year = {2020},
    note = {ADS Bibcode: 2020Sci...367..415J},
    pages = {415--418},
}

@article{nichollSuperluminousSupernovaeDiverse2026,
    title = {Superluminous supernovae: diverse rise times explain diverse spectra},
    volume = {371},
    issn = {0004-640X},
    shorttitle = {Superluminous supernovae},
    url = {https://ui.adsabs.harvard.edu/abs/2026Ap&SS.371....6N},
    doi = {10.1007/s10509-026-04540-0},
    urldate = {2026-07-14},
    journal = {Astrophysics and Space Science},
    publisher = {Springer},
    author = {Nicholl, Matt},
    month = jan,
    year = {2026},
    note = {ADS Bibcode: 2026Ap\&SS.371....6N},
    pages = {6},
}

@article{lucyMassLossHot1970,
    title = {Mass {Loss} by {Hot} {Stars}},
    volume = {159},
    issn = {0004-637X},
    url = {https://ui.adsabs.harvard.edu/abs/1970ApJ...159..879L},
    doi = {10.1086/150365},
    urldate = {2026-07-14},
    journal = {The Astrophysical Journal},
    publisher = {IOP},
    author = {Lucy, L. B. and Solomon, P. M.},
    month = mar,
    year = {1970},
    note = {ADS Bibcode: 1970ApJ...159..879L},
    pages = {879},
}

@article{smithMassLossIts2014,
    title = {Mass {Loss}: {Its} {Effect} on the {Evolution} and {Fate} of {High}-{Mass} {Stars}},
    volume = {52},
    issn = {0066-4146},
    shorttitle = {Mass {Loss}},
    url = {https://ui.adsabs.harvard.edu/abs/2014ARA&A..52..487S},
    doi = {10.1146/annurev-astro-081913-040025},
    urldate = {2026-07-14},
    journal = {Annual Review of Astronomy and Astrophysics},
    author = {Smith, Nathan},
    month = aug,
    year = {2014},
    note = {ADS Bibcode: 2014ARA\&A..52..487S},
    pages = {487--528},
}

@article{woosleyPulsationalPairinstabilitySupernovae2017,
    title = {Pulsational {Pair}-instability {Supernovae}},
    volume = {836},
    issn = {0004-637X},
    url = {https://ui.adsabs.harvard.edu/abs/2017ApJ...836..244W},
    doi = {10.3847/1538-4357/836/2/244},
    urldate = {2026-07-14},
    journal = {The Astrophysical Journal},
    publisher = {IOP},
    author = {Woosley, S. E.},
    month = feb,
    year = {2017},
    note = {ADS Bibcode: 2017ApJ...836..244W},
    pages = {244},
}

@article{laplaceExpansionStrippedenvelopeStars2020,
    title = {The expansion of stripped-envelope stars: {Consequences} for supernovae and gravitational-wave progenitors},
    volume = {637},
    issn = {0004-6361},
    shorttitle = {The expansion of stripped-envelope stars},
    url = {https://ui.adsabs.harvard.edu/abs/2020A&A...637A...6L},
    doi = {10.1051/0004-6361/201937300},
    urldate = {2026-07-14},
    journal = {Astronomy and Astrophysics},
    publisher = {EDP},
    author = {Laplace, E. and Götberg, Y. and de Mink, S. E. and Justham, S. and Farmer, R.},
    month = may,
    year = {2020},
    note = {ADS Bibcode: 2020A\&A...637A...6L},
    pages = {A6},
}

@article{yoonTypeIbIIb2017,
    title = {Type {Ib} and {IIb} {Supernova} {Progenitors} in {Interacting} {Binary} {Systems}},
    volume = {840},
    issn = {0004-637X},
    url = {https://doi.org/10.3847/1538-4357/aa6afe},
    doi = {10.3847/1538-4357/aa6afe},
    language = {en},
    number = {1},
    urldate = {2026-07-14},
    journal = {The Astrophysical Journal},
    publisher = {The American Astronomical Society},
    author = {Yoon, Sung-Chul and Dessart, Luc and Clocchiatti, Alejandro},
    month = apr,
    year = {2017},
    pages = {10},
}

@article{ridleyTimevaryingDoublepeakedEmission2024,
    title = {Time-varying double-peaked emission lines following the sudden ignition of the dormant galactic nucleus {AT2017bcc}},
    volume = {531},
    issn = {0035-8711},
    url = {https://ui.adsabs.harvard.edu/abs/2024MNRAS.531.1905R},
    doi = {10.1093/mnras/stae1129},
    urldate = {2026-07-14},
    journal = {Monthly Notices of the Royal Astronomical Society},
    publisher = {OUP},
    author = {Ridley, E. J. and Nicholl, M. and Ward, C. A. and Blanchard, P. K. and Chornock, R. and Fraser, M. and Gomez, S. and Mattila, S. and Oates, S. R. and Pratten, G. and Runnoe, J. C. and Schmidt, P. and Alexander, K. D. and Gromadzki, M. and Lawrence, A. and Reynolds, T. M. and Smith, K. W. and Wyrzykowski, Ł. and Aamer, A. and Anderson, J. P. and Benetti, S. and Berger, E. and de Boer, T. and Chambers, K. C. and Chen, T.-W. and Gao, H. and Gutiérrez, C. P. and Inserra, C. and Kangas, T. and Leloudas, G. and Magnier, E. A. and Makrygianni, L. and Moore, T. and Müller-Bravo, T. E. and Smartt, S. J. and Sokolovsky, K. V. and Wainscoat, R. and Young, D. R.},
    month = jun,
    year = {2024},
    note = {ADS Bibcode: 2024MNRAS.531.1905R},
    pages = {1905--1930},
}

@article{sanchez-saezSDSS1335+0728Awakening1062024,
    title = {{SDSS1335}+0728: {The} awakening of a ∼106 {M}⊙ black hole},
    volume = {688},
    issn = {0004-6361},
    shorttitle = {{SDSS1335}+0728},
    url = {https://ui.adsabs.harvard.edu/abs/2024A&A...688A.157S},
    doi = {10.1051/0004-6361/202347957},
    urldate = {2026-07-14},
    journal = {Astronomy and Astrophysics},
    publisher = {EDP},
    author = {Sánchez-Sáez, P. and Hernández-García, L. and Bernal, S. and Bayo, A. and Calistro Rivera, G. and Bauer, F. E. and Ricci, C. and Merloni, A. and Graham, M. J. and Cartier, R. and Arévalo, P. and Assef, R. J. and Concas, A. and Homan, D. and Krumpe, M. and Lira, P. and Malyali, A. and Martínez-Aldama, M. L. and Muñoz Arancibia, A. M. and Rau, A. and Bruni, G. and Förster, F. and Pavez-Herrera, M. and Tubín-Arenas, D. and Brightman, M.},
    month = aug,
    year = {2024},
    note = {ADS Bibcode: 2024A\&A...688A.157S},
    pages = {A157},
}

@article{cannizzaroAccretionDiscCooling2021a,
    title = {Accretion disc cooling and narrow absorption lines in the tidal disruption event {AT} 2019dsg},
    volume = {504},
    issn = {0035-8711},
    url = {https://doi.org/10.1093/mnras/stab851},
    doi = {10.1093/mnras/stab851},
    number = {1},
    urldate = {2026-07-14},
    journal = {Monthly Notices of the Royal Astronomical Society},
    author = {Cannizzaro, G and Wevers, T and Jonker, P G and Pérez-Torres, M A and Moldon, J and Mata-Sánchez, D and Leloudas, G and Pasham, D R and Mattila, S and Arcavi, I and Decker French, K and Onori, F and Inserra, C and Nicholl, M and Gromadzki, M and Chen, T-W and Müller-Bravo, T E and Short, P and Anderson, J P and Young, D R and Gendreau, K C and Arzoumanian, Z and Löwenstein, M and Remillard, R and Roy, R and Hiramatsu, D},
    month = jun,
    year = {2021},
    pages = {792--815},
}

@article{trakhtenbrot1ES1927+654AGN2019,
    title = {{1ES} 1927+654: {An} {AGN} {Caught} {Changing} {Look} on a {Timescale} of {Months}},
    volume = {883},
    issn = {0004-637X},
    shorttitle = {{1ES} 1927+654},
    url = {https://ui.adsabs.harvard.edu/abs/2019ApJ...883...94T},
    doi = {10.3847/1538-4357/ab39e4},
    urldate = {2026-07-14},
    journal = {The Astrophysical Journal},
    publisher = {IOP},
    author = {Trakhtenbrot, Benny and Arcavi, Iair and MacLeod, Chelsea L. and Ricci, Claudio and Kara, Erin and Graham, Melissa L. and Stern, Daniel and Harrison, Fiona A. and Burke, Jamison and Hiramatsu, Daichi and Hosseinzadeh, Griffin and Howell, D. Andrew and Smartt, Stephen J. and Rest, Armin and Prieto, Jose L. and Shappee, Benjamin J. and Holoien, Thomas W.-S. and Bersier, David and Filippenko, Alexei V. and Brink, Thomas G. and Zheng, WeiKang and Li, Ruancun and Remillard, Ronald A. and Loewenstein, Michael},
    month = sep,
    year = {2019},
    note = {ADS Bibcode: 2019ApJ...883...94T},
    pages = {94},
}

@article{trakhtenbrotNewClassFlares2019,
    title = {A new class of flares from accreting supermassive black holes},
    volume = {3},
    issn = {2397-3366},
    url = {https://ui.adsabs.harvard.edu/abs/2019NatAs...3..242T},
    doi = {10.1038/s41550-018-0661-3},
    urldate = {2026-07-14},
    journal = {Nature Astronomy},
    author = {Trakhtenbrot, Benny and Arcavi, Iair and Ricci, Claudio and Tacchella, Sandro and Stern, Daniel and Netzer, Hagai and Jonker, Peter G. and Horesh, Assaf and Mejía-Restrepo, Julián Esteban and Hosseinzadeh, Griffin and Hallefors, Valentina and Howell, D. Andrew and McCully, Curtis and Baloković, Mislav and Heida, Marianne and Kamraj, Nikita and Lansbury, George Benjamin and Wyrzykowski, Łukasz and Gromadzki, Mariusz and Hamanowicz, Aleksandra and Cenko, S. Bradley and Sand, David J. and Hsiao, Eric Y. and Phillips, Mark M. and Diamond, Tiara R. and Kara, Erin and Gendreau, Keith C. and Arzoumanian, Zaven and Remillard, Ron},
    month = jan,
    year = {2019},
    note = {ADS Bibcode: 2019NatAs...3..242T},
    pages = {242--250},
}

@article{reuschCandidateTidalDisruption2022,
    title = {Candidate {Tidal} {Disruption} {Event} {AT2019fdr} {Coincident} with a {High}-{Energy} {Neutrino}},
    volume = {128},
    issn = {0031-9007},
    url = {https://ui.adsabs.harvard.edu/abs/2022PhRvL.128v1101R},
    doi = {10.1103/PhysRevLett.128.221101},
    urldate = {2026-07-14},
    journal = {Physical Review Letters},
    publisher = {APS},
    author = {Reusch, Simeon and Stein, Robert and Kowalski, Marek and van Velzen, Sjoert and Franckowiak, Anna and Lunardini, Cecilia and Murase, Kohta and Winter, Walter and Miller-Jones, James C. A. and Kasliwal, Mansi M. and Gilfanov, Marat and Garrappa, Simone and Paliya, Vaidehi S. and Ahumada, Tomás and Anand, Shreya and Barbarino, Cristina and Bellm, Eric C. and Brinnel, Valéry and Buson, Sara and Cenko, S. Bradley and Coughlin, Michael W. and De, Kishalay and Dekany, Richard and Frederick, Sara and Gal-Yam, Avishay and Gezari, Suvi and Giroletti, Marcello and Graham, Matthew J. and Karambelkar, Viraj and Kimura, Shigeo S. and Kong, Albert K. H. and Kool, Erik C. and Laher, Russ R. and Medvedev, Pavel and Necker, Jannis and Nordin, Jakob and Perley, Daniel A. and Rigault, Mickael and Rusholme, Ben and Schulze, Steve and Schweyer, Tassilo and Singer, Leo P. and Sollerman, Jesper and Strotjohann, Nora Linn and Sunyaev, Rashid and van Santen, Jakob and Walters, Richard and Zhang, B. Theodore and Zimmerman, Erez},
    month = jun,
    year = {2022},
    note = {ADS Bibcode: 2022PhRvL.128v1101R},
    pages = {221101},
}

@article{pitikHighenergyNeutrinoEvent2022,
    title = {Is the {High}-energy {Neutrino} {Event} {IceCube}-{200530A} {Associated} with a {Hydrogen}-rich {Superluminous} {Supernova}?},
    volume = {929},
    issn = {0004-637X},
    url = {https://ui.adsabs.harvard.edu/abs/2022ApJ...929..163P},
    doi = {10.3847/1538-4357/ac5ab1},
    urldate = {2026-07-14},
    journal = {The Astrophysical Journal},
    publisher = {IOP},
    author = {Pitik, Tetyana and Tamborra, Irene and Angus, Charlotte R. and Auchettl, Katie},
    month = apr,
    year = {2022},
    note = {ADS Bibcode: 2022ApJ...929..163P},
    pages = {163},
}

@article{nichollExtremelyEnergeticSupernova2020,
    title = {An extremely energetic supernova from a very massive star in a dense medium},
    volume = {4},
    issn = {2397-3366},
    url = {https://ui.adsabs.harvard.edu/abs/2020NatAs...4..893N},
    doi = {10.1038/s41550-020-1066-7},
    urldate = {2026-07-14},
    journal = {Nature Astronomy},
    author = {Nicholl, Matt and Blanchard, Peter K. and Berger, Edo and Chornock, Ryan and Margutti, Raffaella and Gomez, Sebastian and Lunnan, Ragnhild and Miller, Adam A. and Fong, Wen-fai and Terreran, Giacomo and Vigna-Gómez, Alejandro and Bhirombhakdi, Kornpob and Bieryla, Allyson and Challis, Pete and Laher, Russ R. and Masci, Frank J. and Paterson, Kerry},
    month = apr,
    year = {2020},
    note = {ADS Bibcode: 2020NatAs...4..893N},
    pages = {893--899},
}

@article{reynoldsSN2016gsdUnusually2020,
    title = {{SN} 2016gsd: an unusually luminous and linear {Type} {II} supernova with high velocities},
    volume = {493},
    issn = {0035-8711},
    shorttitle = {{SN} 2016gsd},
    url = {https://ui.adsabs.harvard.edu/abs/2020MNRAS.493.1761R},
    doi = {10.1093/mnras/staa365},
    urldate = {2026-07-14},
    journal = {Monthly Notices of the Royal Astronomical Society},
    publisher = {OUP},
    author = {Reynolds, T. M. and Fraser, M. and Mattila, S. and Ergon, M. and Dessart, L. and Lundqvist, P. and Dong, Subo and Elias-Rosa, N. and Galbany, L. and Gutiérrez, C. P. and Kangas, T. and Kankare, E. and Kotak, R. and Kuncarayakti, H. and Pastorello, A. and Rodriguez, O. and Smartt, S. J. and Stritzinger, M. and Tomasella, L. and Chen, Ping and Harmanen, J. and Hosseinzadeh, G. and Howell, D. Andrew and Inserra, C. and Nicholl, M. and Nielsen, M. and Smith, K. and Somero, A. and Tronsgaard, R. and Young, D. R.},
    month = apr,
    year = {2020},
    note = {ADS Bibcode: 2020MNRAS.493.1761R},
    pages = {1761--1781},
}

@article{ivezicLSSTScienceDrivers2019a,
    title = {{LSST}: {From} {Science} {Drivers} to {Reference} {Design} and {Anticipated} {Data} {Products}},
    volume = {873},
    issn = {0004-637X},
    shorttitle = {{LSST}},
    url = {https://doi.org/10.3847/1538-4357/ab042c},
    doi = {10.3847/1538-4357/ab042c},
    language = {en},
    number = {2},
    urldate = {2026-07-14},
    journal = {The Astrophysical Journal},
    publisher = {The American Astronomical Society},
    author = {Ivezi\'{c}, Zeljko and Kahn, Steven M. and Tyson, J. Anthony and Abel, Bob and Acosta, Emily and Allsman, Robyn and Alonso, David and AlSayyad, Yusra and Anderson, Scott F. and Andrew, John and P. Angel, James Roger and Angeli, George Z. and Ansari, Reza and Antilogus, Pierre and Araujo, Constanza and Armstrong, Robert and Arndt, Kirk T. and Astier, Pierre and Aubourg, Éric and Auza, Nicole and Axelrod, Tim S. and Bard, Deborah J. and Barr, Jeff D. and Barrau, Aurelian and Bartlett, James G. and Bauer, Amanda E. and Bauman, Brian J. and Baumont, Sylvain and Bechtol, Ellen and Bechtol, Keith and Becker, Andrew C. and Becla, Jacek and Beldica, Cristina and Bellavia, Steve and Bianco, Federica B. and Biswas, Rahul and Blanc, Guillaume and Blazek, Jonathan and Blandford, Roger D. and Bloom, Josh S. and Bogart, Joanne and Bond, Tim W. and Booth, Michael T. and Borgland, Anders W. and Borne, Kirk and Bosch, James F. and Boutigny, Dominique and Brackett, Craig A. and Bradshaw, Andrew and Brandt, William Nielsen and Brown, Michael E. and Bullock, James S. and Burchat, Patricia and Burke, David L. and Cagnoli, Gianpietro and Calabrese, Daniel and Callahan, Shawn and Callen, Alice L. and Carlin, Jeffrey L. and Carlson, Erin L. and Chandrasekharan, Srinivasan and Charles-Emerson, Glenaver and Chesley, Steve and Cheu, Elliott C. and Chiang, Hsin-Fang and Chiang, James and Chirino, Carol and Chow, Derek and Ciardi, David R. and Claver, Charles F. and Cohen-Tanugi, Johann and Cockrum, Joseph J. and Coles, Rebecca and Connolly, Andrew J. and Cook, Kem H. and Cooray, Asantha and Covey, Kevin R. and Cribbs, Chris and Cui, Wei and Cutri, Roc and Daly, Philip N. and Daniel, Scott F. and Daruich, Felipe and Daubard, Guillaume and Daues, Greg and Dawson, William and Delgado, Francisco and Dellapenna, Alfred and Peyster, Robert de and Val-Borro, Miguel de and Digel, Seth W. and Doherty, Peter and Dubois, Richard and Dubois-Felsmann, Gregory P. and Durech, Josef and Economou, Frossie and Eifler, Tim and Eracleous, Michael and Emmons, Benjamin L. and Neto, Angelo Fausti and Ferguson, Henry and Figueroa, Enrique and Fisher-Levine, Merlin and Focke, Warren and Foss, Michael D. and Frank, James and Freemon, Michael D. and Gangler, Emmanuel and Gawiser, Eric and Geary, John C. and Gee, Perry and Geha, Marla and Gessner, Charles J. B. and Gibson, Robert R. and Gilmore, D. Kirk and Glanzman, Thomas and Glick, William and Goldina, Tatiana and Goldstein, Daniel A. and Goodenow, Iain and Graham, Melissa L. and Gressler, William J. and Gris, Philippe and Guy, Leanne P. and Guyonnet, Augustin and Haller, Gunther and Harris, Ron and Hascall, Patrick A. and Haupt, Justine and Hernandez, Fabio and Herrmann, Sven and Hileman, Edward and Hoblitt, Joshua and Hodgson, John A. and Hogan, Craig and Howard, James D. and Huang, Dajun and Huffer, Michael E. and Ingraham, Patrick and Innes, Walter R. and Jacoby, Suzanne H. and Jain, Bhuvnesh and Jammes, Fabrice and Jee, M. James and Jenness, Tim and Jernigan, Garrett and Jevremović, Darko and Johns, Kenneth and Johnson, Anthony S. and Johnson, Margaret W. G. and Jones, R. Lynne and Juramy-Gilles, Claire and Jurić, Mario and Kalirai, Jason S. and Kallivayalil, Nitya J. and Kalmbach, Bryce and Kantor, Jeffrey P. and Karst, Pierre and Kasliwal, Mansi M. and Kelly, Heather and Kessler, Richard and Kinnison, Veronica and Kirkby, David and Knox, Lloyd and Kotov, Ivan V. and Krabbendam, Victor L. and Krughoff, K. Simon and Kubánek, Petr and Kuczewski, John and Kulkarni, Shri and Ku, John and Kurita, Nadine R. and Lage, Craig S. and Lambert, Ron and Lange, Travis and Langton, J. Brian and Guillou, Laurent Le and Levine, Deborah and Liang, Ming and Lim, Kian-Tat and Lintott, Chris J. and Long, Kevin E. and Lopez, Margaux and Lotz, Paul J. and Lupton, Robert H. and Lust, Nate B. and MacArthur, Lauren A. and Mahabal, Ashish and Mandelbaum, Rachel and Markiewicz, Thomas W. and Marsh, Darren S. and Marshall, Philip J. and Marshall, Stuart and May, Morgan and McKercher, Robert and McQueen, Michelle and Meyers, Joshua and Migliore, Myriam and Miller, Michelle and Mills, David J. and Miraval, Connor and Moeyens, Joachim and Moolekamp, Fred E. and Monet, David G. and Moniez, Marc and Monkewitz, Serge and Montgomery, Christopher and Morrison, Christopher B. and Mueller, Fritz and Muller, Gary P. and Arancibia, Freddy Muñoz and Neill, Douglas R. and Newbry, Scott P. and Nief, Jean-Yves and Nomerotski, Andrei and Nordby, Martin and O’Connor, Paul and Oliver, John and Olivier, Scot S. and Olsen, Knut and O’Mullane, William and Ortiz, Sandra and Osier, Shawn and Owen, Russell E. and Pain, Reynald and Palecek, Paul E. and Parejko, John K. and Parsons, James B. and Pease, Nathan M. and Peterson, J. Matt and Peterson, John R. and Petravick, Donald L. and Petrick, M. E. Libby and Petry, Cathy E. and Pierfederici, Francesco and Pietrowicz, Stephen and Pike, Rob and Pinto, Philip A. and Plante, Raymond and Plate, Stephen and Plutchak, Joel P. and Price, Paul A. and Prouza, Michael and Radeka, Veljko and Rajagopal, Jayadev and Rasmussen, Andrew P. and Regnault, Nicolas and Reil, Kevin A. and Reiss, David J. and Reuter, Michael A. and Ridgway, Stephen T. and Riot, Vincent J. and Ritz, Steve and Robinson, Sean and Roby, William and Roodman, Aaron and Rosing, Wayne and Roucelle, Cecille and Rumore, Matthew R. and Russo, Stefano and Saha, Abhijit and Sassolas, Benoit and Schalk, Terry L. and Schellart, Pim and Schindler, Rafe H. and Schmidt, Samuel and Schneider, Donald P. and Schneider, Michael D. and Schoening, William and Schumacher, German and Schwamb, Megan E. and Sebag, Jacques and Selvy, Brian and Sembroski, Glenn H. and Seppala, Lynn G. and Serio, Andrew and Serrano, Eduardo and Shaw, Richard A. and Shipsey, Ian and Sick, Jonathan and Silvestri, Nicole and Slater, Colin T. and Smith, J. Allyn and Smith, R. Chris and Sobhani, Shahram and Soldahl, Christine and Storrie-Lombardi, Lisa and Stover, Edward and Strauss, Michael A. and Street, Rachel A. and Stubbs, Christopher W. and Sullivan, Ian S. and Sweeney, Donald and Swinbank, John D. and Szalay, Alexander and Takacs, Peter and Tether, Stephen A. and Thaler, Jon J. and Thayer, John Gregg and Thomas, Sandrine and Thornton, Adam J. and Thukral, Vaikunth and Tice, Jeffrey and Trilling, David E. and Turri, Max and Berg, Richard Van and Berk, Daniel Vanden and Vetter, Kurt and Virieux, Francoise and Vucina, Tomislav and Wahl, William and Walkowicz, Lucianne and Walsh, Brian and Walter, Christopher W. and Wang, Daniel L. and Wang, Shin-Yawn and Warner, Michael and Wiecha, Oliver and Willman, Beth and Winters, Scott E. and Wittman, David and Wolff, Sidney C. and Wood-Vasey, W. Michael and Wu, Xiuqin and Xin, Bo and Yoachim, Peter and Zhan, Hu},
    month = mar,
    year = {2019},
    pages = {111},
}

@article{vernetXshooterNewWide2011,
    title = {X-shooter, the new wide band intermediate resolution spectrograph at the {ESO} {Very} {Large} {Telescope}},
    volume = {536},
    issn = {0004-6361},
    url = {https://ui.adsabs.harvard.edu/abs/2011A&A...536A.105V},
    doi = {10.1051/0004-6361/201117752},
    urldate = {2026-07-14},
    journal = {Astronomy and Astrophysics},
    publisher = {EDP},
    author = {Vernet, J. and Dekker, H. and D'Odorico, S. and Kaper, L. and Kjaergaard, P. and Hammer, F. and Randich, S. and Zerbi, F. and Groot, P. J. and Hjorth, J. and Guinouard, I. and Navarro, R. and Adolfse, T. and Albers, P. W. and Amans, J.-P. and Andersen, J. J. and Andersen, M. I. and Binetruy, P. and Bristow, P. and Castillo, R. and Chemla, F. and Christensen, L. and Conconi, P. and Conzelmann, R. and Dam, J. and de Caprio, V. and de Ugarte Postigo, A. and Delabre, B. and di Marcantonio, P. and Downing, M. and Elswijk, E. and Finger, G. and Fischer, G. and Flores, H. and François, P. and Goldoni, P. and Guglielmi, L. and Haigron, R. and Hanenburg, H. and Hendriks, I. and Horrobin, M. and Horville, D. and Jessen, N. C. and Kerber, F. and Kern, L. and Kiekebusch, M. and Kleszcz, P. and Klougart, J. and Kragt, J. and Larsen, H. H. and Lizon, J.-L. and Lucuix, C. and Mainieri, V. and Manuputy, R. and Martayan, C. and Mason, E. and Mazzoleni, R. and Michaelsen, N. and Modigliani, A. and Moehler, S. and Møller, P. and Norup Sørensen, A. and Nørregaard, P. and Péroux, C. and Patat, F. and Pena, E. and Pragt, J. and Reinero, C. and Rigal, F. and Riva, M. and Roelfsema, R. and Royer, F. and Sacco, G. and Santin, P. and Schoenmaker, T. and Spano, P. and Sweers, E. and Ter Horst, R. and Tintori, M. and Tromp, N. and van Dael, P. and van der Vliet, H. and Venema, L. and Vidali, M. and Vinther, J. and Vola, P. and Winters, R. and Wistisen, D. and Wulterkens, G. and Zacchei, A.},
    month = dec,
    year = {2011},
    note = {ADS Bibcode: 2011A\&A...536A.105V},
    pages = {A105},
}

@article{appenzellerSuccessfulCommissioningFORS11998,
    title = {Successful commissioning of {FORS1} - the first optical instrument on the {VLT}.},
    volume = {94},
    issn = {0722-6691},
    url = {https://ui.adsabs.harvard.edu/abs/1998Msngr..94....1A},
    urldate = {2026-07-14},
    journal = {The Messenger},
    author = {Appenzeller, I. and Fricke, K. and Fürtig, W. and Gässler, W. and Häfner, R. and Harke, R. and Hess, H.-J. and Hummel, W. and Jürgens, P. and Kudritzki, R.-P. and Mantel, K.-H. and Meisl, W. and Muschielok, B. and Nicklas, H. and Rupprecht, G. and Seifert, W. and Stahl, O. and Szeifert, T. and Tarantik, K.},
    month = dec,
    year = {1998},
    note = {ADS Bibcode: 1998Msngr..94....1A},
    pages = {1--6},
}

@article{dopitaWideFieldSpectrograph2007,
    title = {The {Wide} {Field} {Spectrograph} ({WiFeS})},
    volume = {310},
    issn = {0004-640X},
    url = {https://ui.adsabs.harvard.edu/abs/2007Ap&SS.310..255D},
    doi = {10.1007/s10509-007-9510-z},
    urldate = {2026-07-14},
    journal = {Astrophysics and Space Science},
    publisher = {Springer},
    author = {Dopita, Michael and Hart, John and McGregor, Peter and Oates, Patrick and Bloxham, Gabe and Jones, Damien},
    month = aug,
    year = {2007},
    note = {ADS Bibcode: 2007Ap\&SS.310..255D},
    pages = {255--268},
}

@article{dopitaWideFieldSpectrograph2010,
    title = {The {Wide} {Field} {Spectrograph} ({WiFeS}): performance and data reduction},
    volume = {327},
    issn = {0004-640X},
    shorttitle = {The {Wide} {Field} {Spectrograph} ({WiFeS})},
    url = {https://ui.adsabs.harvard.edu/abs/2010Ap&SS.327..245D},
    doi = {10.1007/s10509-010-0335-9},
    urldate = {2026-07-14},
    journal = {Astrophysics and Space Science},
    publisher = {Springer},
    author = {Dopita, Michael and Rhee, Jonghwan and Farage, Catherine and McGregor, Peter and Bloxham, Gabe and Green, Anthony and Roberts, Bill and Neilson, Jon and Wilson, Greg and Young, Peter and Firth, Peter and Busarello, Gianni and Merluzzi, Paola},
    month = jun,
    year = {2010},
    note = {ADS Bibcode: 2010Ap\&SS.327..245D},
    pages = {245--257},
}

@article{priceConvertingANU232024,
    title = {Converting the {ANU} 2.3 telescope to fully automated operation},
    volume = {41},
    issn = {1323-3580},
    url = {https://ui.adsabs.harvard.edu/abs/2024PASA...41...57P},
    doi = {10.1017/pasa.2024.75},
    urldate = {2026-07-14},
    journal = {Publications of the Astronomical Society of Australia},
    author = {Price, Ian and Nielsen, Jon and Lidman, Chris and Soon, Jamie and Travouillon, Tony and Sharp, Rob},
    month = sep,
    year = {2024},
    note = {ADS Bibcode: 2024PASA...41...57P},
    pages = {e057},
}

@article{childressPyWiFeSRapidData2014,
    title = {{PyWiFeS}: a rapid data reduction pipeline for the {Wide} {Field} {Spectrograph} ({WiFeS})},
    volume = {349},
    issn = {0004-640X},
    shorttitle = {{PyWiFeS}},
    url = {https://ui.adsabs.harvard.edu/abs/2014Ap&SS.349..617C},
    doi = {10.1007/s10509-013-1682-0},
    urldate = {2026-07-14},
    journal = {Astrophysics and Space Science},
    publisher = {Springer},
    author = {Childress, Michael J. and Vogt, Frédéric P. A. and Nielsen, Jon and Sharp, Robert G.},
    month = feb,
    year = {2014},
    note = {ADS Bibcode: 2014Ap\&SS.349..617C},
    pages = {617--636},
}

@misc{youngSherlockContextualClassification2023,
    title = {Sherlock. {Contextual} classification of astronomical transient sources},
    url = {https://zenodo.org/records/8289325},
    doi = {10.5281/zenodo.8289325},
    urldate = {2026-07-14},
    publisher = {Zenodo},
    author = {Young, David},
    month = aug,
    year = {2023},
}

@ARTICLE{Price:2024pywifes,
       author = {{Price}, Ian and {Nielsen}, Jon and {Lidman}, Chris and {Soon}, Jamie and {Travouillon}, Tony and {Sharp}, Rob},
        title = "{Converting the ANU 2.3 telescope to fully automated operation}",
      journal = {\pasa},
         year = 2024,
        month = sep,
       volume = {41},
          eid = {e057},
        pages = {e057},
          doi = {10.1017/pasa.2024.75},
archivePrefix = {arXiv},
       eprint = {2409.19842},
 primaryClass = {astro-ph.IM},
       adsurl = {https://ui.adsabs.harvard.edu/abs/2024PASA...41...57P}
}

@article{rothXRAYOPTICALFLUXES2016,
    title = {{THE} {X}-{RAY} {THROUGH} {OPTICAL} {FLUXES} {AND} {LINE} {STRENGTHS} {OF} {TIDAL} {DISRUPTION} {EVENTS}},
    volume = {827},
    issn = {0004-637X},
    url = {https://doi.org/10.3847/0004-637X/827/1/3},
    doi = {10.3847/0004-637X/827/1/3},
    language = {en},
    number = {1},
    urldate = {2026-07-14},
    journal = {The Astrophysical Journal},
    publisher = {The American Astronomical Society},
    author = {Roth, Nathaniel and Kasen, Daniel and Guillochon, James and Ramirez-Ruiz, Enrico},
    month = aug,
    year = {2016},
    pages = {3},
}

@article{bonningAccretionDiskTemperatures2007,
    title = {Accretion {Disk} {Temperatures} and {Continuum} {Colors} in {QSOs}},
    volume = {659},
    issn = {0004-637X, 1538-4357},
    url = {https://iopscience.iop.org/article/10.1086/510712},
    doi = {10.1086/510712},
    language = {en},
    number = {1},
    urldate = {2026-07-14},
    journal = {The Astrophysical Journal},
    author = {Bonning, E. W. and Cheng, L. and Shields, G. A. and Salviander, S. and Gebhardt, K.},
    month = apr,
    year = {2007},
    pages = {211--217},
}

@article{firthRisingLightCurves2015,
    title = {The rising light curves of {Type} {Ia} supernovae},
    volume = {446},
    issn = {0035-8711},
    url = {https://ui.adsabs.harvard.edu/abs/2015MNRAS.446.3895F},
    doi = {10.1093/mnras/stu2314},
    urldate = {2026-07-15},
    journal = {Monthly Notices of the Royal Astronomical Society},
    publisher = {OUP},
    author = {Firth, R. E. and Sullivan, M. and Gal-Yam, A. and Howell, D. A. and Maguire, K. and Nugent, P. and Piro, A. L. and Baltay, C. and Feindt, U. and Hadjiyksta, E. and McKinnon, R. and Ofek, E. and Rabinowitz, D. and Walker, E. S.},
    month = feb,
    year = {2015},
    note = {ADS Bibcode: 2015MNRAS.446.3895F},
    pages = {3895--3910},
}

@article{gonzalez-gaitanRisetimeTypeII2015,
    title = {The rise-time of {Type} {II} supernovae},
    volume = {451},
    issn = {0035-8711},
    url = {https://ui.adsabs.harvard.edu/abs/2015MNRAS.451.2212G},
    doi = {10.1093/mnras/stv1097},
    urldate = {2026-07-15},
    journal = {Monthly Notices of the Royal Astronomical Society},
    publisher = {OUP},
    author = {González-Gaitán, S. and Tominaga, N. and Molina, J. and Galbany, L. and Bufano, F. and Anderson, J. P. and Gutierrez, C. and Förster, F. and Pignata, G. and Bersten, M. and Howell, D. A. and Sullivan, M. and Carlberg, R. and de Jaeger, T. and Hamuy, M. and Baklanov, P. V. and Blinnikov, S. I.},
    month = aug,
    year = {2015},
    note = {ADS Bibcode: 2015MNRAS.451.2212G},
    pages = {2212--2229},
}

@article{suzukiSystematicStudyRise2020,
    title = {A {Systematic} {Study} on the {Rise} {Time}-{Peak} {Luminosity} {Relation} for {Bright} {Optical} {Transients} {Powered} by {Wind} {Shock} {Breakout}},
    volume = {899},
    issn = {0004-637X},
    url = {https://ui.adsabs.harvard.edu/abs/2020ApJ...899...56S},
    doi = {10.3847/1538-4357/aba0ba},
    urldate = {2026-07-15},
    journal = {The Astrophysical Journal},
    publisher = {IOP},
    author = {Suzuki, Akihiro and Moriya, Takashi J. and Takiwaki, Tomoya},
    month = aug,
    year = {2020},
    note = {ADS Bibcode: 2020ApJ...899...56S},
    pages = {56},
}




\appendix

\begin{table*}
\centering
\caption{Log of spectroscopic observations of AT\,2025agpz. Phases are given in the rest frame relative to the first DECam detection (MJD 61\,003.19).}
\label{tab:speclog}
\begin{tabular}{lccccccc}
\hline
UT Date & MJD & Phase (d) & Telescope/Instrument & Configuration & Wavelength (\AA) & Resolution ($R$) & Exp. Time (s)\\
\hline
2026 Jan 20 & 61060 & +49 & VLT/FORS2 & 300V+10 + GG435, 1.3\arcsec\ slit & 4450--8650 & 440 & 1200\\

2026 Feb 07 & 61078 & +65 & VLT/FORS2 & 300V+10 + GG435, 1.3\arcsec\ slit & 4450--8650 & 440 &  1200\\

2026 Feb 28 & 61099 & +82 & ANU 2.3\,m/WiFeS & RT560 + B3000/R3000 & 3400--9500 & 3000 & 600\\

2026 Mar 14 & 61113 & +95 & ANU 2.3\,m/WiFeS & RT560 + B3000/R3000 & 3400--9500 & 3000 & 600 \\

2026 Mar 21 & 61120 & +101 & VLT/X-shooter & UVB/VIS/NIR; 1.0/0.9/0.9\arcsec\ slits & 3000--25000 & 5400/8900/5600 & 1400/1463/3$\times$480\\

2026 Mar 28 & 61127 & +107 & ANU 2.3\,m/WiFeS & RT560 + B3000/R3000 & 3400--90500 & 3000 & 600\\

2026 Apr 06 & 61136 & +115 & VLT/FORS2 & 300V+10 + GG435, 1.3\arcsec\ slit & 4450--8650 & 440 & 1200 \\
\hline
\end{tabular}
\end{table*}

\bsp	
\label{lastpage}
\end{document}